\documentclass{article}
\usepackage{e,epsfig}
\begin{document}
\branch{A}
\DOI{123}                       
\idline{A}{1, 1--11}{1}         
\editorial{}{}{}{}              
\newcommand{\pom}{{\rm I\! P}}
\newcommand{\reg}{{\rm I\! R}}
\newcommand{\lsim}{\raisebox{-.6ex}{${\textstyle\stackrel{<}{\sim}}$}}
\newcommand{\gsim}{\raisebox{-.6ex}{${\textstyle\stackrel{>}{\sim}}$}}
\newcommand{\ptmiss}{\mbox{$\not\hspace{-1mm}\pt$}} 
\newcommand{\ptmisso}{\mbox{$\not\hspace{-1mm}\pt^{~outer}$}} 
\newcommand{\ecalP}{\mbox{$\frac{Q^2}{Q^2+M_Z^2}$}} 
\newcommand{\MWtwo}{\mbox{$M^2_{\tiny{W}}$}} 
\newcommand{\MZtwo}{\mbox{$M^2_{\tiny{Z}}$}} 
\newcommand{\MZ}{\mbox{$M_{\tiny{Z}}$}} 
\newcommand{\eplus}{\mbox{$\rm e^+$}} 
\newcommand{\emin}{\mbox{$\rm e^-$}} 
\newcommand{\ee}{\eplus \emin} 
\newcommand{\Jpsi}{\mbox{$J/ \psi$}} 
\def\coll{Collaboration}
\def\etal{et al.}
\title{Lectures on HERA physics\footnote{Delivered at 7th Hellenic School on Elementary Particle Physics, Corfu Summer Institute, September 2001}} 
\author{B. Foster\inst{1,2}}
\institute{H.H. Wills Physics Laboratory, University of Bristol, Tyndall Avenue,
Bristol, BS8 1TL, U.K. \\
e-mail: b.foster@bris.ac.uk
\and
DESY, Notkestrasse 85, 22607 Hamburg, Germany.}
%
%
\maketitle
\begin{abstract}
In these lectures I introduce the basics of HERA physics and give a
survey of the major aspects, discussing in somewhat
more depth the subject of low $x$ physics.

\end{abstract}

%
\section{Introduction}
\label{sec-int}

The study of deep inelastic lepton-proton scattering has produced some of the major underpinnings of the Standard Model. For example, the quark-parton model took shape in the light of the deep inelastic scattering (DIS) 
experiments~\cite{TaylorRS} begun at SLAC in the late 1960s. Going even further back,  
the scattering of energetic ``simple'' $\alpha$ particles from 
the nuclei in a thin gold foil, carried 
out by Geiger and Marsden in
Manchester in 1909, led to the concept of the nuclear 
atom~\cite{pm:21:669,prslon:82:495} and is clearly analogous to deep
inelastic lepton scattering in modern particle physics. 

In my lectures I first gave an overview before covering one particular area in more detail. In the interests of producing a more coherent write-up, I have changed the original order. I first outline the HERA accelerator and detectors before discussing the theoretical techniques used to derive
the basic formulae used in the study of electron-quark scattering. 
Then I give an overview of the main areas of HERA physics, 
going into somewhat more detail in one particular area, 
that of electron-proton scattering
when the interacting quark has a low fraction of the original proton's momentum, so-called ``low-$x$'' physics, and related areas such as diffraction.
Here, recent theoretical developments and model building are changing
our perception and improving our understanding of the very rich phenomenology
arising from the many different but related channels that can be
explored at HERA. I conclude with
a discussion of the recent upgrades to both the accelerator and
experiments in the HERA II programme and the main areas of physics that
they will address. 

\section{Introduction to the HERA machine and experiments}
\label{sec-HERAI}
\begin{figure}[ht]
\begin{center}
\epsfig{file=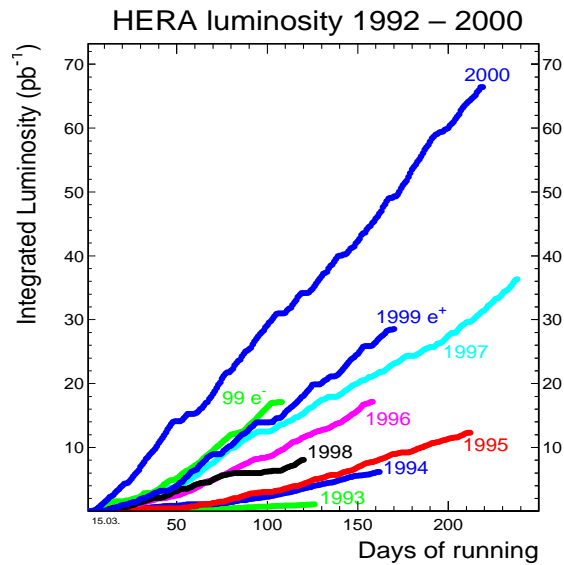,%
      width=8cm,%
      height=8cm%
        }
\end{center}
\caption{The luminosity delivered to the ZEUS detector as a function of
days of the run, shown separately for each year of running. A continuous
improvement in the performance of HERA is evident, except for 1998, in which
electrons were used rather than positrons.}
\label{fig:HERAlumi}
\end{figure}

HERA is a unique facility, colliding beams of electrons or positrons with protons at high energy. The protons are accelerated and stored in a ring of superconducting magnets; until 1998 the protons were accelerated to 820 GeV and subsequently to 920 GeV. The electron or positron ring is normal conducting and
beams are stored at 27.5 GeV. HERA began operation in 1992 and continuously improved its
performance in successive years, as illustrated
in Fig.~\ref{fig:HERAlumi}, which
shows the luminosity delivered to the ZEUS experiment in each
year of running. Because of difficulties with the electron lifetime,
the great majority of data has been taken with positron beams; only
32 pb$^{-1}$ of electron data has been collected by ZEUS compared to
a total of 148  pb$^{-1}$ of positron data. 

\begin{figure}[ht]
\begin{center}
\epsfig{file=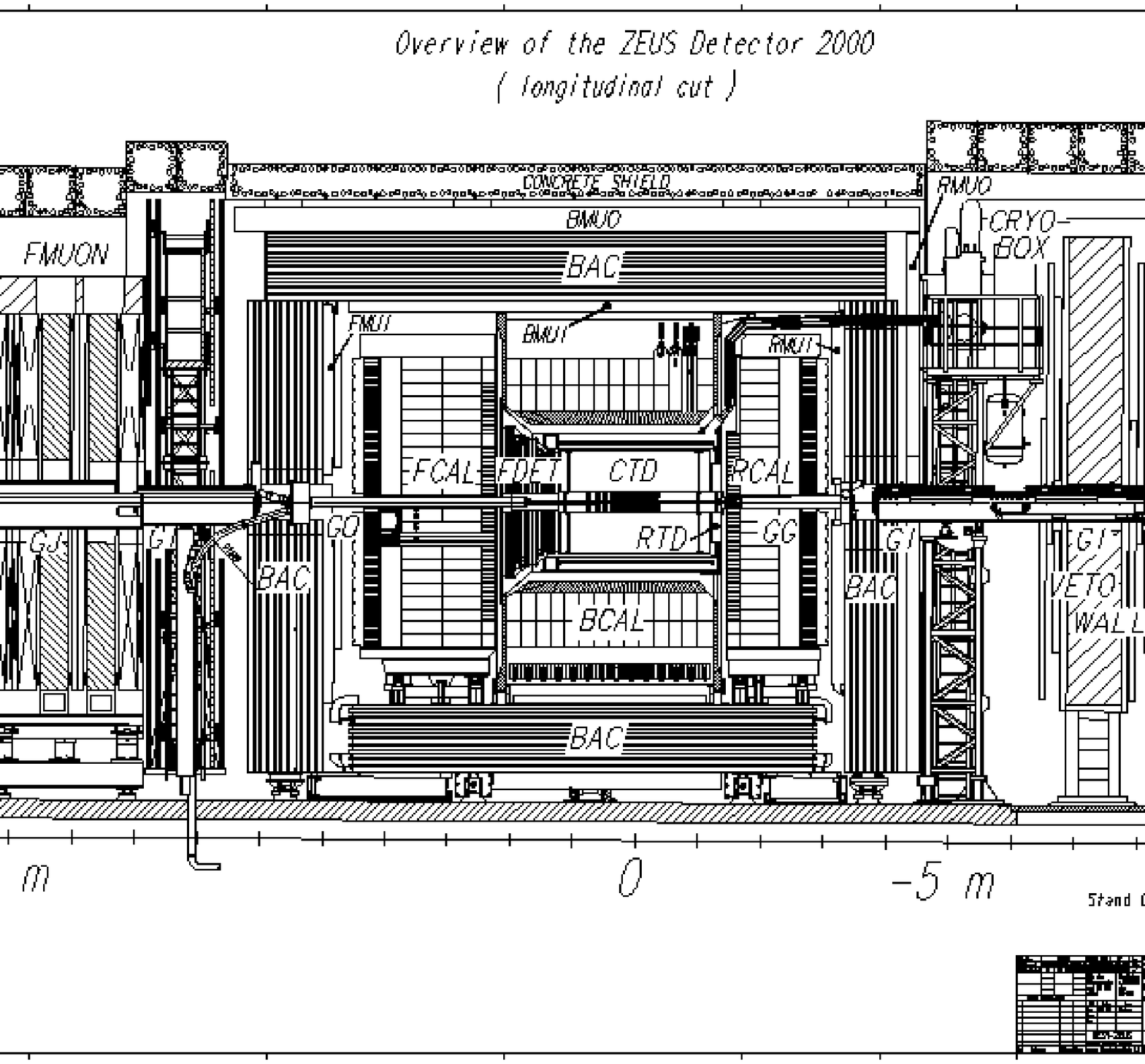,%
      width=12cm,%
      height=8cm,%
      clip=%
       }
\end{center}
\caption{A vertical section through the ZEUS detector. The black cylinders
in the very centre represent the silicon microvertex detector, which was installed in 2000. Surrounding this are the tracking detectors (CTD, FDET and RTD) and surrounding these are the uranium/scintillator calorimeters (FCAL, RCAL
and BCAL). These are surrounded by an iron-scintillator backing calorimeter (BAC) which also acts as a flux return for the superconducting solenoid which surrounds the CTD. Muon chambers (FMUON, with associated toroidal magnets, BMUO and RMUO), a Veto Wall to veto off-momentum protons and a concrete shield complete the detector.}
\label{fig:ZEUSdetector}
\end{figure}

There are two ``general-purpose'' detectors at HERA, H1 and ZEUS. In addition the HERMES experiment uses a gas target to examine polarised electron or positron-polarised proton scattering, and the HERA-B detector, designed to study CP violation in the $B$ sector. The latter two are not discussed further in these lectures due to lack of time. Both H1 and ZEUS have a rather
similar configuration, as far as possible enclosing the full solid angle with tracking detectors surrounded
by calorimetry. Because of the large asymmetry between the proton and positron beam energies, the energy flow
is predominantly in the proton, or ``forward'' direction, so that the detectors are
asymmetric, with thicker calorimetery and a higher density of tracking detectors in the forward
direction. Figure~\ref{fig:ZEUSdetector} shows a diagram of the ZEUS detector, illustrating
that its general structure, with the exception of the more complex forward instrumentation, is very
typical of modern $4\pi$ detectors, such as those at LEP and the Tevatron.   

\begin{figure}[ht]
\begin{center}
\epsfig{file=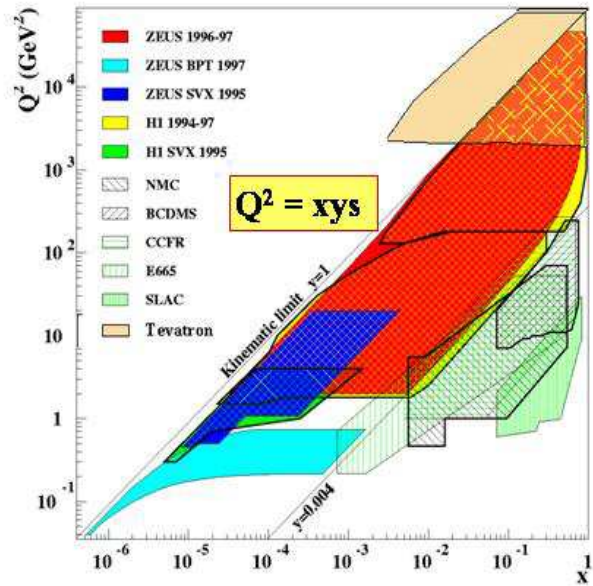,%
      width=8cm,%
      height=8cm%
        }
\end{center}
\caption{The kinematic plane in $x$ and $Q^2$ for experiments probing
the parton distribution of the proton. The regions explored by each
experiment are shown in a variety of shadings as shown in the legend.
Hadron-hadron collisions are also able to measure the proton structure,
predominantly at high $x$ and high $Q^2$.}
\label{fig:q2xkinreg}
\end{figure}

With the advent of HERA, the accessible phase space
in the kinematic invariants $Q^2$ (the virtuality of the exchanged virtual 
photon) and $x$ increased by approximately three orders of magnitude in each variable compared to 
what was available at earlier fixed-target experiments 
(see Fig.~\ref{fig:q2xkinreg}).
This extension in kinematic range has opened up qualitatively new fields of
study, both at high and low $Q^2$.

Since HERA is a high-energy lepton-hadron collider, it clearly gives access to hard
processes in both the strong and electroweak interactions. HERA is very sensitive
to the production of new particles that can be formed by the fusion of
leptons and quarks, e.g.\ leptoquarks or many of the particles
predicted by R-parity-violating supersymmetry. It is also
very sensitive to any small changes in the pattern of the electroweak
interaction predicted by the Standard Model. Since the colliding leptons
are point-like, HERA allows complete control over the conditions of the collision
by varying the $Q^2$, and thereby the size, of the probe. At high $Q^2$, HERA
is a probe of the complex structure of the proton via the point-like coupling 
of the photon. At low $Q^2$, the photon becomes large and evolves its own
complex structure which can be probed using the point-like interactions
between its parton constituents and those of the proton. The hadronic
nature of the photon under these conditions gives rise to hadron-hadron
interactions with large cross sections; by also analysing diffractive interactions, in 
which the proton can be violently struck but remain intact, the rich structure
and phenomenology of the strong interactions can be explored. 

One thing that HERA physics cannot do is be simple. Unlike the situation
in electron-positron annihilation, energy and quantum numbers are transferred
between the colliding particles, each of which has its own conserved quantum numbers of lepton and baryon number. This means
that the single annihilation energy necessary to describe most of 
electron-positron
or high-energy hadron-hadron collisions is insufficient; two invariants are required.
These can be picked from several different possibilities, the most
common of which are $x$ and $Q^2$. Others include 
$W^2$, the square of the energy of the hadronic final
state, $s$, the squared centre-of-mass energy of the electron-proton system, or $y$, the inelasticity,
which in the rest frame of the proton is the energy transferred from the electron to the proton. Only two of these variables are independent; their definitions are
given in Eq.~1 in terms of the initial- and final-state 
four-vectors of the electron, $k$ and $k'$ respectively, and the same quantities, $p$
and $p'$, for the proton. 
\newpage
\begin{eqnarray} 
s &=& (p+k)^2 \nonumber \\ 
Q^2 &=& -q^2 = -(k' - k)^2 \nonumber \\ 
y &=& \frac{p \cdot q}{p \cdot k} \\
\label{eq:invariantdef}
W^2 &=& (p')^2 = (p+q)^2 \nonumber
\end{eqnarray} 
Energy-momentum conservation implies that  
\begin{equation} 
x = \frac{Q^2}{2 p\cdot q} 
\end{equation}
so that, ignoring the masses of the lepton and proton:  
\begin{eqnarray}
y &=& \frac{Q^2}{sx}
\label{eq:yq2sx} \\
W^2 &=& Q^2\frac{1-x}{x} 
\label{eq:w2q2x}
\end{eqnarray} 

Another thing that one cannot expect from HERA is to discover the Higgs. HERA physics is dominated by the collisions of
the lightest quarks with the lightest leptons, so that it is the worst place to look for new particles whose couplings are proportional to mass. Neither can HERA produce new particles with mass close to the centre-of-mass energy unless they have the quantum numbers of leptoquarks; a large proportion of the energy must be carried off by the final-state lepton and baryon in the $t$ channel and is not available for the production of new particles. 

Finally, it cannot be expected that HERA physics will be simple. 
Many of the simplest problems in QCD have already been studied in detail. The
remaining ones, such as the problem of confinement, are of the highest importance but extremely difficult to study in practice, since they manifest
themselves in regimes in which the strong interaction is really strong.

\section{The formalism of Deep Inelastic Scattering}
It is instructive to remind oneself that the basic formalism of DIS
can be relatively easily derived\footnote{In my lecturers I spent considerable time in deriving many of the standard
DIS equations from first principles. My approach was based on the use
of the Mandelstam variables and follows closely the treatment in Chapters
6 -- 10 of Halzen and Martin~\cite{HalzenMartin}. Only an outline of the main
points is reproduced here.} from the QED treatment of spin-$\frac{1}{2}$ --
spin-$\frac{1}{2}$ scattering. It is convenient to work in a frame in which 
the interaction of the virtual photon with the constituents in the proton
can be considered as incoherent, i.e. the characteristic time of the 
$\gamma^* q$ interactions is much shorter than any interactions between the partons. A suitable frame is the infinite-momentum frame of the proton, which 
at HERA can be approximated by the lab. or centre-of-mass frame. 
In such a frame, Lorentz contraction reduces the proton
to a ``pancake'' and time dilation increases the
lifetimes of the fluctuating partons so that the
proton constituents are effectively ``frozen''.
Provided that the quarks have negligible effective mass,
i.e. have small rest mass and are "asymptotically free",
and that $Q^2 \ll k_T^2$, then the interactions can be 
considered incoherent~\cite{BjorkenPaschos}.

The basic process of an electron scattering incoherently from
a quark is now identical to the classic QED calculation of
electron-muon scattering. Writing the initial and final four-momenta
of the electron as $k, k'$ and those of the proton as $p, p'$, the
standard Feyman rules allow us to write the matrix element as
\begin{equation}
{\cal M} = -e^2\overline{u}(k')\gamma^\nu u(k) \frac{1}{q^2}
 \overline{u}(p')\gamma_\nu u(p)
\end{equation}
Using standard trace techniques and ignoring mass terms leads to
\begin{equation}
\left| {\cal M} \right|^2 = \frac{8e^4}{(k-k')^4} \left[(k' \cdot  
p')(k \cdot p) + [(k' \cdot  p)(k \cdot p')\right]
\label{eq:Msq}
\end{equation}
It is very convenient to use the Mandelstam variables, $s,t$ and $u$,
since not only can they easily be evaluated in any frame, but also there
are several useful relations between them and the more usual DIS variables of
Eq.~1 that simplify the algebra. The Mandelstam variables
are defined in terms of the four-vectors as:
\begin{eqnarray}
s &\equiv& (k+p)^2 \simeq 2k\cdot p \simeq 2k' \cdot p', \\
\label{eq:Mandels}
t &\equiv& (k-k')^2 \simeq -2k\cdot k'  \simeq -2p \cdot p', \\
\label{eq:Mandelt}
u &\equiv& (k-p')^2 \simeq -2k\cdot p'  \simeq -2 k' \cdot p,
\label{eq:Mandelu}
\end{eqnarray}
so that Eq.~\ref{eq:Msq} simplifies to become
\begin{equation}
\left| {\cal M} \right|^2 = \frac{2e^4}{t^2}(s^2 + u^2)
\label{eq:Msq-Mand}
\end{equation}
Note that crossing, the replacement of $s$ by $t$
and vice-versa, leads to the well-known formula for $e^+e^- \rightarrow \mu^+\mu^-$.

The matrix element can be converted to a cross section by using the standard
formula for 2 $\rightarrow$ 2 scattering, 
\begin{equation}
\frac{d\sigma}{d\hat{t}} = \frac{1}{16\pi\hat{s}^2}\left| {\cal M} \right|^2, \nonumber
\end{equation} 
to give
\begin{equation}
\frac{d\sigma}{d\hat{t}} = \frac{e^4}{8\pi\hat{s}^2\hat{t}^2}(\hat{s}^2 +
\hat{u}^2),
\label{eq:dsigma-dthat}
\end{equation}
where the hatted variables represent the Mandelstam variables for the subprocess
in question, which for $e-\mu$ scattering are identical to the unhatted
variables, but which for $eq$ scattering are not. Considering now $eq$
scattering, the Mandelstam variables satisfy
\begin{equation}
\hat{s}+\hat{t}+\hat{u} = 0,
\label{eq:stueq0}
\end{equation}
which, assuming that the quark brings a fraction $x$ of the proton's energy
into the $eq$ collision can be written as
\begin{equation}
x(s+u)+ t = 0.
\label{eq:xstueq0}
\end{equation}
Equation~\ref{eq:dsigma-dthat} can be converted into a double-differential
cross section using the appropriate $\delta$ function to give
\begin{equation}
\frac{d^2\sigma}{d\hat{t}d\hat{u}}= \frac{e^4}{8\pi\hat{s}^2\hat{t}^2}(\hat{s}^2 + \hat{u}^2)\delta(\hat{s}+\hat{t}+\hat{u}), \nonumber
\end{equation}
which, using the appropriate Jacobian, can be written in terms of $s,t$ and $u$ as
\begin{equation}
\frac{d^2\sigma}{dtdu}= 
\frac{2\pi \alpha^2 x e_q^2}{s^2t^2}(s^2+u^2)\delta(t+x(s+u)),
\label{eq:d2sigma-dtdu}
\end{equation}
where $\alpha$ is the fine-structure constant, $\alpha = e^2/4\pi$ and $e_q$
is the charge of the struck quark in units of the electron charge. We can then
write the total $ep$ cross section as the incoherent sum of all possible $eq$
scatters, i.e.
\begin{equation}
\left( \frac{d^2\sigma}{dtdu}\right)_{ep \rightarrow eX} =
\sum_i \int f_i(x) \left( \frac{d^2\sigma}{dtdu}\right)_{eq_i \rightarrow eq_i} dx,
\label{eq:epsigma}
\end{equation}
where $f_i(x)$ is the density distribution inside the proton 
of quark $i$ between $x$ and $x + dx$. 

\begin{figure}[ht]
\begin{center}
\epsfig{file=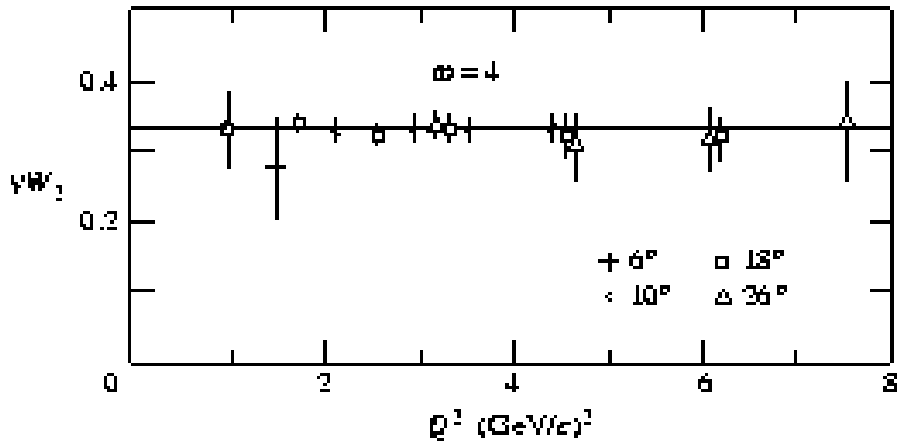,%
     width=12cm,%
      height=6cm,%
      clip=%
        }
\end{center}
\caption{The $\nu W_2 (\equiv F_2)$ structure function at $\omega = 1/x = 4$ as a function of $Q^2$ as measured by the SLAC-MIT group~\cite{SLACscaling}. Data taken at four different scattering angles are shown. All data is consistent with
being independent of $Q^2$.}
\label{fig:SLACscale}
\end{figure}

\begin{figure}[ht]
\begin{center}
\epsfig{file=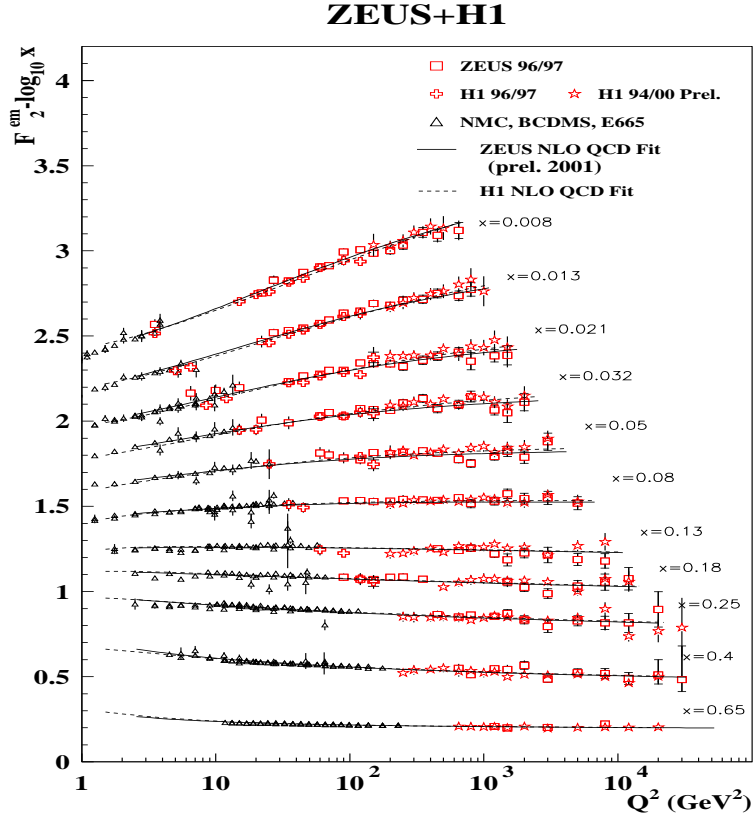,%
      width=11cm,%
      height=11cm%
        }
\end{center}
\caption{The $F_2$ structure function as measured by the H1 and ZEUS
experiments for bins at high $x$ as a function of $Q^2$. The bins centred around $x=0.25$ are where scaling
was originally observed in the SLAC experiments. Clear scaling violation
is observed in the HERA data outside this region,
particularly at lower $x$.}

\label{fig:HERA-F2-xbins}
\end{figure}

We can also treat the overall inclusive DIS process $ep \rightarrow eX$
from first principles using the Feynman rules provided that we take
cognizance of the fact that the proton is not a point-like particle. To
do that, we parameterise the proton vertex contribution to the matrix element
in the most general way possible in terms of a hadronic tensor, $W_{\mu\nu}$,
given by
\begin{equation}
W_{\mu\nu} = \left ( -g_{\mu \nu} \; + \; 
                                  \frac{q_\mu q_\nu}{q^2} \right ) \; 
                                  F_1 (x, Q^2)  +  \frac{\hat{p}_\mu 
                                  \hat{p}_\nu}{p \cdot q} \; F_2 (x, Q^2) 
                                  - \; i \varepsilon_{\mu \nu \alpha \beta}
                                  \; \frac{q^\alpha q^\beta}{2 p \cdot q} \; 
                                  F_3 (x, Q^2)
\label{eq:Wmunu}
\end{equation}
where 
\begin{equation}
\hat{p}_\mu \; = \; p_\mu \; - \; \frac{p \cdot q}{q^2} \; q_\mu, \nonumber 
\end{equation}
and $F_1,F_2$ and $F_3$ are ``structure functions'' describing,
in the most general way compatible with relativistic invariance, the 
unknown structure of the proton. The antisymmetric $\varepsilon$ tensor 
shows that the $F_3$ structure function is parity violating; we will
ignore it for the moment, restricting the discussion to low-$Q^2$
neutral current events where the effects of $W$ and $Z$ exchange can be neglected. Contracting $W_{\mu\nu}$ with the leptonic tensor used to
obtain Eq.~\ref{eq:Msq} leads to 
\begin{equation}
\frac{d^2\sigma}{dt du} = \frac{4\pi\alpha^2}{s^2t^2(s+u)}\left[(s+u)^2
xF_1(x,Q^2) - suF_2(x,Q^2)\right],
\label{eq:d2sigmasf}
\end{equation}
which can now be compared with Eq.~\ref{eq:epsigma}.
Substituting Eq.~\ref{eq:d2sigma-dtdu} and evaluating the integral
using the $\delta$ function leads to 
\begin{equation}
\left( \frac{d^2\sigma}{dtdu}\right)_{ep \rightarrow eX} =
\sum_i f_i(x) \frac{2\pi x  \alpha^2
 e_{q_i^2}}{s^2t^2}\frac{(s^2 + u^2)}{s+u}.
\label{eq:epsigma2}
\end{equation}
Comparison with the right-hand side of Eq.~\ref{eq:d2sigmasf} shows that,
since $s$ and $u$ are continuous variables, the two equations can only
be consistent if the coefficients of the $s^2 + u^2$ and $su$ terms
are equal, i.e.\
\begin{equation}
2xF_1(x,Q^2) = \sum_i f_i(x)xe^2_{q_i} = F_2(x,Q^2).
\label{eq:CallanGross}
\end{equation}
This relation between $F_1$ and $F_2$ is known as the 
Callan-Gross~\cite{Callan-Gross} relation. Equation~\ref{eq:CallanGross}
also implies that $F_2$ is a function of $x$ only, a phenomenon
known as ``scaling''. This was clearly observed in the original SLAC
experiments, as shown in
Fig.~\ref{fig:SLACscale} 
and is also clearly visible in the HERA data shown in 
Fig.~\ref{fig:HERA-F2-xbins} at similar values of $x$.
However, when one looks at other values of $x$, it is clear that scaling
becomes progressively more and more violated.  

The phenomenon of scaling violation is one of the clearest manifestations
of Quantum Chromodynamics and is caused by gluon radiation from the struck
quark. This radiation is accompanied by a transfer of energy to the emitted
gluon, which leads to a shift of the average quark $x$ to lower values.
The emitted gluon can also split into further quark-antiquark pairs
which are also at low $x$. Thus, since the gluon bremsstrahlung
depends on $Q^2$, the cross section develops a strong $Q^2$ and
$x$ dependence. This can conveniently be taken into account by
re-writing Eq.~\ref{eq:d2sigmasf} as a function of $x$ and $Q^2$
using the Jacobian, $(s+u)/x$, as 
\begin{eqnarray}
\frac{d^2\sigma}{dx dQ^2} &=& \frac{2\pi\alpha^2}{xQ^4}\left[2\frac{(s+u)}{s}^2
xF_1(x,Q^2) - 2\frac{u}{s} F_2(x,Q^2)\right] \nonumber \\
&=& \frac{2\pi\alpha^2}{xQ^4}\left[ 2xy^2F_1 + 2(1-y)F_2\right],
\label{eq:xyFs}
\end{eqnarray}
where we have used the identity 
\begin{equation}
y \equiv \frac{p\cdot q}{p \cdot k} = \frac{s+u}{s}. \nonumber
\end{equation}
Rearranging and introducing the longitudinal structure function $F_L = F_2 -
2xF_1$ gives
\begin{eqnarray}
\frac{d^2\sigma}{dx dQ^2} &=& \frac{2\pi\alpha^2}{xQ^4}\left[-y^2F_L + \{1+(1-y)^2\} F_2\right].
\label{eq:FL}
\end{eqnarray}
The longitudinal structure function is zero in the quark-parton model
since the quarks have zero transverse momentum. Gluon bremsstrahlung however
develops a non-zero $p_t$, leading to non-zero values of $F_L$.

Consolidating, we can re-introduce the parity-violating $xF_3$ term and
write down the most general spin-averaged form for the cross-section
as 
\begin{eqnarray}  
\frac{d^2 \sigma} 
{dx dQ^2} & = & \frac{2\pi \alpha^2}{x Q^4} (1 + \delta)   
 \left[   
Y_+ \cdot
 F_2(x, Q^2) \right. \nonumber \\ 
& \; - & \left.  {y^2} F_L(x, Q^2) \pm Y_- \cdot x F_3(x, Q^2) 
\right], 
\label{eq:Fl:sigma} 
\end{eqnarray}
where the $\pm$ before $xF_3$
is taken as positive for electron scattering and negative for 
positron scattering, $Y_{\pm}$ are kinematic factors given by 
\begin{equation}
Y_{\pm} = 1 \pm (1-y)^2,
\label{eq:Y}
\end{equation}
and $\delta$ is the QED radiative correction. 

The $F_2$ structure function can be expressed, in the 
``DIS scheme'' of renormalization~\cite{DISscheme} in
a particularly simple way as
\begin{eqnarray} 
 F_2(x, Q^2) & = & \sum_{i=u,d,s,c,b} A_i(Q^2) \left[ xq_i(x,Q^2) + 
   x\overline{q}_i(x,Q^2). \right]  
\label{eq:F2:qpm}  
\end{eqnarray} 
The parton distributions $q_i(x,Q^2)$ and $\overline{q}_i(x,Q^2)$ refer  
to quarks and antiquarks of type 
$i$. For $Q^2 \ll M_Z^2$, where $M_Z$ is the mass of
the $Z^0$ boson, the quantities
$A_i(Q^2)$ are given by the square of the electric charge of quark or 
antiquark $i$. Similarly,  
\begin{eqnarray} 
 xF_3(x, Q^2) & = & \sum_{i=u,d,s,c,b} B_i(Q^2) \left[ xq_i(x,Q^2) - 
   x\overline{q}_i(x,Q^2). \right]  
\label{eq:xF3:qpm}  
\end{eqnarray} 
The full forms for the $A$ and $B$ terms are:
\begin{eqnarray}
  A_i(Q^2) & = & e_i^2 - 2e_i c_V^e c_V^i P_Z +  
                 ({c_V^e}^2+{c_A^e}^2)({c_V^i}^2+{c_A^i}^2){P_Z}^2, \\ 
  B_i(Q^2) & = & \ \ \ - 2e_ic_A^ec_A^iP_Z +  
                 4c_V^ec_V^ic_A^ec_A^i{P_Z}^2,
\end{eqnarray}
where 
\begin{eqnarray}
P_Z  &=&  \frac{Q^2}{Q^2+M_Z^2}\\ 
c_V^i & = & T_3^i - 2 e_i \sin^2 \theta_W\\ 
c_A^i & = & T_3^i\\ 
T_3^i & = & 
+ \frac{1}{2} ~{\rm for} \ i = \nu, u, c, t\\ 
& = & - \frac{1}{2} ~{\rm for} \ i = e, d, s, b 
\end{eqnarray} 

\section{The HERA DIS data}
\label{sec:HERADIS}

\subsection{The $F_2$ structure function at medium and high $Q^2$}
\label{sec:F2medium}
\begin{figure}[ht]
\begin{center}
\epsfig{file=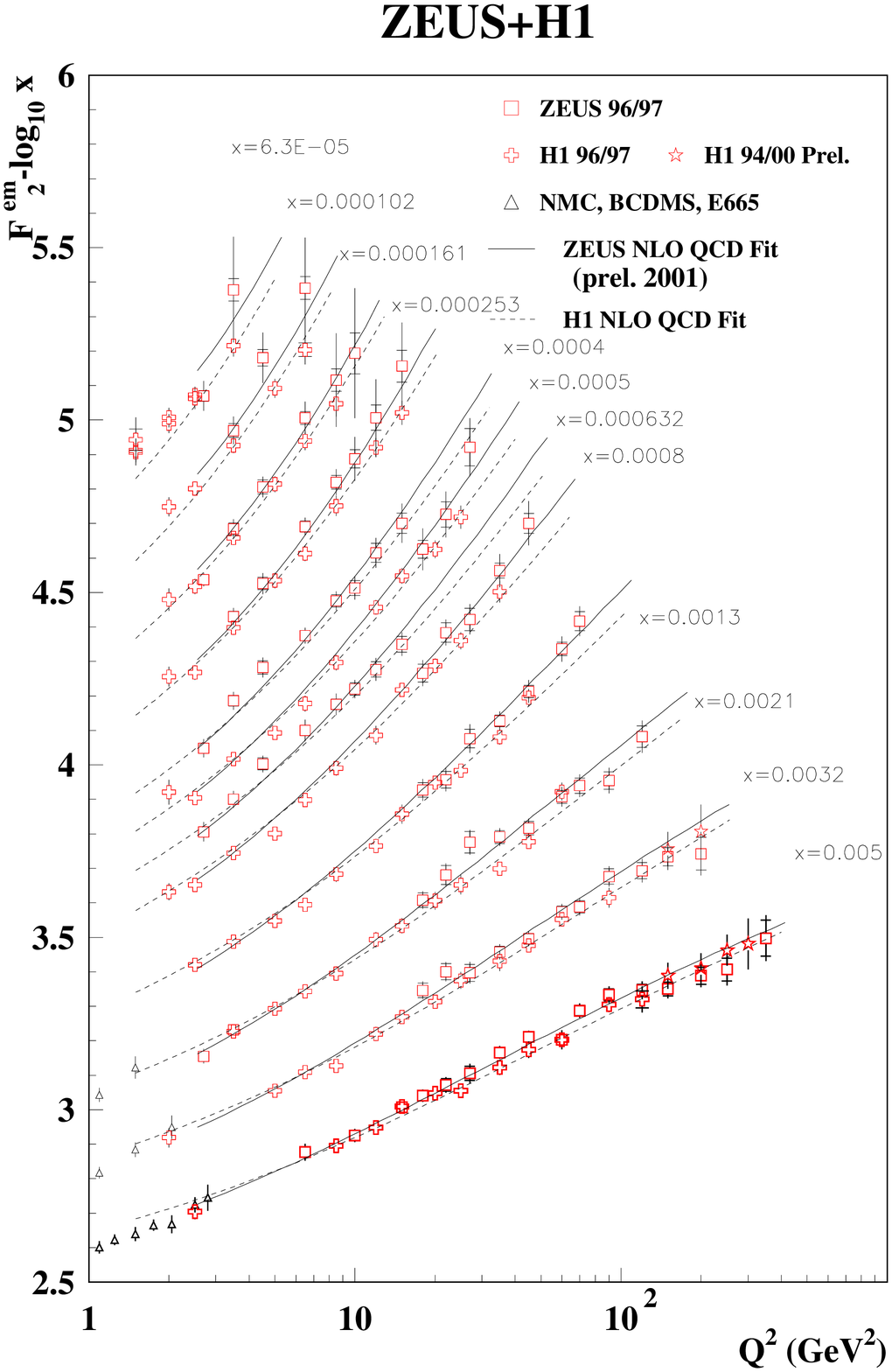,%
      width=11cm,%
      height=11cm%
        }
\end{center}
\caption{The $F_2$ structure function as measured by the H1 and ZEUS
experiments for bins at low $x$ as a function of $Q^2$.}
\label{fig:HERA-F2-xbins2}
\end{figure}

\begin{figure}[ht]
\vskip1.5cm
\begin{center}
\epsfig{file=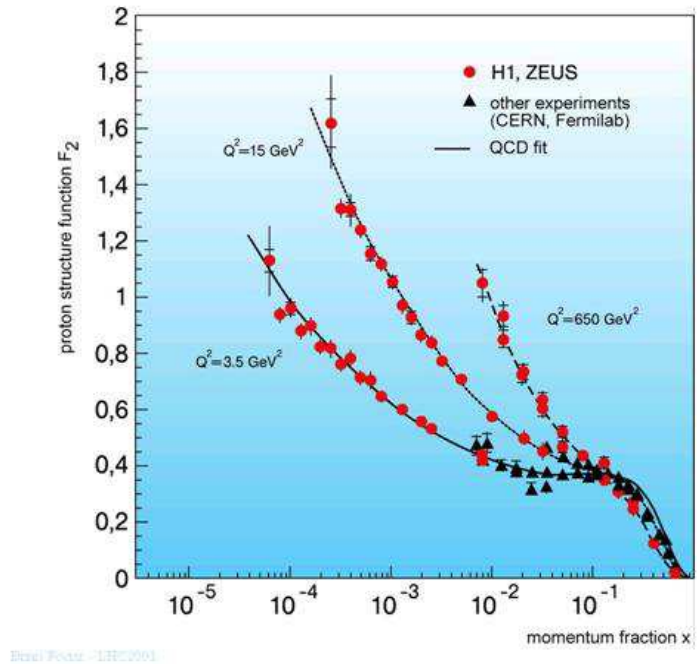,%
      width=9cm,%
      height=9cm%
        }
\end{center}
\caption{H1 and ZEUS data on the $F_2$ structure function shown in three
bins of $Q^2$ as a function of $x$. The steep rise of the structure
function at low $x$ is clearly apparent.}
\label{fig:HERA-F2-q2bins}
\end{figure}

The determination of the structure functions of the proton is a delicate
and painstaking process requiring an excellent understanding of the H1
and ZEUS detector response. This understanding has progressed to the 
extent that the accuracy of the HERA 
data~\cite{H1F2-9697a, H1F2-9697b, ZEUSF2-9697}  is equal to that
of the fixed-target experiments in the kinematic range explored by
them. Of course, the HERA data also extends into a much larger region of
$x$ and $Q^2$ and matches well onto the fixed-target data in the region
of overlap. This has already been exhibited in Fig.~\ref{fig:HERA-F2-xbins}
for the higher $x$ bins, and is further illustrated in 
Fig.~\ref{fig:HERA-F2-xbins2} for the lower $x$ bins, where the large
scaling violations are even more obvious.
The rapid
rise of the structure function at low $x$ can be clearly seen
in Fig.~\ref{fig:HERA-F2-q2bins}, which shows $F_2$ in three
$Q^2$ bins as a function of $x$.

\subsection{Next-to-leading-order QCD fits}
\label{NLOQCDfit}

\begin{figure}[ht]
\begin{center}
\epsfig{file=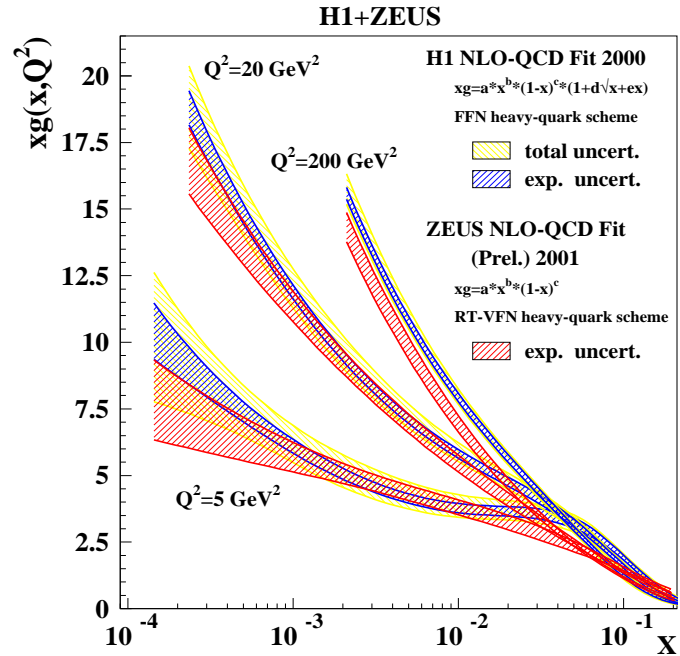,%
      width=9cm,%
      height=9cm%
        }
\end{center}
\caption{The gluon density in the proton as measured by ZEUS (red shaded
band) and H1~\protect{\cite{H1F2-9697b}} (yellow and blue shaded bands) as a function
of $x$ in three bins of $Q^2$. The functional form used by the two
collaborations in the gluon fit is somewhat different
and is shown in the legend.} 
\label{fig:gluon}
\end{figure}

The precise measurement of the proton
structure at low $x$ at HERA is very sensitive both to the details 
of the evolution in QCD of the density of gluons and to the
value of the strong coupling constant, $\alpha_s$, which determines
the probability of gluon emission.
This sensitivity has
been exploited by both ZEUS and H1. Each experiment has made a global QCD fit to
its own data plus some or all of the fixed-target DIS data. There is
reasonably good agreement in general terms between the experiments,
although each experiment has a rather different
fit procedure as well as a different choice of the fixed-target 
data. The quality of the
fits in both experiments is excellent, as demonstrated by
the curves shown in Figs.~\ref{fig:HERA-F2-xbins} - \ref{fig:HERA-F2-q2bins}.
The results for the
density of the gluon are
shown in Fig.~\ref{fig:gluon}. 

\begin{figure}[ht]
\begin{center}
\epsfig{file=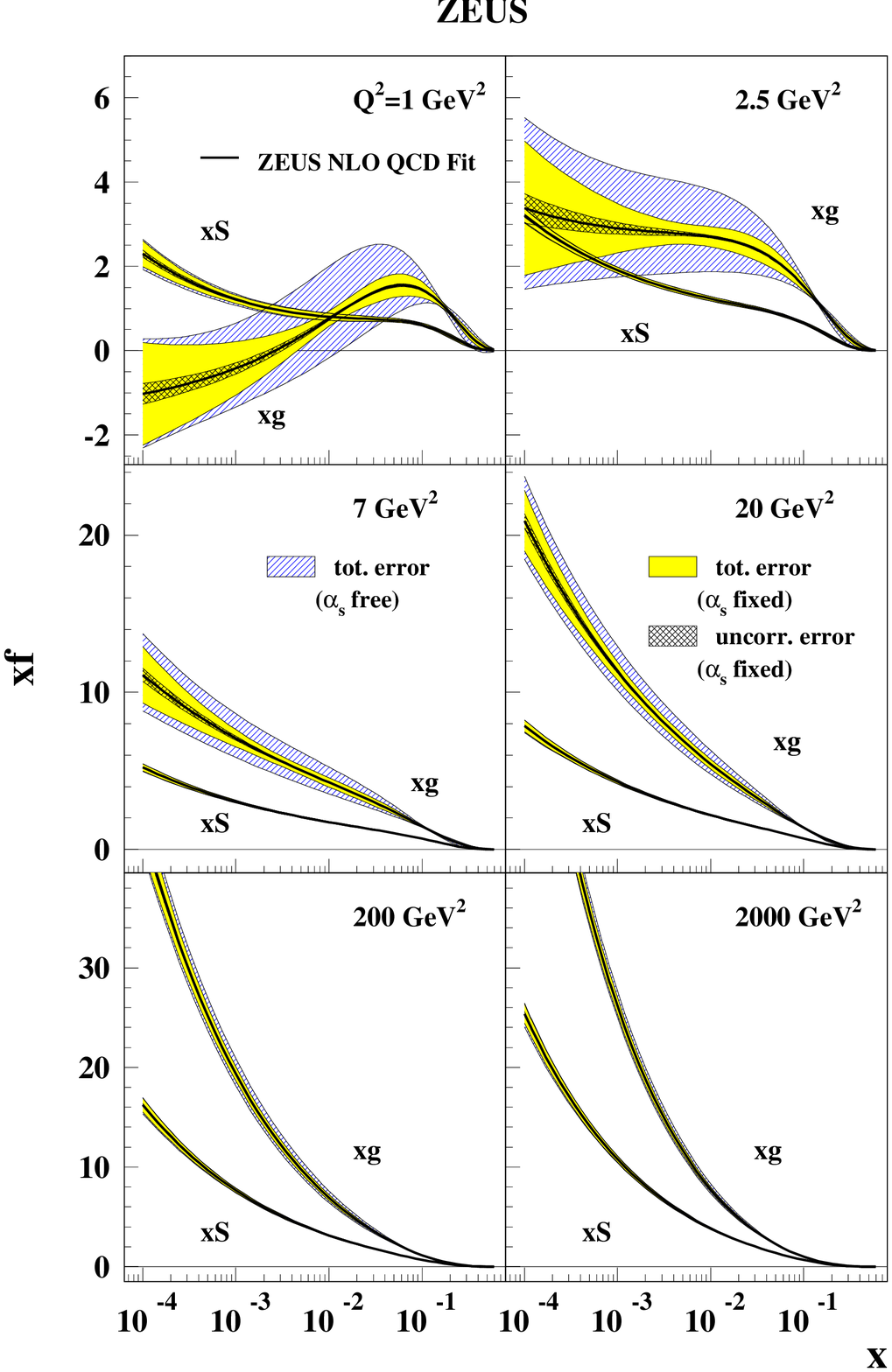,%
      width=11cm,%
      height=11cm%
        }
\end{center}
\caption{The gluon density in the proton compared to that of the
quark-antiquark sea in bins of $Q^2$ as a function of $x$.} 
\label{fig:ZEUSNLO-gandsea}
\end{figure}

I now use the ZEUS NLOQCD fit to illustrate some points of interest. 
Figure~\ref{fig:ZEUSNLO-gandsea} illustrates the evolution of the gluon
density and that of the sea as a function of $Q^2$.
While at medium $Q^2$ the sea density lies below that of the gluon and
follows its shape, at low $Q^2$ it is higher than the gluon. This makes
the normal interpretation, that the sea is driven by gluon splitting,
rather difficult to maintain. 

\begin{figure}[ht]
\begin{center}
\epsfig{file=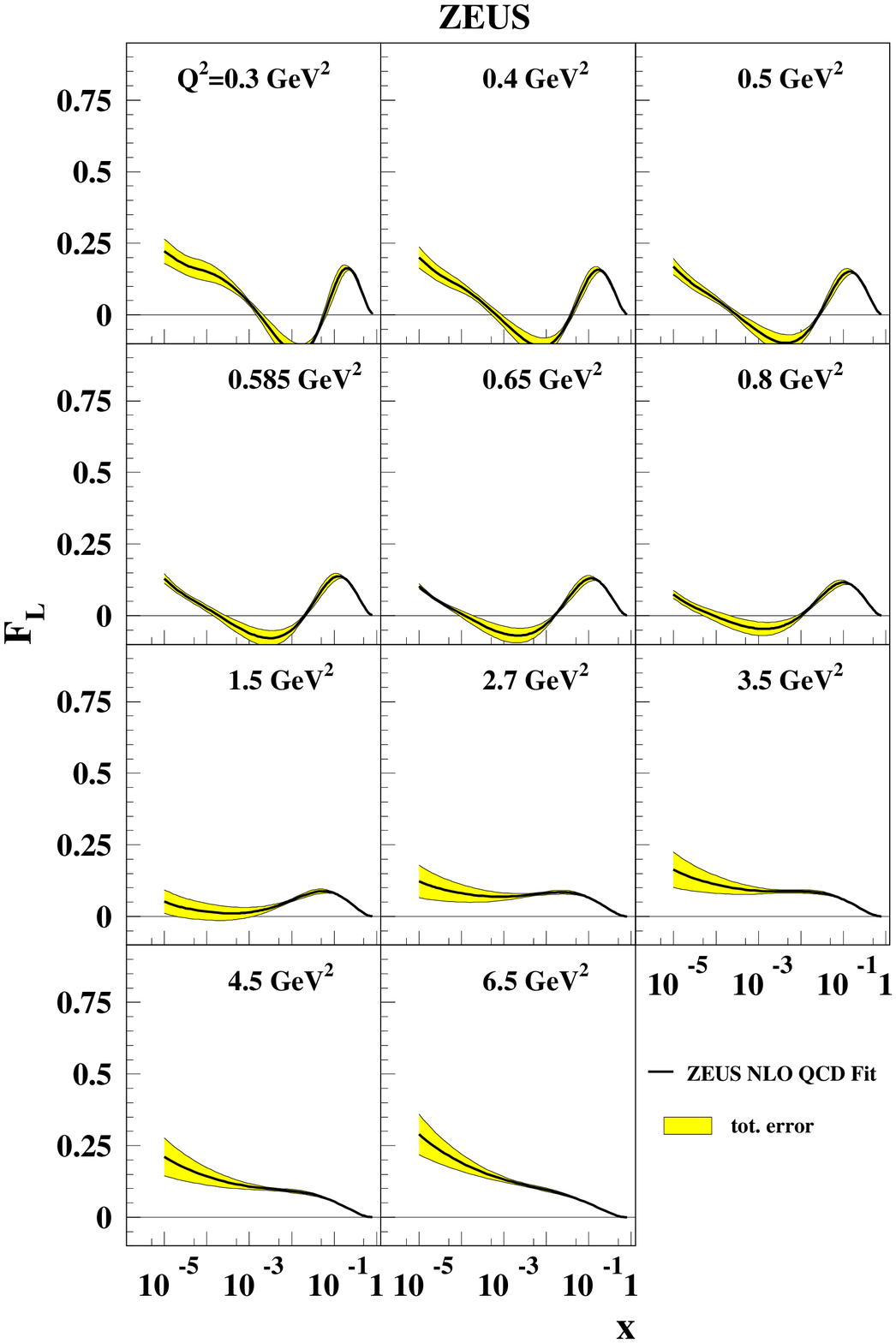,%
      width=11cm,%
      height=11cm%
        }
\end{center}
\caption{The $F_L$ structure function as predicted by the ZEUS
NLOQCD fit as a function of $x$ in bins of $Q^2$.} 
\label{fig:FL}
\end{figure}

\begin{figure}[ht]
\begin{center}
\epsfig{file=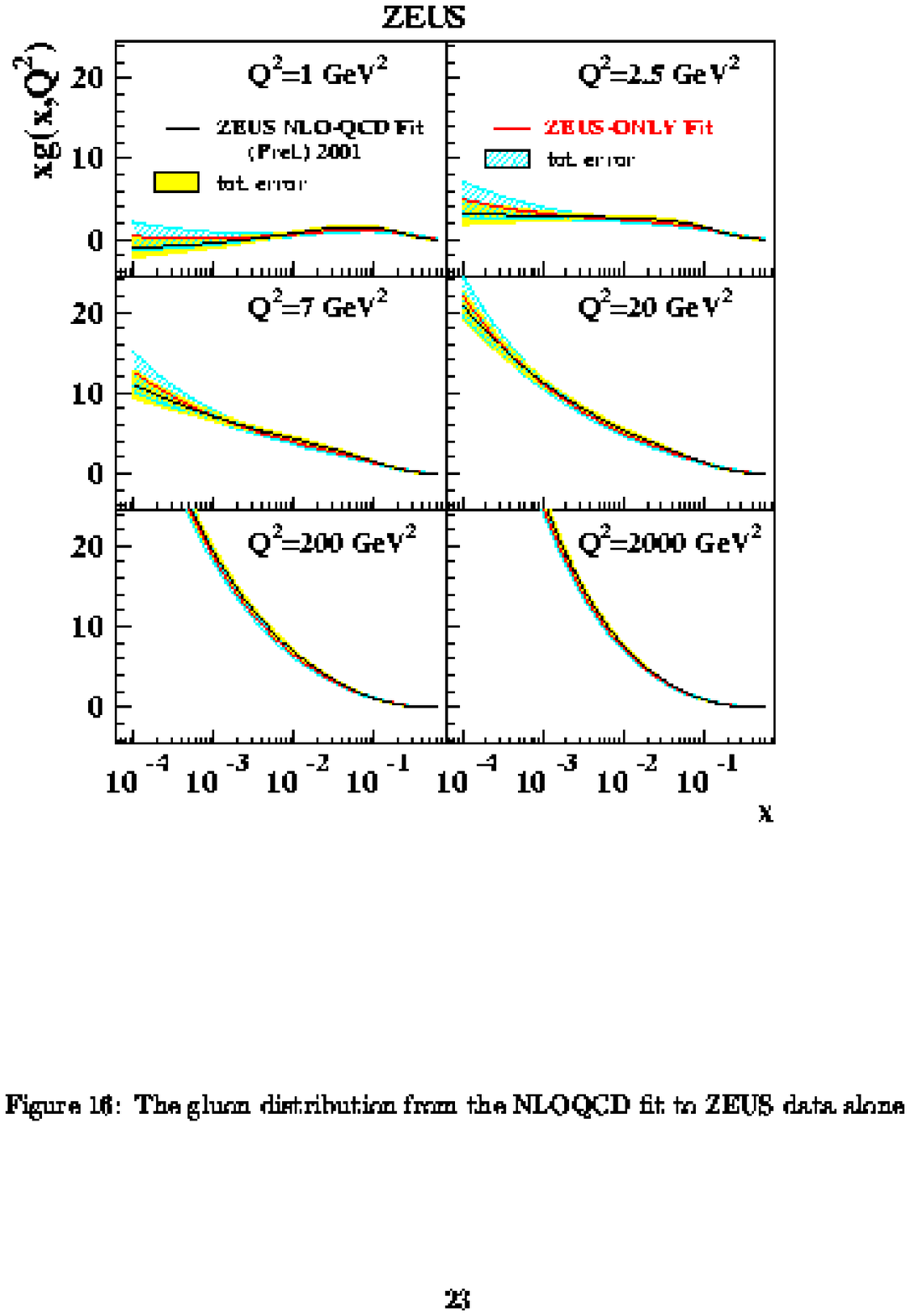,%
      width=10cm,%
      height=10cm,%
      clip=%
         }
\end{center}
\caption{The gluon density as a function of $x$ in $Q^2$ bins from an
NLOQCD fit using only ZEUS data.} 
\label{fig:NLOQCD_Zonly}
\end{figure}

Figure~\ref{fig:FL} shows that $F_L$ also begins to behave strangely
at low $Q^2$, becoming very flat and at the lowest values of $Q^2$
becoming negative, although the size of the uncertainties still allow
it in principle to remain positive. The gluon density, which
is directly related to $F_L$ in QCD, certainly becomes negative, which
is somewhat difficult to interpret in QCD. However, it is not an observable,
whereas $F_L$ is, so that the tendency for $F_L$ to become negative at
low $Q^2$ implies a break-down in the QCD paradigm. We will return to
this discussion in Section~\ref{sec:global}. 
The quality of the HERA data is now so high that it alone can give
constraints on the parton densities scarcely less good than the fits that include
also fixed-target and other data. However, it is necessary to make

some simplifying assumptions, particularly for the high-$x$ valence
behaviour, in such fits. The results are illustrated in Fig.~\ref{fig:NLOQCD_Zonly},
which shows the gluon density arising from such a fit, with its associated uncertainty band,
in comparison with the result from the standard fit; the general behaviour is similar, with a somewhat larger uncertainty, particularly at low $Q^2$. 
 
\subsection{The determination of $\alpha_s$ at HERA}
\label{sec:alphas}
 
\begin{figure}[ht]
\begin{center}
\epsfig{file=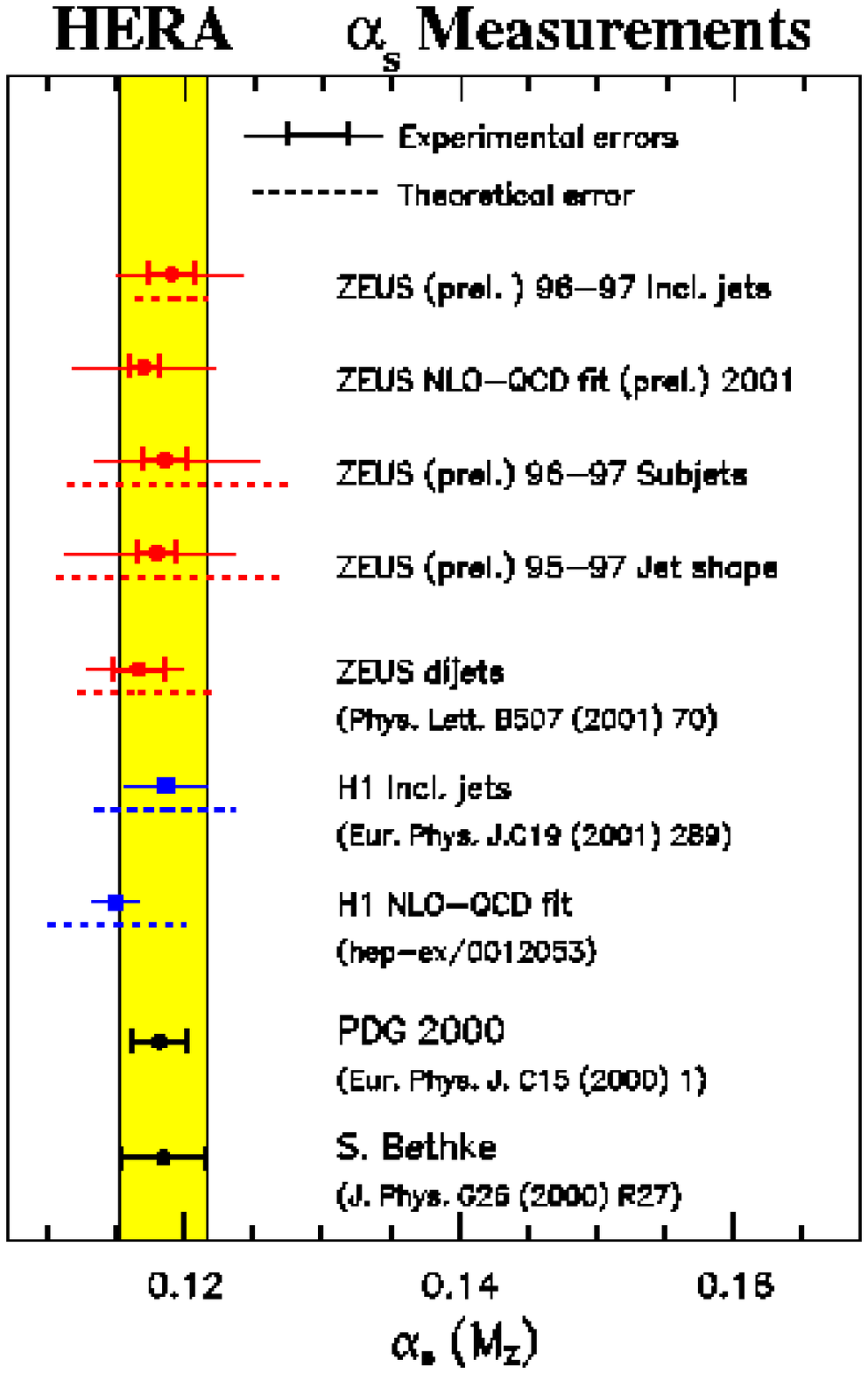,%
      width=9cm,%
      height=9cm,%
      clip=%
          }
\end{center}
\caption{Values of the strong coupling constant as determined at HERA. Each
different measurement is displaced vertically for ease of visibility;
each value arises from a different method as briefly indicated in the
legend. The reference for published results is shown below the method
label. The world
average as calculated by the Particle Data Group\protect{\cite{PDG}} and by Bethke~\protect{\cite{Bethke}} are
shown at the bottom of the figure.}
\label{fig:alphas}
\end{figure}

Another output of the NLOQCD fit is a
value of $\alpha_s$; the results from the two experiments are shown in
Fig.~\ref{fig:alphas}, labelled as ``NLO-QCD fit''. The value obtained 
by ZEUS is
\[
\alpha_s(M_Z^2) = 0.1166 \pm 0.008 \pm 0.0032 \pm 0.0018,
\]
where the uncertainties derive from statistical and other uncorrelated experimental uncertainties, correlated experimental uncertainties, normalisation uncertainties and the error related to omissions and simplifications in the NLOQCD model. The value H1 obtain from their fit to their and the BCDMS data
is
\[
\alpha_s(M_Z^2) = 0.1150 \pm 0.0017 ^{+0.0009}_{-0.0005} \pm 0.005,
\]
where the first error source takes account of all experimental uncertainties, the second takes account of the construction of the NLOQCD model and the final
uncertainty results from the variation in the factorisation and renormalisation
scale. The values of $\alpha_s$ obtained by the two experiments are in good agreement. 

Also shown in Fig.~\ref{fig:alphas} are a variety
of other high-precision measurements of $\alpha_s$ that can be
made at HERA using a variety of techniques. These include classic methods
such as the rate of dijet + proton-remnant production compared to that
of single jet plus remnant, the subjet-multiplicity evolution
inside jets and the shape of jets. Many of these give excellent
precision, comparable to the world average~\cite{PDG, Bethke}. 
The dominant uncertainty
is usually theoretical and arises from the lack of predictions at next-to-next-to-leading order. 

\subsection{The charm-quark structure function, $F_2^c$}
\label{sec:f2charm}

\begin{figure}[ht]
\begin{center}
\epsfig{file=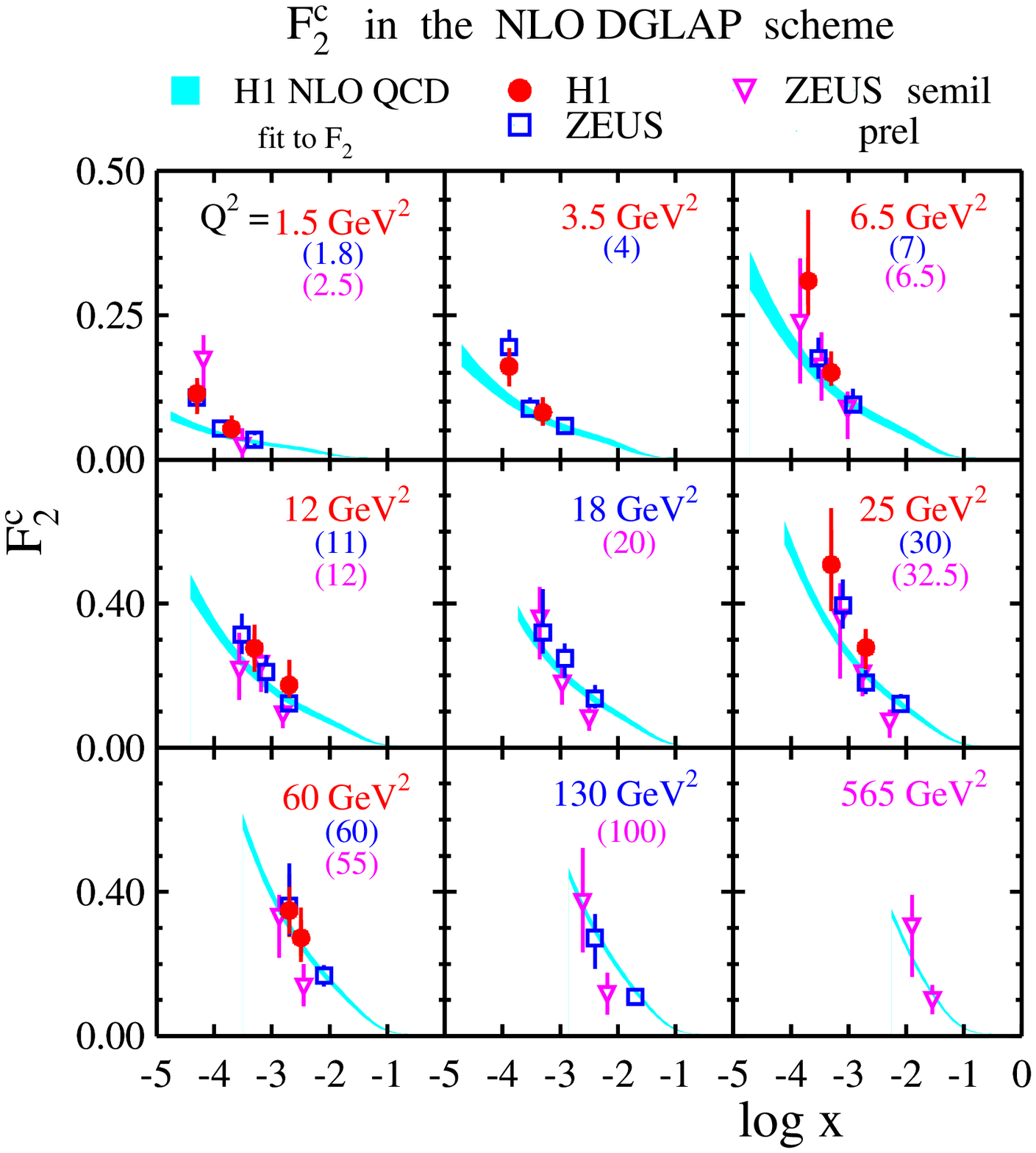,%
      width=8cm,%
      height=10cm%
        }
\end{center}
\caption{Values of the charm structure function, $F_2^c$ from the H1
and ZEUS experiments in bins of $Q^2$ as a function of $\ln x$. The
blue squares show the ZEUS using the $D^* \rightarrow K\pi\pi_s$ decay
mode, whereas the purple triangles show a ZEUS determination using the
semileptonic decay of the $D$. The 
shaded curves show the predictions from the NLO QCD fit to the inclusive
$F_2$ data by H1.}
\label{fig-H1ZEUS-F2C}
\end{figure}

In addition to the fully inclusive structure functions discussed above, both
ZEUS and H1 can identify that fraction of $F_2$ that arises from charm production, $F_2^c$. This is achieved by looking for the decay mode
$D^* \rightarrow D\pi$, in which, because the mass difference between
the $D^*$ and $D$ is only just larger than the pion mass, the daughter pion has a very small momentum. The mass difference between the $D^*$ candidate and the $D$ candidate can therefore be measured very accurately, allowing sufficient suppression of the combinatorial background that the charm signal can be cleanly identified. The structure function can then be unfolded from the measured 
differential cross section using models to correct for the unmeasured parts of
the phase space. The measurements of the semi-inclusive charm structure function, $F_2^c$, made
by both experiments~\cite{ZEUS-F2c,H1-F2c} using this technique are shown in Fig.~\ref{fig-H1ZEUS-F2C}. 

The data is still of rather limited statistical precision. Since the charm quark is produced predominantly via boson-gluon fusion,
$F^c_2$ is driven by the gluon density and thus rises steeply as $x$ falls.
\subsection{The $F_L$ structure function}
\label{sec:Fl}
Since in principle both $F_2$ and $F_L$
are unknown functions that depend on $x$ and $Q^2$, the only way in which they
can be separately determined is to measure the differential cross section at
fixed $x,Q^2$ and at different values of $y$, since as shown in 
Eq.~\ref{eq:Fl:sigma}, the effect of $F_L$ is weighted by $y^2$
whereas $F_2$ is weighted by $1 + (1-y)^2$. 
However, since $Q^2 = sxy$, fixed $x$ and $Q^2$ implies taking measurements at different values of $s$. This can certainly
in principle be accomplished by reducing the beam energies in HERA.
However, the
practical difficulties for the experiments and the accelerator
inherent in
reducing either the proton or electron beam energy, or both,
by a factor sufficient to permit an accurate measurement of $F_L$ 
mean that it has not to date been attempted. An alternative way to
achieve the same end is to isolate those events in which the incoming
lepton radiates a hard photon in advance of the deep inelastic scattering,
thereby reducing the effective collision energy. Unfortunately, the
acceptance of the luminosity taggers typically used to detect such photons
is sufficiently small and understanding the acceptance sufficiently difficult
that no result has been obtained as yet.

In the absence of any direct determination, the H1 collaboration has used 
events at very large values of $y$ to
make an indirect measurement of $F_L$. The determinations of $F_2$
rely on the fact that most of the measurements are
made at values of $y$ sufficiently small that the
effects of $F_L$ are negligible; at higher $y$, a QCD estimate of $F_L$, which is normally a small fraction of $F_2$, is subtracted. The H1 collaboration inverts this procedure by isolating kinematic regions in which the contribution of $F_L$ is maximised and then subtracts off an estimate of $F_2$ extrapolated
from lower $y$. 

The method used by H1 employs the derivatives of the
reduced cross section with respect to $\ln y$. The
reduced cross section can be expressed as
\begin{eqnarray}  
\frac{x Q^4}{2\pi \alpha^2 Y_+ \cdot (1 + \delta) }\frac{d^2 \sigma} 
{dx dQ^2} & =& \sigma_r \nonumber \\ \nopagebreak
&=& F_2(x, Q^2) - \frac{y^2}{Y_+} F_L(x, Q^2),
\label{eq:redsigma}
\end{eqnarray}
which, when differentiated leads to 
\begin{equation}
\frac{\partial \sigma_r}{\partial \ln y} = \frac{\partial F_2}{\partial \ln y}
- \frac{2y^2(2-y)}{Y^2_+}F_L - \frac{y^2}{Y_+}\cdot 
\frac{\partial F_L}{\partial \ln y}
\label{eq:dsigmardlny}
\end{equation}
which gives improved sensitivity to $F_L$ via the stronger $y$ dependence
at the cost of involving derivatives of $\sigma_r, F_2$ and $F_L$, the
quantity to be measured. It is instructive to consider various 
restrictions:
\begin{itemize}
\item Small $y$ - here $\partial \sigma_r/\partial \ln y \sim  
\partial F_2/\partial \ln y$. For low $x$, $F_2$ can be well approximated
by: 
\begin{eqnarray}
F_2 &\propto& x^{-\lambda} \propto y^\lambda 
\label{eq:f2xlambda} 
\end{eqnarray}
so that: 
\begin{eqnarray}
\frac{\partial F_2}{\partial \ln y} &=& \lambda y^\lambda 
\label{eq:df2dy}
\end{eqnarray}
which can be expanded as:
\begin{equation}
\frac{\partial F_2}{\partial \ln y} \propto  \lambda e^{\lambda \ln y} \sim
\lambda(1 + \lambda \ln y \ldots )
\label{eq:df2dlny:exp}
\end{equation}
provided $\lambda \ln y$ is small. From this it is clear that $\partial \sigma_r/\partial \ln y$ is linear in
$\ln y$; 
\item $F_L = 0$ - for all $y$, $\partial \sigma_r/\partial \ln y$ is linear in
$\ln y$ for the same reason as above; 
\item $F_L \neq 0$ and large $y$ - $\partial \sigma_r/\partial \ln y$ is 
non-linear in $\ln y$ and the deviations are proportional to $F_L$ and
its logarithmic derivative; 
\end{itemize}
\begin{figure}[ht]
\begin{center}
\epsfig{file=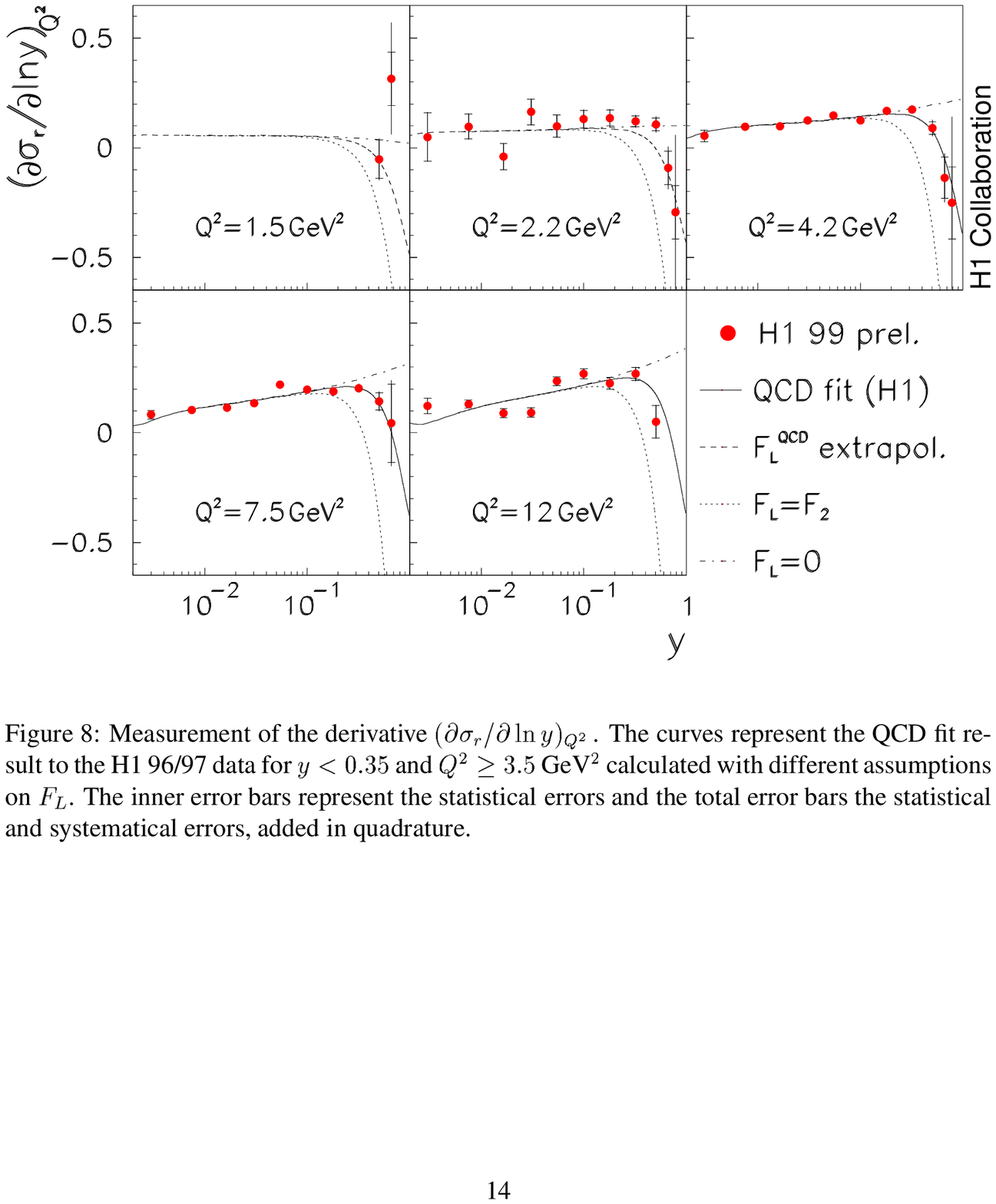,%
      width=10cm,%
      height=10cm,%
      clip=%
        }
\end{center}
\caption{The logarithmic derivative $\frac{\partial \sigma_r}{\partial \ln y}$
as a function of $y$ in $Q^2$ bins. The curves represent the results
of the H1 NLO QCD fit with differing assumptions about $F_L$ as shown in
the legend.}
\label{fig:H1:dsigmardlny}
\end{figure} 

These features can be seen in the preliminary H1 data of 
Fig.~\ref{fig:H1:dsigmardlny}. 
At the largest values of $y$, the
deviation from linearity implies that $F_L$ is non-zero. Although
it is in principle possible to solve the differential equation
for $F_L$ implied by Eq.~\ref{eq:dsigmardlny}, in practice the
data are insufficiently precise and the QCD expectation is
that the derivative of $F_L$ is negligible. The
uncertainty in this assumption is included in the
systematic error. 

\begin{figure}[ht]
\begin{center}
\epsfig{file=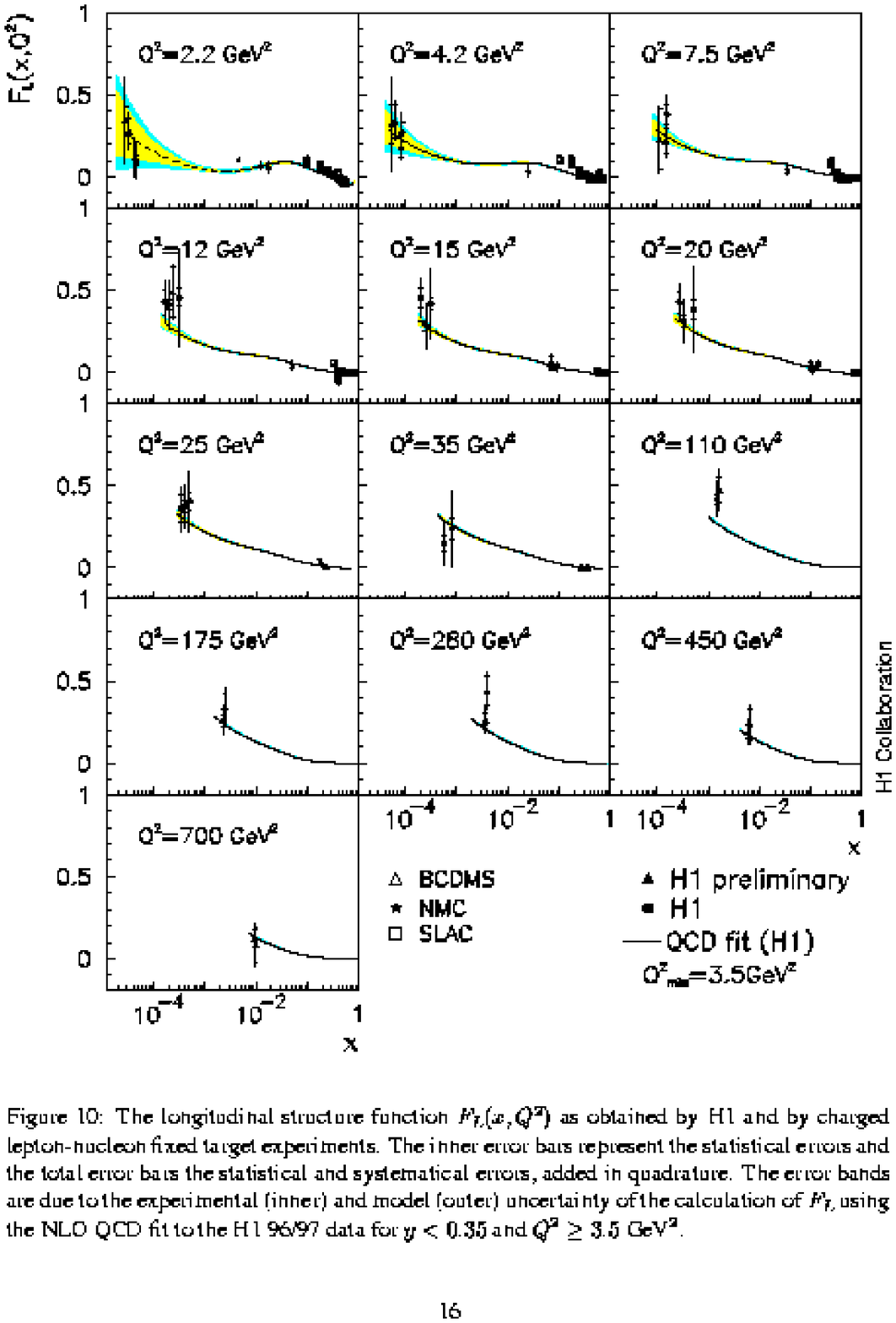,%
      width=11cm,%
      height=11cm,%
      clip=%
        }
\end{center}
\caption{Preliminary H1 estimate of $F_L$. The
$F_L$ values obtained are plotted in $Q^2$ bins as a function of $x$.
Also shown are earlier bins at higher $x$ from the SLAC
and NMC experiments. The curves with error bands are the predictions of the H1
NLO QCD fit.}
\label{fig:H1:fl}
\end{figure} 
The results
are shown in Fig.~\ref{fig:H1:fl}, together with earlier determinations
from SLAC~\cite{pl:b282:475}, NMC~\cite{np:b483:3}
and BCDMS~\cite{pl:b223:485, pl:b237:592}.
The curve is the result
of an NLO QCD fit to the H1 data deriving from the $F_2$ determination,
i.e. by deriving the 
gluon and quark distributions from scaling violations and then 
calculating $F_L$ using QCD. The
QCD prediction is in good agreement with the H1 estimate.

\section{Deep inelastic scattering at low $x$}
\label{sec:transition}

Until now we have concentrated on hard processes, in which $Q^2$ has been large
and where QCD has shown itself to be applicable. However, the data, in
particular the ZEUS ``BPT'' data~\cite{ZEUS-BPT}, give access
to very low $Q^2$ and $x$ regimes, in which the strong interaction becomes very strong and perturbative QCD would be expected to break down. This kinematic region has traditionally been understood in terms of Regge theory. The study of how and where the transition between these two regimes occurs is very interesting, not only intrinsically but also because of the insight it gives us into links between the apparently rather different processes of diffraction and deep inelastic scattering. In addition, the access given to very low values of
$x$ at these small $Q^2$ in principle gives sensitivity to the mechanism of
QCD evolution. Given the steep rise in the parton densities as $x$ falls, the
data at the lowest $x$ values may also be sensitive to high-density effects, such as parton recombination, sometimes known as saturation.
\subsection{QCD evolution}
\label{sec:QCDevolution}
We assume that the parton distribution functions,
$f$,  
satisfy the schematic equation:
\begin{equation}
\frac{\partial f}{\partial \ln \mu^2} \sim
\frac{\alpha_s(\mu^2)}{2\pi}
\cdot \left(f \otimes {\cal P} \right)
\label{eq:RGE}
\end{equation}
where $\mu$ represents the renormalisation scale and ${\cal P}$ is a 
`splitting 
function' that describes the probability of a given parton splitting
into two others. This equation is known
as the Dokshitzer-Gribov-Lipatov-Altarelli-Parisi (DGLAP) 
Equation~\cite{sovjnp:20:95,sovjnp:15:438,np:b126:298,jetp:46:641}.
There are four distinct Altarelli-Parisi (AP) splitting functions 
representing the 4 possible $1 \rightarrow 2$ splittings and referred to as 
$P_{qq}, P_{gq}, P_{qg}$ and $P_{gg}$.  
The calculation of the splitting functions in perturbative QCD in
Eq.~\ref{eq:RGE} requires approximations, both in 
order of terms which can be taken into account as well as the most important
kinematic variables. The generic form for the splitting functions can be
shown to be~\cite{ellis:1996:qcd}:
\begin{equation}
x{\cal P}(x,\alpha_s) = \sum^\infty_{n=0} \left(\frac{\alpha_s}{2\pi}\right)^n
\left[\sum^n_{m=0} A^{(n)}_m \left\{\ln \left(\frac{1}{x}\right)\right\}^m
+ x\overline{\cal P}^{(n)}(x)\right]
\label{eq:general-splitting-fn}
\end{equation}
where $\overline{\cal P}^{(n)}(x)$
are the $x$-finite parts of the
AP splitting functions and $A^{(n)}_m$ are 
numerical coefficients which can in principle be calculated 
for each splitting function. 
In the axial gauge, leading $\ln Q^2$ terms arise
from evolution along the parton chain that is strongly
ordered in transverse momentum, i.e.
\[
Q^2\gg k_{t,n}^2\gg k_{t,n-1}^2\gg \ldots 
\]
Leading-order DGLAP evolution sums up $(\alpha_s \ln Q^2)^n$ terms,
while NLO sums up $\alpha_s(\alpha_s \ln Q^2)^{n-1}$ terms, which
arise when two adjacent transverse momenta become comparable, losing
a factor of $\ln Q^2$. 

In some kinematic regions, and in particular at low $x$, it must become
essential to sum leading terms in $\ln 1/x$ independent of the value of
$\ln Q^2$. This is done by the 
Balitsky-Fadin-Kuraev-Lipatov~\cite{pl:b60:50,jetp:44:433,jetp:45:199,sovjnp:28:822} 
(BFKL) equation,
which governs the evolution in $x$ at fixed $Q^2$.
The leading-order terms in $(\alpha _s\ln 1/x)^n$ arise
from strong ordering in $x$, i.e.
\[
x\ll x_n\ll x_{n-1}\ll \ldots 
\]
One of
the most important goals of HERA physics is the search for 
experimental effects that can be unambiguously attributed to BFKL evolution.

Generally, however, QCD coherence implies angular ordering. To
see the implications of this it is more convenient to work with
unintegrated parton density functions, $f(x,k_t^2,\mu^2)$, where
$\mu$ is the scale of the probe. There are
now two hard scales, $k_t$ and 
complicated QCD evolution, known as the 
Ciafaloni, Catani, Fiorani and 
Marchesini~\cite{np:b296:49,pl:b234:339,np:b336:18}
(CCFM) evolution equation. The DGLAP and BFKL equations can
then be seen to be two 
limits of angular ordering. In the DGLAP collinear approximation, the
branching angle, $\theta$, where $\theta \sim k_t/k_l$, grows since $k_t$ 
grows; while for BFKL evolution, $\theta$ grows because $k_l \propto x$ falls. 
 
\begin{figure}[ht]
\begin{center}
\epsfig{file=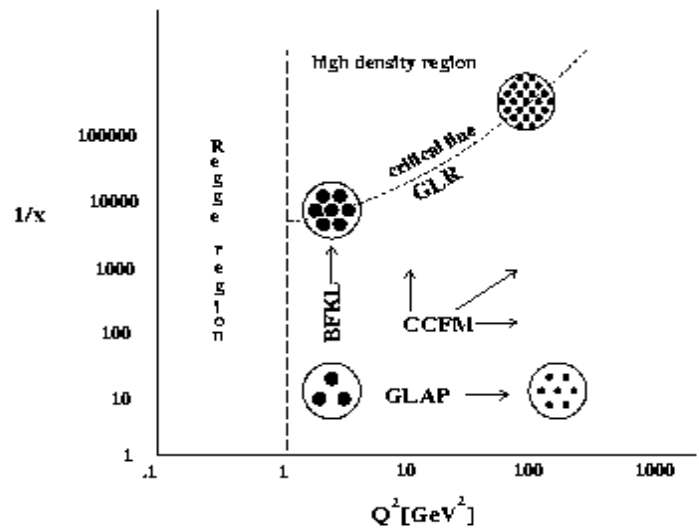,%
      width=10cm,%
      height=8cm,%
      clip=%
        }
\end{center}
\caption{Schematic diagram showing different regions of the $\ln 1/x$ and
$\ln Q^2$ plane and the evolution equations expected to hold therein.
The line marked `saturation' represents the boundary between GLAP
evolution and evolution governed by the GLR equation. The `size' of
partons is also indicated in differing kinematic regions.}
\label{fig:evolution}
\end{figure}

Figure~\ref{fig:evolution} shows the $\ln 1/x$ - $\ln Q^2$ plane at HERA, together
with schematic indications of the directions in which GLAP, BFKL and CCFM evolution is expected to be most applicable. 
Also indicated on the 
figure are schematic
indications of both the `size'  
and density of partons in the proton in different kinematic regions.

The transverse size of the partons that can be resolved by a probe
with virtuality $Q^2$ is proportional to $1/Q$, so that the area of the
partonic `dots' in Fig.~\ref{fig:evolution} falls as $Q^2$ rises.
For particular combinations of parton size and density, the proton will
eventually become `black' to probes, or, equivalently, the component gluons
will become so dense that they will begin to recombine.
The dotted line labelled `Critical line - GLR' refers to the boundary
beyond which it is expected that such parton saturation effects will become
important, i.e. the region in which partons become so densely crowded
that interactions between them reduce the growth in parton density predicted by 
the linear GLAP and BFKL evolution equations. The parton evolution in this 
region can be described by the 
Gribov-Levin-Riskin~\cite{prep:100:1,np:b268:427} 
equation, which explicitly takes into account an absorptive term in the gluon 
evolution equation. Naively, it can be assumed~\cite{quadt:phd:1997} that 
the gluons inside the proton each occupies on average a 
transverse area of
$\pi Q^{-2}$, so that the total transverse area occupied by gluons is
proportional to the number density multiplied by this area, i.e.\
$\pi Q^{-2}xg(x,Q^2)$. Since the gluon density
increases quickly as $x$ falls, and the gluon `size' increases as $Q^{-1}$,
in the region in which both $x$ and $Q^2$ are small, saturation
effects ought to become important. This should occur when the size
occupied by the partons becomes similar to the size of the proton:
\begin{equation}
xg(x,Q^2)\frac{\pi}{Q^2} = \pi R^2
\label{eq:satlimit}
\end{equation}
where $R$ is the radius of the proton, ($\sim 1$~fm $\sim 5$~GeV$^{-1}$). 
The measured values of $xg(x,Q^2)$ imply that saturation ought to be observable at HERA~\cite{proc:ringberg:1999:levin} at low $x$ and $Q^2$, 
although the values of $Q^2$ which satisfy Eq.~\ref{eq:satlimit}
are sufficiently small that possible non-perturbative and 
higher-twist effects certainly complicate the situation. Of course,

it is also possible that the assumption of homogenous gluon density
is incorrect; for example, the gluon density may be larger in the
close vicinity of the valence quarks, giving rise to
so-called `hot spots'~\cite{pr:189:267}, which could lead to 
saturation being observable at smaller distances and thereby larger $Q^2$.

\subsection{Interpretation and Models}
\label{sec:interpretation}
The region of low $Q^2$ and low $x$ is one in 
which perturbative QCD meets and competes with a large variety 

of other approaches, some based on QCD, others either on older paradigms such as 
Regge theory or essentially {\it ad-hoc} phenomenological models. In a previous
article~\cite{BFrsreview} I gave a quite detailed review of these models, and given that there have not been major developments here, I refer the reader to that for
further information. 

\subsection{QCD fits}
\label{sec:global}
The extension of the kinematic range and the high-precision data on $F_2$
from HERA provided a substantial impetus to the determination of parton
distribution functions via global fits to a wide variety of data. The major
current approaches are due to the CTEQ group~\cite{epj:c12:375} and Martin et al. (MRST)~\cite{proc:dis99:1999:105}. In general
both groups fit to data from fixed target muon and neutrino deep inelastic
scattering data, the HERA DIS data from HERMES, H1 and ZEUS, the
$W$-asymmetry data from the Tevatron as well as to
selected process varying from group to group such as prompt photon data from
Fermilab as well as high-$E_T$ jet production at the Tevatron. The different data sets give different sensitivity to the proton distributions, depending
on the kinematic range, but together constrain them across almost the whole
kinematic plane, with the possible exception of the very largest values of $x$,
where significant uncertainties still remain~\cite{epj:c13:241}. 

The approaches of CTEQ and MRST are basically similar, although they differ both
in the data sets used as well as in the fitting procedure and the technical details of the theoretical tools used, e.g. the treatment of heavy quarks in DIS. In their latest fits, CTEQ prefer to omit the prompt photon data because of
the uncertainties in scale dependence and the appropriate value for the
intrinsic $k_T$ required to fit the data. Instead they use single-jet inclusive
$E_T$ distributions to constrain the gluon distribution at large $x$.
In contrast, until their most recent publications, MRST retained
the prompt photon data, giving alternative PDFs
depending on the value for the prompt-photon intrinsic $k_T$ used. Both
groups parameterise the parton distributions
in terms of powers of $x$ and $(1-x)$ leading to fits with many
free parameters. The MRST NLO parameterisation of the gluon is
shown below as an example:

\begin{eqnarray}
xg &=& A_{g}x^{-\lambda_g} (1-x)^{\eta_g}
(1+\epsilon_{g}\sqrt{x}+\gamma_{g}x)
\label{eq:MRST:NLOg}
\end{eqnarray}
where $A_{g}, \lambda_g, \eta_g, \epsilon_g$ and $\gamma_g$
are free parameters in the fit.

\begin{figure}[ht]
\begin{center}
\epsfig{file=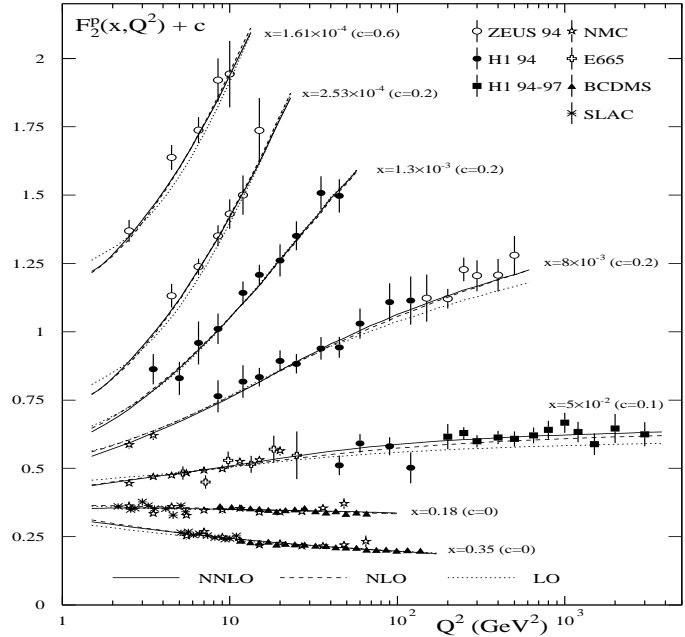,%
      width=10cm,%
      height=10cm%
        }
\end{center}
\caption{The MRST `central' NNLO fit to DIS data. The solid line
shows the NNLO fit, while the NLO fit is shown by the dashed line
and the LO fit by the dotted line. The data are from H1, ZEUS
and the fixed target experiments and are plotted 
in $x$ bins as a function of $Q^2$ with an additive constant
added to the data of each $x$ bin to improve visibility.}
\label{fig:MRST:NNLOfit}
\end{figure}

The seminal work of Botje in producing PDFs with associated
error matrices for the first time~\cite{epj:c14:285}
has led to similar fits being produced
by other groups. CTEQ~\cite{CTEQ-errors} have
produced a global fit with associated errors, while both H1~\cite{H1QCDfit}
and ZEUS~\cite{ZEUSQCDfit} have produced their own fits using DIS
data only, as discussed in  Section~\ref{NLOQCDfit}. 

These fits allow one to see very graphically the salient features
of the QCD evolution that have been discussed above. For example, the ZEUS
NLOQCD fit shown in Fig.~\ref{fig:ZEUSNLO-gandsea} has already been discussed
in Section~\ref{NLOQCDfit}. Here we saw that the strange behaviour both of the gluon vs. sea densities, as well as $F_L$, as a function of $Q^2$ showed that the normal QCD interpretation may well be breaking down at $Q^2$ around 1 GeV$^2$.

One possible reason for the problems with the interpretation of the NLO QCD fits could be the need for higher-order QCD fits i.e. next-to-next-to-leading-order (NNLO) fits. The first steps in implementing such fits have already begun;
some moments of the NNLO splitting functions
have been calculated~\cite{np:b492:338}. Using this with other
available information, van Neerven and Vogt~\cite{np:b568:263,hep-ph-0006154} have produced analytical expressions for the splitting functions which represent
the slowest and fastest evolution consistent with the currently available
information. The MRST group has recently used this information to investigate
NNLO fits to the available data~\cite{hep-ph-0007099}. Such an analysis requires
some changes to the parameterisations used, so that for example the NLO
parameterisation of the gluon of Eq.~\ref{eq:MRST:NLOg} becomes:
\begin{equation}
 xg (x, Q_0^2) \; = \; A_g \: x^{-\lambda_g} \: (1 - x)^{\eta_g} \:  
(1 + \varepsilon_g \sqrt{x} +  
\gamma_g x) \: - \: A_g^\prime \: x^{- \lambda_g^\prime} \:  
(1 - x)^{\eta_g^\prime}, 
\label{eq:MRST:NNLOg}
\end{equation}
primarily in order to facilitate a negative gluon density at low $x$ and
low $Q^2$, which, as we have seen, is preferred by the fits, even at NLO. The results of the `central' fit, between the extremes of the van~Neerven-Vogt parameterisation, is shown in Fig.~\ref{fig:MRST:NNLOfit}.

\begin{figure}[ht]
\begin{center}

\epsfig{file=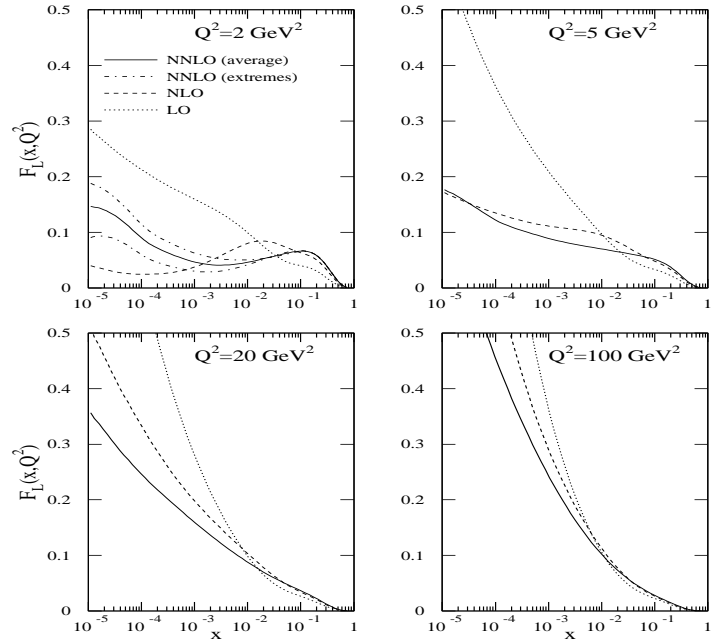,%
      width=10cm,%
      height=10cm%
        }
\end{center}
\caption{The $F_L$ structure function from the MRST fits, taking into account
part of the NNLO corrections in four bins of $Q^2$ as a function of $x$.
The solid line shows the 'average' of the parameter space available
to choose the NNLO parameters, while the dashed-dotted lines show the two extreme possibilities. The NLO fit is indicated by the dashed 
line while the LO fit is indicated by the dotted line.}
\label{fig:MRST:NNLOPDFs}
\end{figure}

There are also changes of the LO and NLO fits with respect to earlier
publications, in as much as MRST now follow CTEQ in using the Tevatron
high-$E_T$ data rather than the prompt-photon data, and HERA
$F_2$ data has been included in the fit. There is a marked improvement
in the quality of the fit in the progression LO $\rightarrow$ NLO
$\rightarrow$ NNLO, in particular in terms of the NMC data. The size
of higher-twist contributions at low $x$ also decreases, so that at
NNLO is it essentially negligible. The effect of going to NNLO on the
PDFs themselves is highly non-trivial. This is illustrated in
Fig.~\ref{fig:MRST:NNLOPDFs}, where the quite major changes in 
$F_L$, particularly at low $x$, are evident. 
There is also a large variation depending on the choices made
in the parameter space allowed by the partial NNLO
{\it ansatz}.  Indeed, the
GLAP approach is not convergent for $Q^2 < 5$ GeV$^2$, which may well
be due to the neglect of important $\ln 1/x$ contributions. 
However, the instability seen at low $Q^2$ soon
vanishes at higher $Q^2$.

Thorne has investigated the question of incorporating $\ln 1/x$
terms in the splitting functions by incorporating the solution of the 
NLO BFKL kernel using a running coupling 
constant~\cite{pl:b474:373,misc:DIS2000:Thorne}.
The inclusion of the BFKL terms does indeed give an improved fit compared to
the `central' NNLO fit, particularly at the lowest $Q^2$ and $x$. This
may be one of the first unambiguous 
indications of the importance of BFKL evolution.

\subsection{Data in the transition region}
\label{sec:transdata}
\begin{figure}[ht]
\begin{center}
\epsfig{file=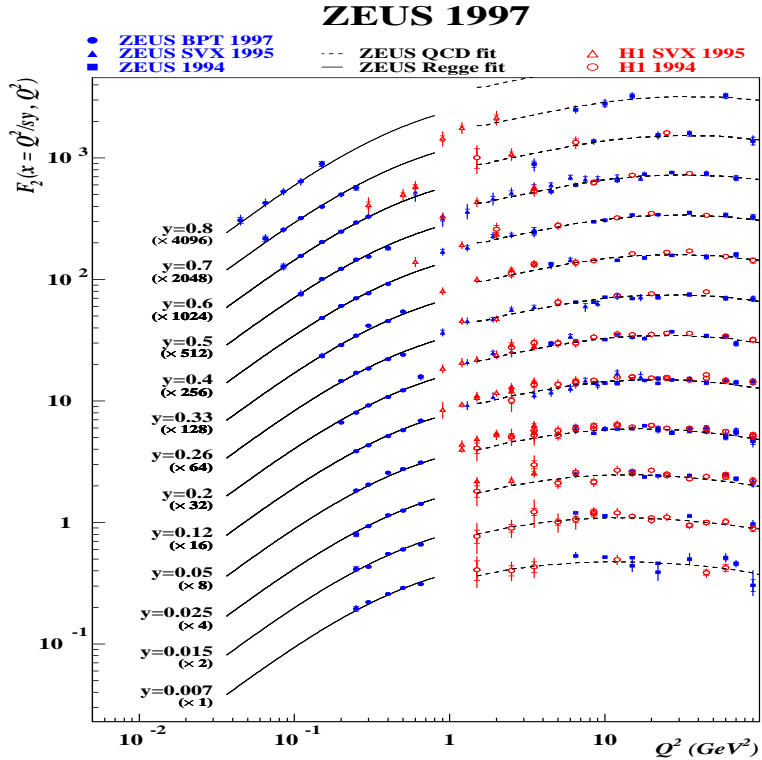,%
      width=10cm,%
      height=10cm%
        }
\end{center}
\caption{ZEUS BPT data on $F_2$ in bins of $y$ as
a function of $Q^2$. Also shown are earlier
ZEUS data as well as data from H1.
The solid line shows the results of the ZEUS Regge fit to
the form of Eq.~\protect\ref{eq:GVDM+REGGE}, while 
the dotted line shows the result of the ZEUS NLO QCD fit.}
\label{fig:ZEUSBPT:ybins}
\end{figure}

The approximate position of the transition between
data that can be described by pertrubative QCD evolution
and that which require the Regge approach can be seen in Fig.~\ref{fig:ZEUSBPT:ybins}.
 
For $Q^2~\gsim~1$
GeV$^2$, the data are roughly independent of $Q^2$, whereas at lower $Q^2$
they fall rapidly, approaching the $Q^{-2}$ dependence that would be expected in
the limit $Q^2 \rightarrow 0$ from conservation of the electromagnetic current. 
Although QCD gives a good fit to the data down to $Q^2 \sim 1$ GeV$^2$,
below that it is necessary to use a Regge-based fit of 
the form 
\begin{equation}
F_2(x,Q^2) = \left(\frac{Q^2}{4\pi^2\alpha} \right)
\left( \frac{M_0^2}{M_0^2+Q^2}\right) \left( A_\reg
\left( \frac{Q^2}{x} \right)^{\alpha_\reg-1}+A_\pom
\left(\frac{Q^2}{x}\right)^{\alpha_\pom-1}\right),
\label{eq:GVDM+REGGE}
\end{equation}
where $A_\reg, A_\pom$ and $M_0$ are constants and $\alpha_\reg $ and
$\alpha_\pom$ are the Reggeon and Pomeron intercepts, respectively.
Regge theory is expected to apply at asymptotic energies. The
appropriate energy here is $W$, the
centre-of-mass energy of the virtual photon-proton system, 
given by Eq.~\ref{eq:w2q2x}.
Since, at low $x$, $W^2 \sim 1/x$, it would be expected that Regge fits would be 
applicable at very low $x$ and $Q^2$. 

\begin{figure}[t]
\begin{center}
\epsfig{file=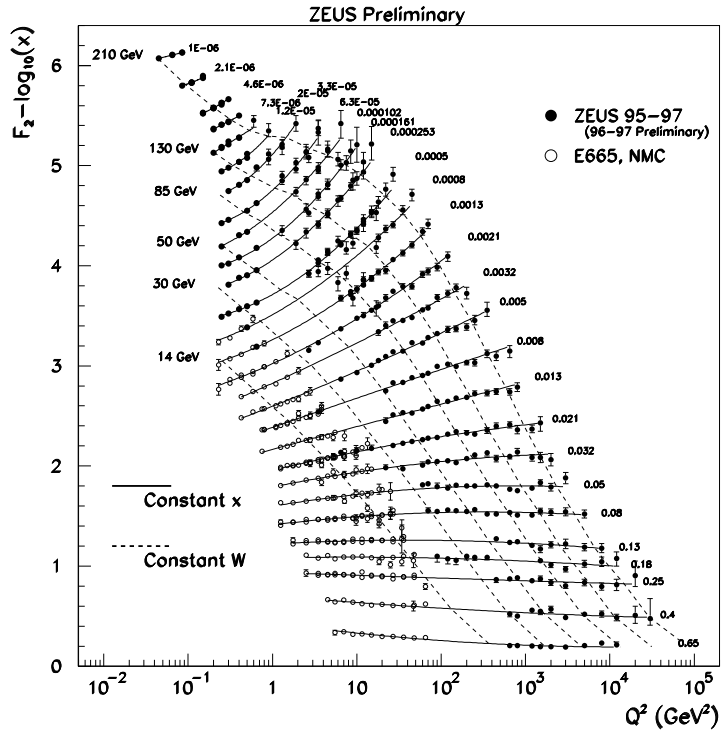,%
      width=11cm,%
      height=11cm%
        }
\end{center}
\caption{Compilation of ZEUS $F_2$ data in $x$ bins as a function of $Q^2$. 
Each $x$ bin is shifted by an additive constant for ease of visibility.
Data from NMC and E665 are also shown.
The dotted lines show lines of constant $W$, while the solid lines are
fits to the form of Eq.~\protect\ref{eq:F2param}.} 
\label{fig:ZEUS:6OF2}
\end{figure}

The complete ZEUS data over 
six orders of magnitude in $x$ and $Q^2$ 
are shown in $x$ bins as a function of $\ln Q^2$ 
in Fig.~\ref{fig:ZEUS:6OF2}, together with fixed target data
from NMC and E665, which extends the range in the direction of medium
$x$ and $Q^2$. 

The availability of this very wide range of precise data makes possible qualitatively new investigations of models that describe $F_2$. Since the logarithmic derivative of 
$F_2$ is directly
proportional to the gluon density in leading-order QCD, which in turn is 
the dominant parton density at small $x$, its behaviour as a function of both
$x$ and $Q^2$ is important. The solid curves
on the figure correspond to fits to a polynomial in $\ln Q^2$ of the form

\begin{equation}
F_2 = A(x) + B(x) \left(\log_{10} Q^2\right) + C(x) 
\left(\log_{10}Q^2\right)^2,  
\label{eq:F2param}
\end{equation}
which gives a good fit to the data through the entire kinematic range. The dotted lines 
on Fig.~\ref{fig:ZEUS:6OF2} are lines of constant $W$. 
The curious `bulging' shape of these contours in the
small-$x$ region immediately implies that something interesting is
going on there. Indeed, simple inspection of Fig.~\ref{fig:ZEUS:6OF2} shows
that the slope of $F_2$ at constant $W$ begins flat in the scaling region,
increases markedly as the gluon grows and drives the evolution of $F_2$
and then flattens off again at the lowest $x$. 

\begin{figure}[ht]
\begin{center}
\epsfig{file=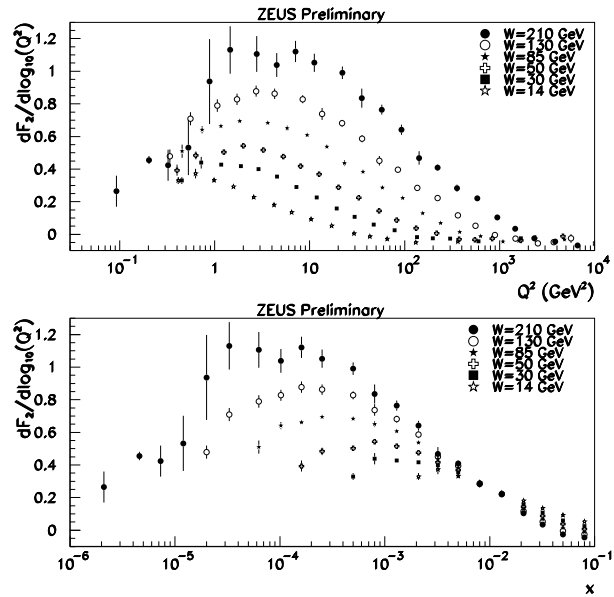,%
      width=9cm,%
      height=9cm%
        }
\end{center}
\caption{The logarithmic derivative of the ZEUS $F_2$ data in six bins
of $W$, plotted as a function of $Q^2$ and $x$.}
\label{fig:ZEUS:logder}
\end{figure}

\begin{figure}[ht]
\begin{center}
\epsfig{file=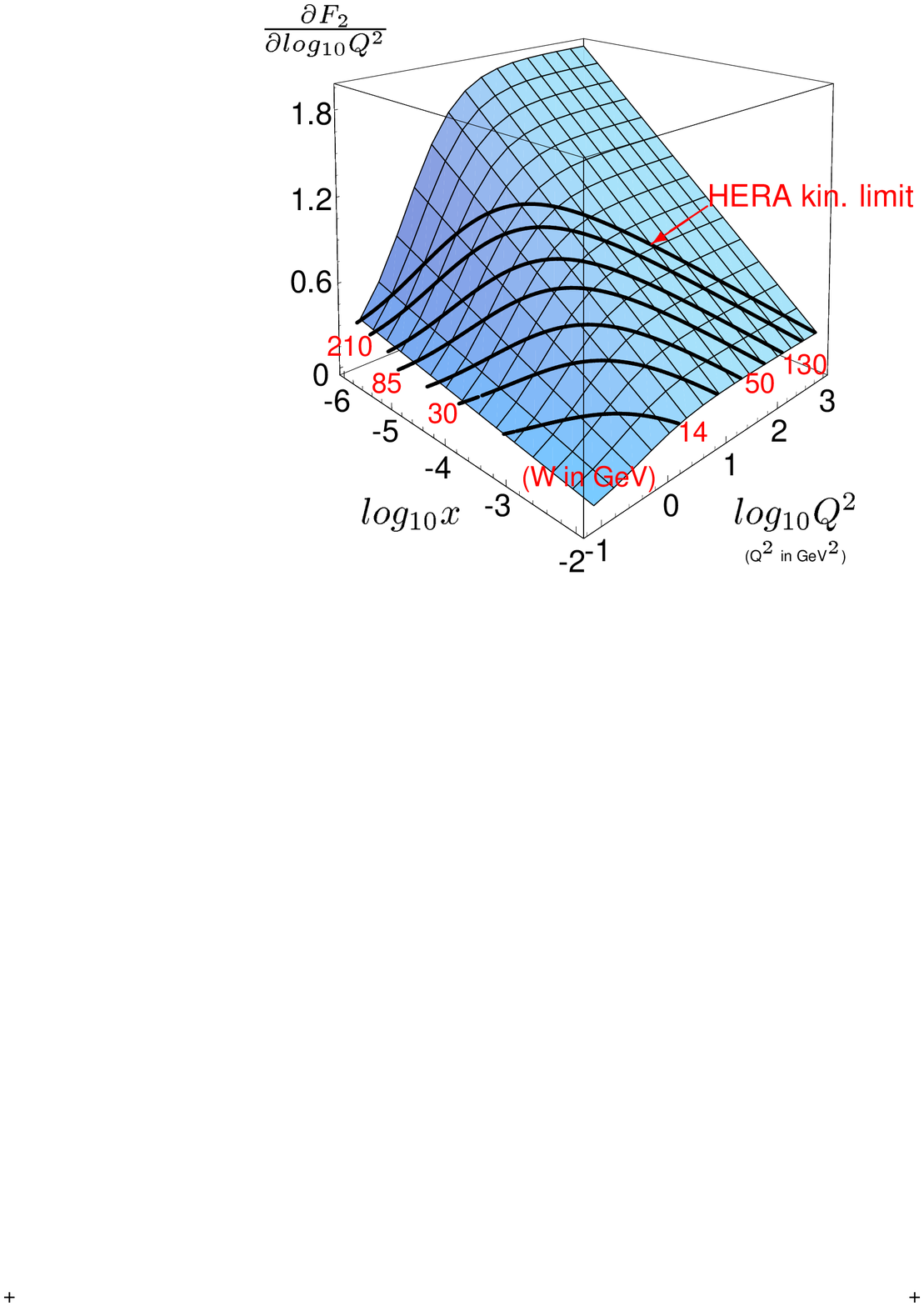,%
      width=7cm,%
      height=7cm%
        }
\end{center}
\caption{The logarithmic derivative of the ZEUS $F_2$ data plotted as a surface in three dimensions versus both $Q^2$ and $x$. The curves show lines of constant $W$.}
\label{fig:ZEUS:3Dlogder}
\end{figure}

Figure~\ref{fig:ZEUS:logder}
shows the logarithmic derivative evaluated at $(x, Q^2)$ points
along the contours of fixed $W$
shown on Fig.~\ref{fig:ZEUS:6OF2} according to the derivative of
Eq.~\ref{eq:F2param}, {\it viz.}:
\begin{equation}
\frac{\partial F_2}{\partial \log_{10} Q^2} = B(x) + 2C(x) \log_{10} Q^2,  
\label{eq:F2deriv}
\end{equation}
where the data are plotted separately as functions of $\ln Q^2$ and $\ln x$.
The turn-over in the derivatives in all $W$ bins is marked. Within the framework of pQCD, the interpretation of such an effect is that the growth of the gluon 
density at low $x$ is tamed as $Q^2$ and $x$ fall. This behaviour can also be
seen more clearly in Fig.~\ref{fig:ZEUS:3Dlogder}, which shows
a three-dimensional plot of the derivatives as a function of both $\ln x$ and $\ln Q^2$, obtained
from a parameterisation of the DIS data.  

\begin{figure}[ht]
\begin{center}

\epsfig{file=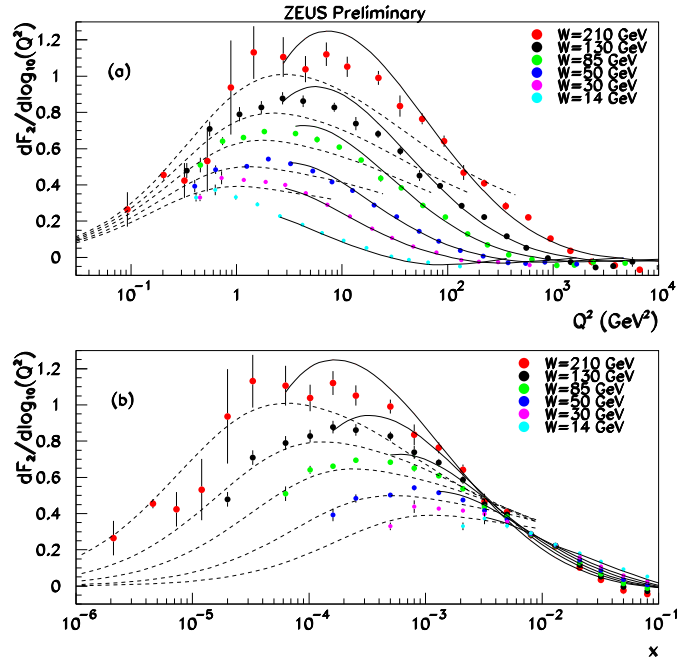,%
      width=10cm,%
      height=10cm%
        }
\end{center}
\caption{The logarithmic derivatives of the ZEUS $F_2$ data as shown 
in Fig.\protect~\ref{fig:ZEUS:logder}. The dotted curves show the
predictions of the BGW model for the five highest $W$ bins; since
these predictions do not include QCD evolution effects, they should
become progressively less accurate as $Q^2$ increases. The solid
curves show the predictions of the ZEUS NLO QCD fit for all $W$ bins;
they are only
shown down to $Q^2 \sim 3$ GeV$^2$.} 
\label{fig:ZEUS:logderGBW}
\end{figure}

Such an effect
is by no means necessarily an indication of deviations from the 
standard DGLAP evolution. 
It can be seen from Fig.~\ref{fig:ZEUS:3Dlogder} that, for
example, no turnover effect occurs in bins of constant $Q^2$.
Nevertheless, the features of Figs.~\ref{fig:ZEUS:logder} and \ref{fig:ZEUS:3Dlogder} can be explained as a natural consequence of parton saturation or shadowing. 
These effects can be naturally discussed
in ``dipole models''~\cite{BFrsreview}, which often explicitly
take into account parton-saturation effects. 
In such models, the ``standard'' picture of deep 
inelastic scattering in the infinite-momentum frame of the proton is
replaced by an equivalent picture produced by a Lorentz boost into
the proton rest frame. In this frame, the virtual photon undergoes

time dilation and develops structure far upstream of the interaction
with the proton. The dominant configurations of this structure are
$q\overline{q}$ and $q\overline{q}g$ Fock states, which interact with
the proton as a colour dipole. The higher the $Q^2$ of the interaction,
the smaller the transverse size of the dipole. For small $x$, the deep inelastic process can be considered semi-classically as the coherent interaction of the dipole with the
stationary colour field of the proton a long time after
the formation of the dipole. As an example, the model of Golec-Biernat
and W\"{u}sthoff (GBW)~\cite{pr:d59:014017,pr:d60:114023} is shown in
Fig.~\ref{fig:ZEUS:logderGBW}, together with the results of the 
ZEUS NLOQCD fit. 

It can be seen that the GBW model reproduces the basic
features of the logarithmic derivative plot reasonably well.  However,
so does the ZEUS NLOQCD fit, so that no firm conclusion can be drawn on
the existence or otherwise of saturation effects.

\section{Diffraction}
\label{sec:diffraction}
Diffractive DIS
is that subset
characterised by a hard interaction between the proton and the exchanged virtual photon that nevertheless leaves the proton intact. Such interactions are normally thought of in Regge theory as proceeding via the exchange of a colourless particle with the quantum numbers of the vacuum, known as the Pomeron. The Pomeron under some circumstances can be considered to develop its own partonic structure, analogous to the proton, which can be parameterised using diffractive DIS data. 

\begin{figure}[ht]
\begin{center}
\epsfig{file=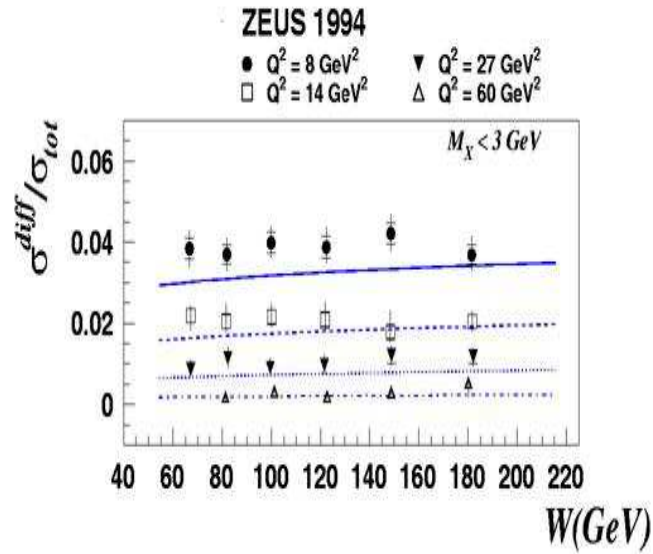,%
      width=9cm,%
      height=8cm,%
      clip=%
        }
\end{center}
\caption{The ratio of the diffractive to total cross section in four $Q^2$ bins as a function of $W$. The curves show the predictions from the
Golec-Biernat \& Wusthoff model.}
\label{fig:diff-tot-ratio}
\end{figure}

One of the most attractive features of dipole models such as
that of GBW discussed in the previous section is 
the rather natural way in which they can lead to a unified description of diffraction and 
deep inelastic scattering. The QCD interpretation of the Pomeron is that it is equivalent to the exchange of two gluons in a colour-singlet state.  In this picture, therefore, diffraction can be considered as a subset of fully inclusive DIS, which sums over all possible exchanges between the dipole and the proton, dominantly one- and two-gluon exchange in a colour octet, in contrast to the colour-singlet exchange that dominates inclusive DIS. 
This deep connection between these two processes leads to non-trivial predictions which do indeed seem to be at 
least qualitatively in agreement
with the data. This is illustrated in Fig.~\ref{fig:diff-tot-ratio}.
This figure is surprising for several reasons. It demonstrates that the diffractive cross section
has the same $W$ dependence as the total cross section. To the extent to
which the diffractive cross section can be related to the elastic cross section,
one would have expected from the Optical Theorem that the
ratio would have a power-law dependence on $W$, as indeed would also be
expected from Regge theory via the exchange of a Pomeron. A strong $W$ 
($\sim 1/x$) dependence is also expected in QCD models, since the total cross section is dominated by single-gluon
exchange, whereas diffraction is dominated by
two-gluon exchange. The other surprise is the fact that the GBW model gives 
a rather good qualitative representation of the data. 

\begin{figure}[ht]
\begin{center} 
\epsfig{file=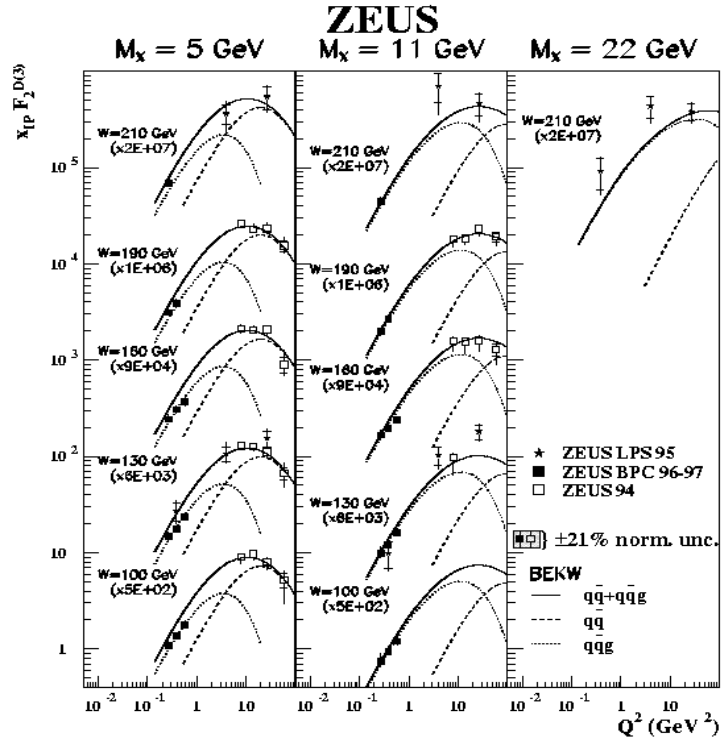,%
      width=10cm,%
      height=10cm%
        }
\end{center}
\caption{$x_\pom F_2^{D(3)}$ in bins of $W$ and $M_X$ as a function of $Q^2$. The curves refer to the model of Bartels et al.~\protect\cite{BEKW}}
\label{fig:ZEUS:lowQ2diff}
\end{figure}

ZEUS has also investigated the behaviour of this
ratio as $Q^2 \rightarrow 0$~\cite{ZEUS-lowQ2diff}. 
Figure~\ref{fig:ZEUS:lowQ2diff} shows the diffractive structure
function, $x_\pom F_2^{D(3)}$ (the analogue to $F_2$, integrated over $t$) multiplied
by $x_\pom$, the fraction of the proton's momentum carried by the Pomeron, 
as a function of $Q^2$ in various bins of $W$ and in two bins of $M_X$, the mass of the hadronic system other than the proton. The $F_2^{D(3}$ data points are determined using two methods,
one of which requires the observation of a large rapidity gap in the proton-beam direction, while the other uses the ZEUS
Leading Proton Spectrometer (LPS)~\cite{LPSref}. This device is an array of
six stations of silicon-strip detectors placed in Roman pots downstream of the interaction point in the proton beam direction. It uses the HERA beam elements
to form a magnetic spectrometer to analyse the leading proton from diffractive interactions. Although its acceptance is of necessity small, it avoids the
low-mass proton dissociative background endemic with other forms of identifying
diffractive interactions. The similarity of
Fig.~\ref{fig:ZEUS:lowQ2diff} to Fig.~\ref{fig:ZEUSBPT:ybins} is striking, again
illustrating the $Q^{-2}$ falloff enforced by electromagnetic current conservation. 
\subsection{Vector meson production}
\label{sec:VM}

\begin{figure}[ht]
\begin{center} 
\epsfig{file=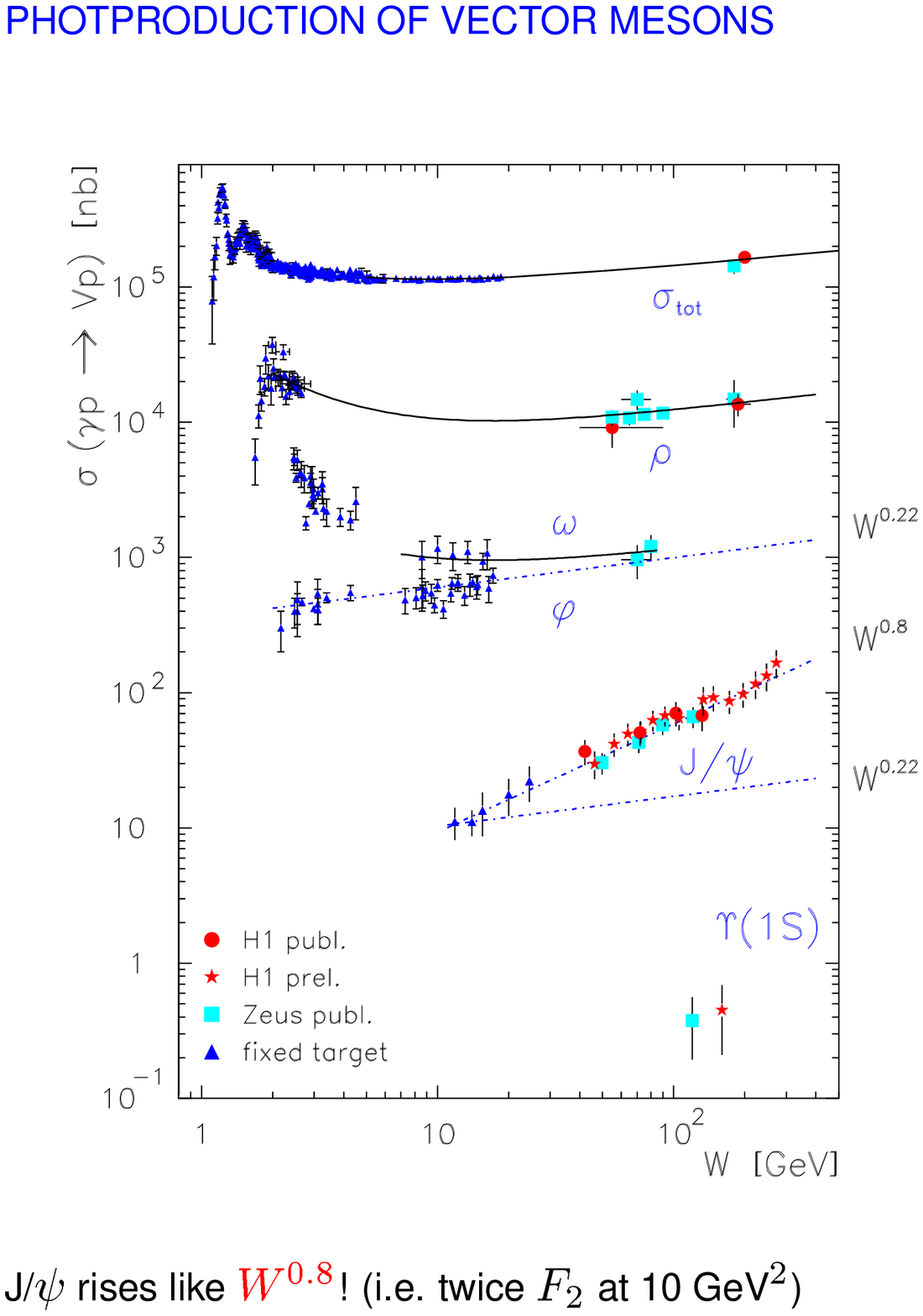,%
      width=10cm,%
      height=10cm,%
      clip=%
          }
\end{center}
\caption{The total photoproduction cross section together with the photoproduction cross sections for a variety of vector mesons as a function of $W$ from H1, ZEUS and several fixed-target experiments. The dotted lines show $W$ to the indicated powers.}
\label{fig:VMcomp}
\end{figure}

\begin{figure}[ht]
\begin{center} 
\epsfig{file=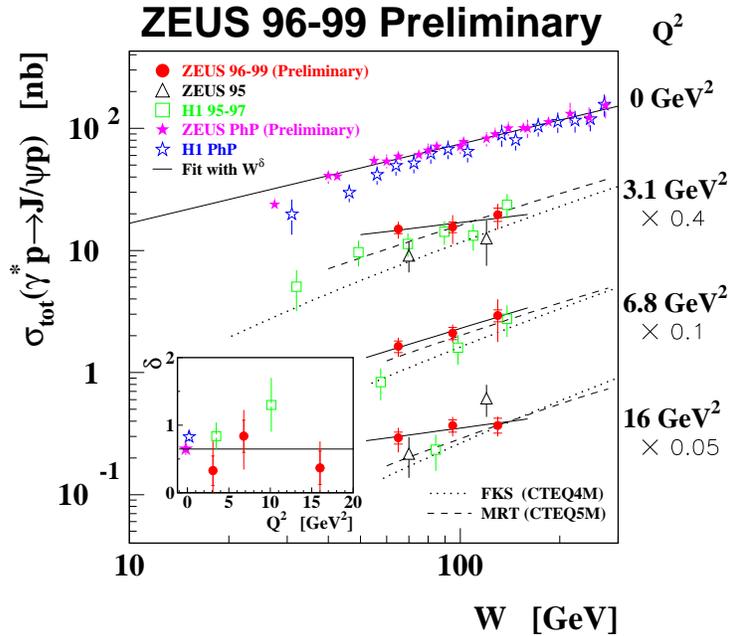,%
      width=10cm,%
      height=10cm%
        }
\end{center}
\caption{The ZEUS and H1 $J/\psi$ cross section for several bins of $Q^2$ as a function of $W$. The dotted lines show the predictions of the model of Frankfurt, Koepf and Strikman~\protect\cite{FKS}, the dashed lines show the predictions
of Martin, Ryskin and Teubner~\protect\cite{MRT} and the solid lines show fits to the form of $W^\delta$. The inset shows the $\delta$ obtained from these fits as a function of $Q^2$.}
\label{fig:ZEUS:jpsi}
\end{figure}

The exclusive production of vector mesons is both a very simple laboratory to study many aspects of diffraction as well as a process in which dipole models are likely to be particularly appropriate. Figure~\ref{fig:VMcomp} shows the 
photoproduction cross sections for a variety of vector mesons as a function of $W$, as well as
the total cross section. The relatively slow rise of the total cross section
with $Q^2$ is indicative of the dominance of soft processes. The $J/\psi$ 
cross section clearly has a much steeper rise with $Q^2$. This is more clearly seen in Fig.~\ref{fig:ZEUS:jpsi}, where the cross section for several different $Q^2$ is shown as a function of $W$. It would appear that the mass of the $J/\psi$ is sufficiently large that it gives rise to a hard scale even at
$Q^2 \sim 0$.

\begin{figure}[ht]
\begin{center} 
\centerline{\epsfig{file=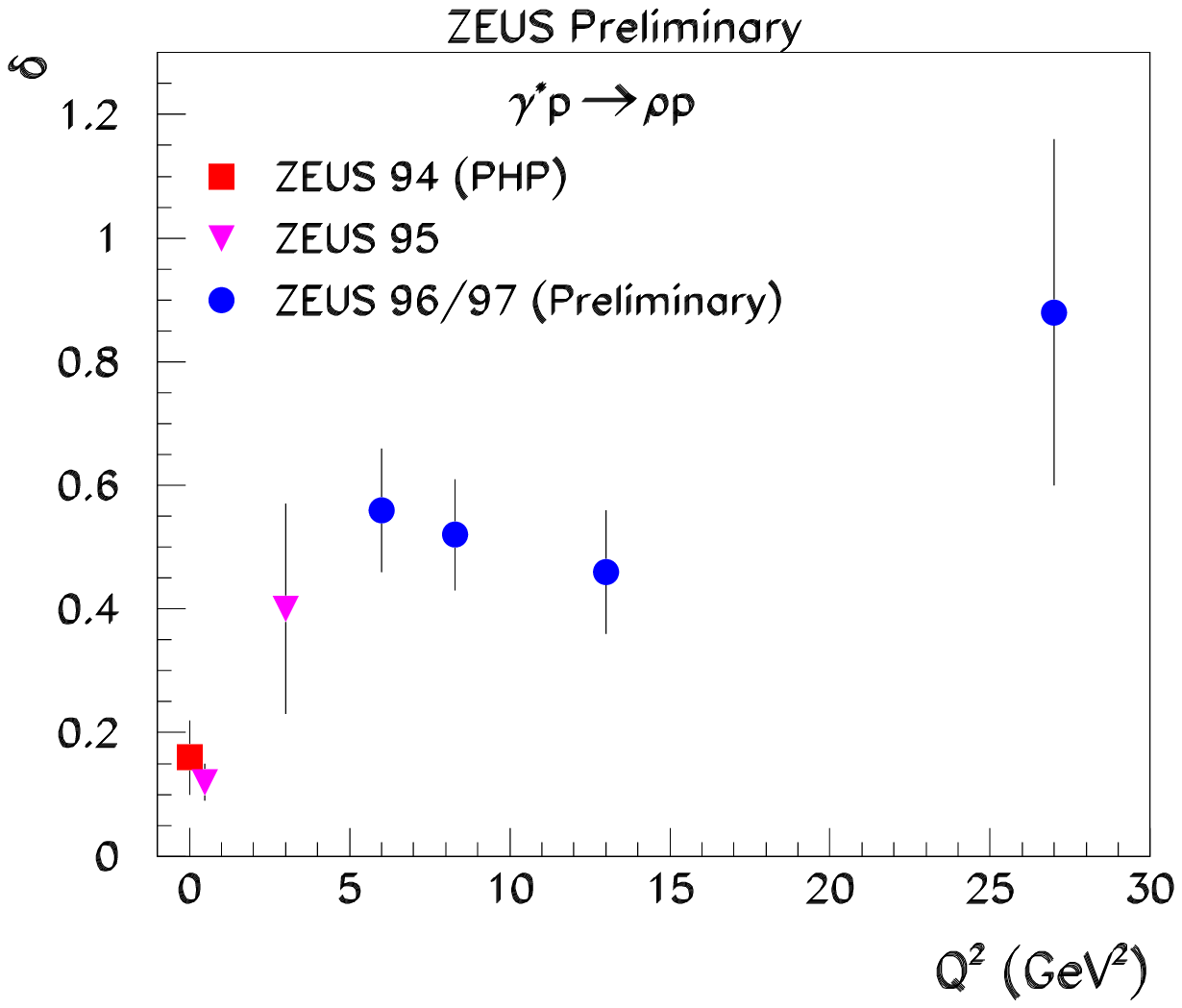,%
      width=7cm,%
      height=7cm%
        }
\epsfig{file=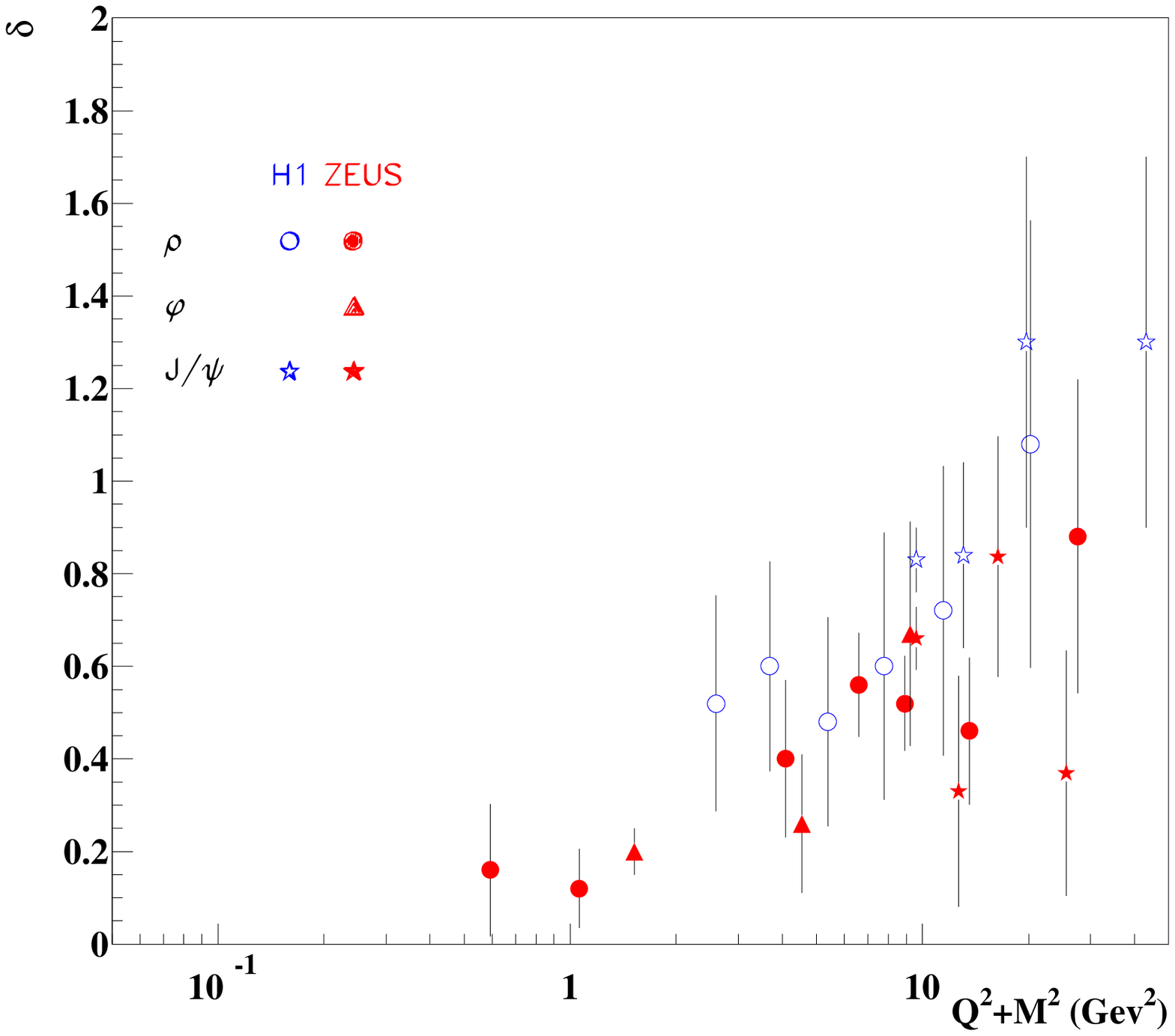,%
      width=5cm,%
      height=5cm,%
      clip=%
        }}
\end{center}
\caption{a)Results of a fit to the ZEUS $\rho$ data of the form $W^\delta$.
The value of the exponent is plotted as a function of $Q^2$. b) The exponent obtained from fits of the form $W^\delta$ to the vector meson data from ZEUS and H1. The value of the exponent $\delta$ is plotted against $Q^2 + M^2$.}
\label{fig:ZEUS:rhodelta}
\end{figure}

In contrast, the mass of the $\rho$ is small and quite large values of $Q^2$ need to be reached before the $W$ dependence rises to the values associated with
hard processes. This can be seen in Fig.~\ref{fig:ZEUS:rhodelta}, which shows the values of the fit to a power law in $W$ as a function of $Q^2$. An appropriately hard scale seems to pertain for $Q^2 > 5$. 

This behaviour, in which either $Q^2$ or the mass can act as a hard scale, leads
to the obvious question of whether a combination of these two quantities can also give a hard scale. 
Figure~\ref{fig:ZEUS:rhodelta} shows the data on $\rho$, $\phi$
and $J/\psi$ production from H1 and ZEUS plotted against $Q^2 + M^2$. There
is indeed a tendency for the data for all the vector mesons to lie on a universal curve. However, other more detailed 
comparisons~\cite{aharonsEPSpaper} show that some differences do remain 
between the different species even when plotted against $Q^2 + M^2$.

\begin{figure}[ht]
\begin{center} 
\centerline{\epsfig{file=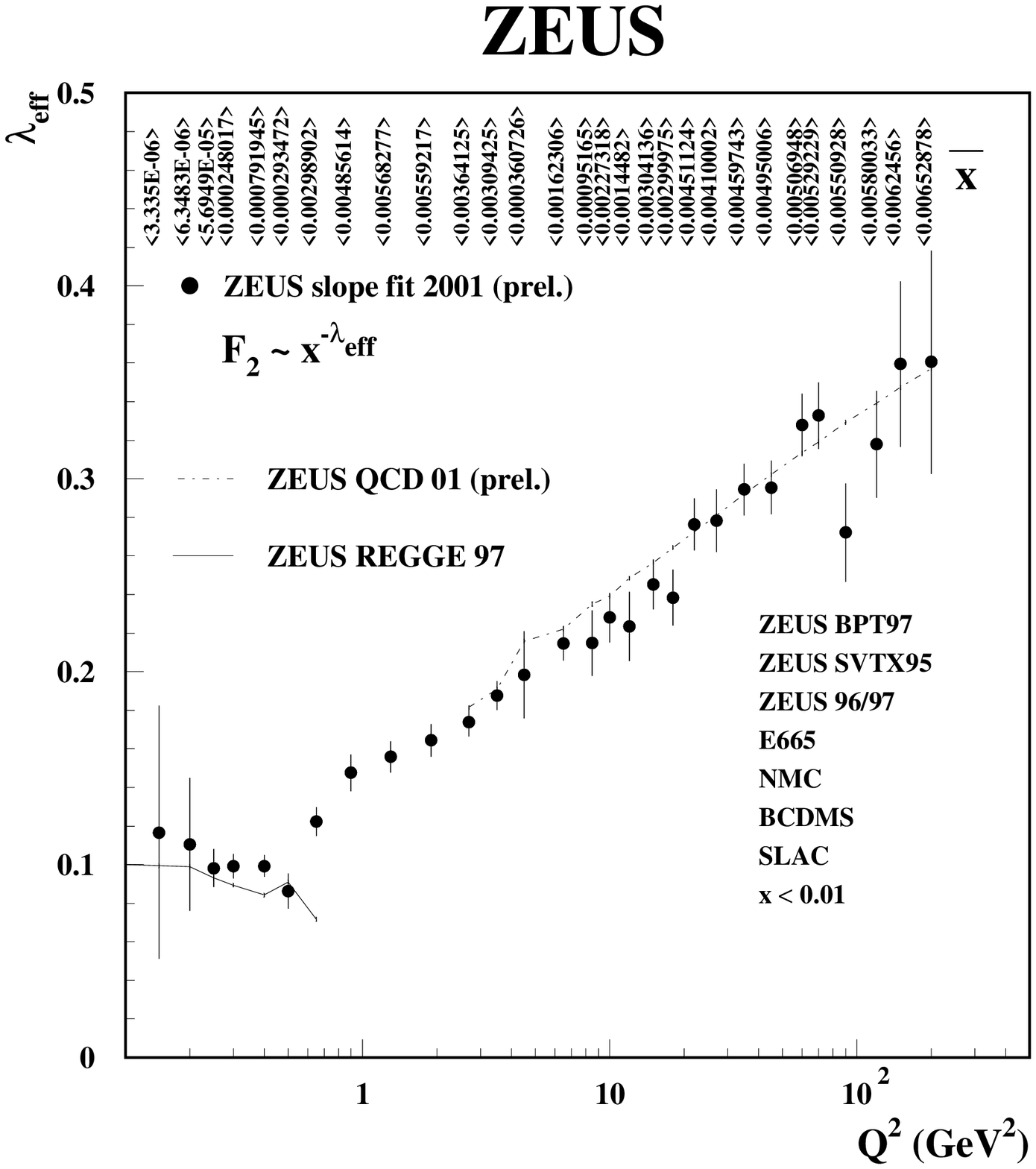,%
      width=6cm,%
      height=6cm%
        }\epsfig{file=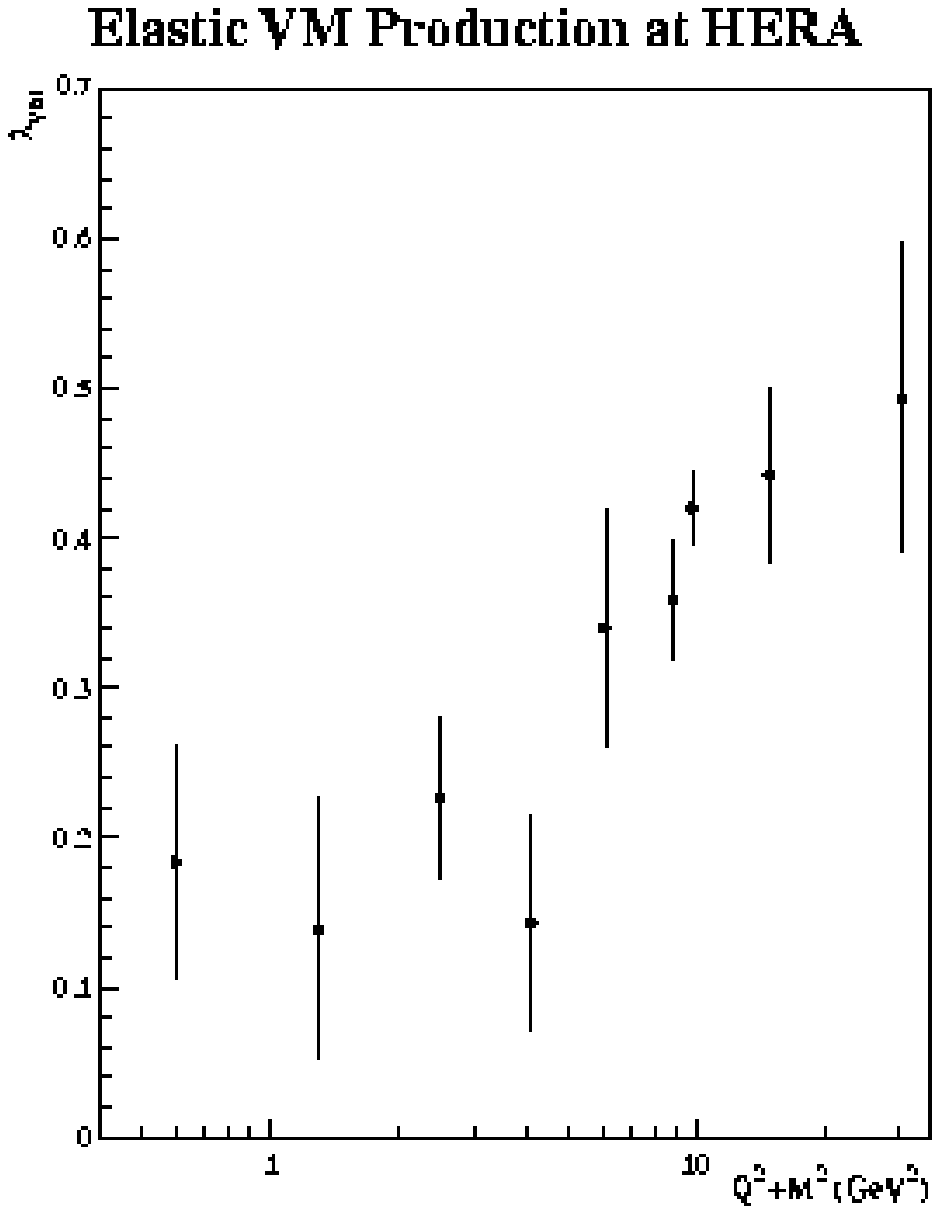,%
      width=6cm,%
      height=6cm,%
      clip=%
        }}
\end{center}
\caption{a) Values of the exponent in fits of the form $W^\delta$ to the $W$
dependence of the inclusive DIS total cross section, as a function of $Q^2$.
b) Values of the exponent in fits of the form $W^\delta$ to the $W$
dependence of the cross section for a variety of vector mesons, as a function of $Q^2+M^2$.}
\label{fig:DISdelta}
\end{figure}

To the extent that we can model diffraction by the exchange of a colourless two-gluon state, we would expect a difference between the $W$ dependence of vector
meson production and inclusive DIS. This is illustrated in Fig.~\ref{fig:DISdelta} shows the values of
fits to the $W$-dependence of the inclusive DIS and vector-meson cross sections
against $Q^2$ and $Q^2 + M^2$, respectively. For $Q^2 + M^2$ greater
than about 5 GeV$^2$, the value of $\delta$ is indeed about twice that at the same value of $Q^2$ in inclusive DIS, as would be expected in the simple picture of two-gluon exchange.  

\begin{figure}[ht]
\begin{center} 
\centerline{\epsfig{file=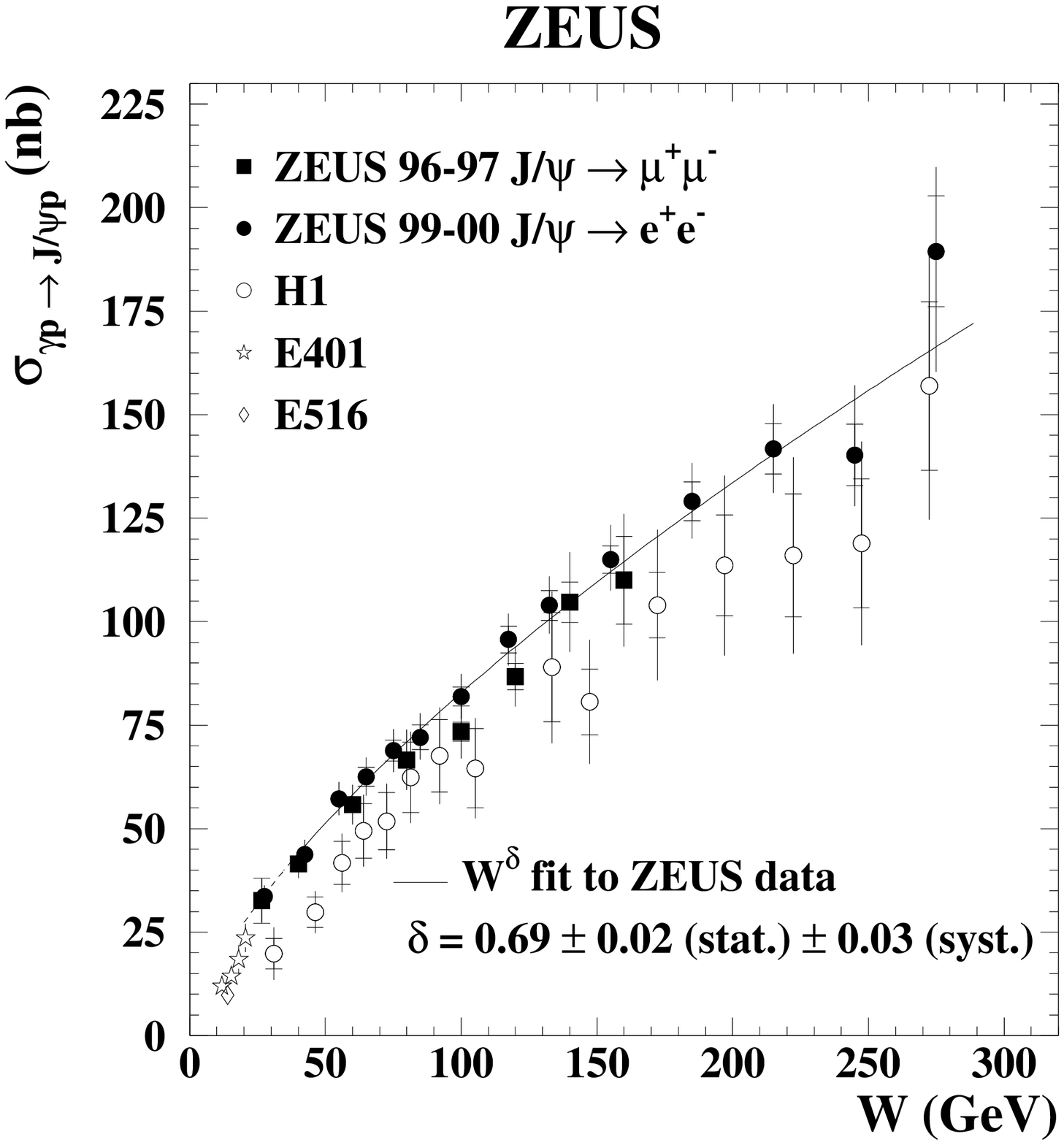,%
      width=6cm,%
      height=6cm%
        }
\epsfig{file=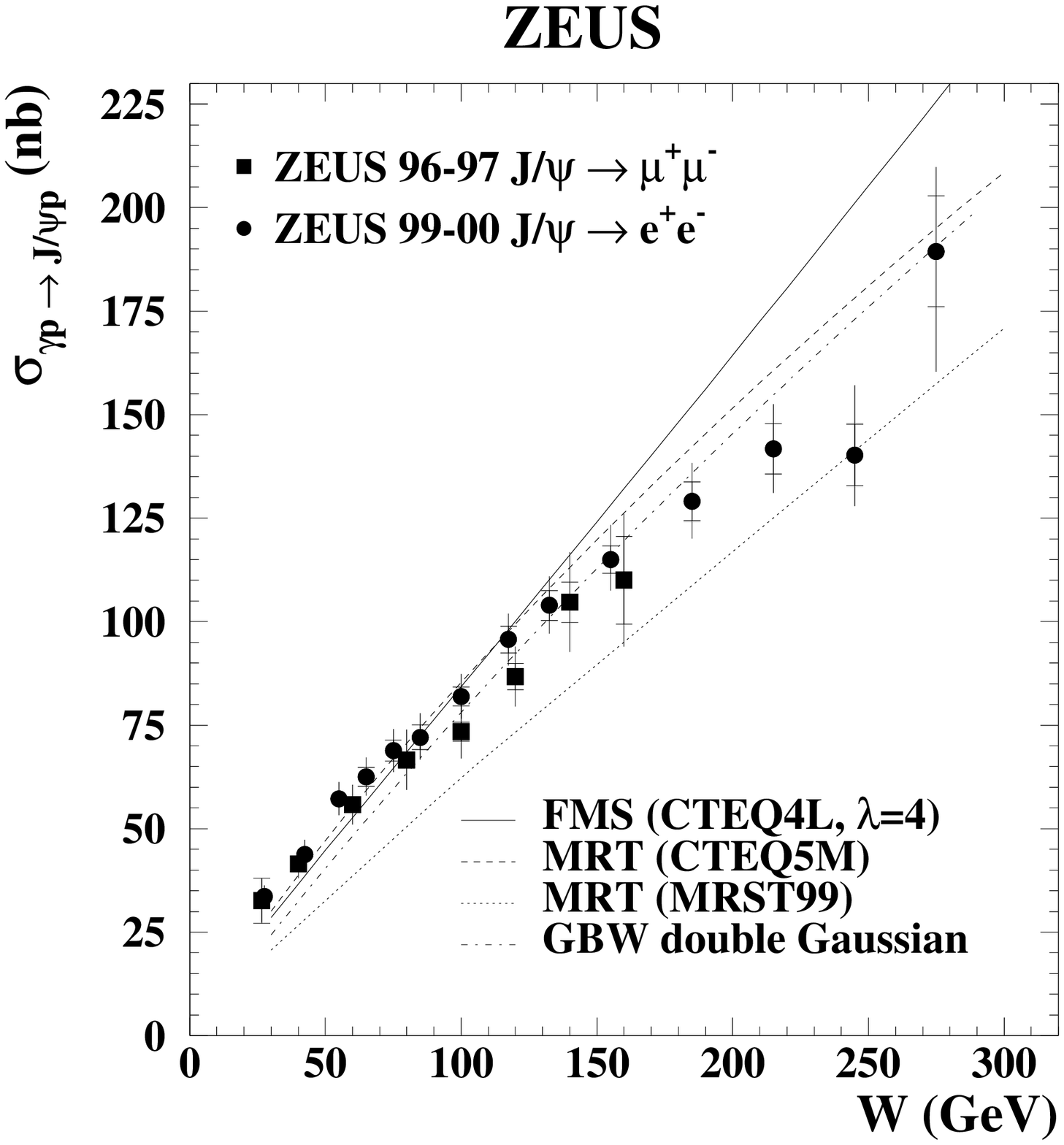,%
      width=6cm,%
      height=6cm,%
      clip=%
        }}
\end{center}
\caption{a) The ZEUS and H1 data on $J/\psi$ photoproduction, together
with the data from two fixed target experiments as a function of
$W$. The curve shows a fit to the form $W^\delta$. b) The ZEUS
data together with curves showing the predictions using the stated gluon pdfs
from the models of Frankfurt, McDermott and Strikman~\protect\cite{FMS} and Martin, Ryskin and Teubner~\protect\cite{MRT}.}
\label{fig:JpsivsW}
\end{figure}

Provided that we have a scale sufficiently hard that pQCD is applicable, 
we can also use vector meson production to probe the gluon density in the proton. Since the cross section is proportional to the gluon density squared,
such a determination should in principle be much more sensitive to the gluon density than, for example, scaling violations in deep inelastic scattering. 
This is illustrated in Fig.~\ref{fig:JpsivsW}, which shows ZEUS and H1 data
for $J/\psi$ photoproduction as a function of $W$. We have seen already that
the $J/\psi$ mass is sufficiently large to guarantee that pQCD is applicable
even in photoproduction. The quality of the data is sufficiently high that
it is in principle sensitive to the gluon density. Unfortunately, the wave-function of the
$J/\psi$ must be modelled, which leads to substantial uncertainty 
in the model predictions, so that this data has not as yet been used in
global fits to constrain the gluon density. The good agreement between
at least some version of the models and the data, 
as shown in Fig.~\ref{fig:JpsivsW}, does however indicate that
the gluon determined in the global fits to DIS data is indeed also 
able to explain the dynamics of this completely different diffractive process.  

\begin{figure}[ht]

\begin{center} 
\epsfig{file=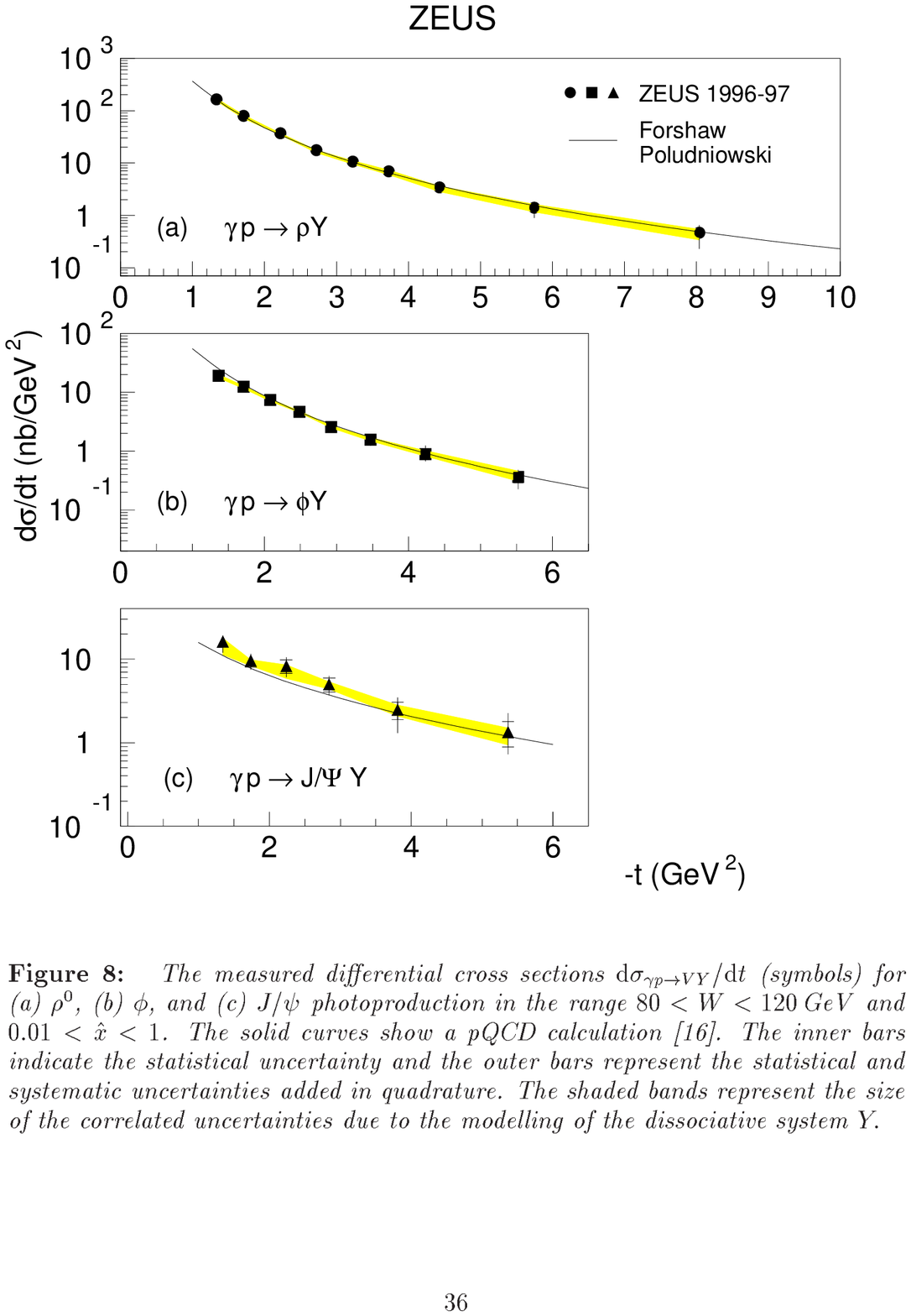,%
      width=11cm,%
      height=8cm,%
      clip=%
        }
\end{center}
\caption{The ZEUS data on vector-meson production at high $-t$.
The curve shows a fit to a BFKL-model by Forshaw and Poludniowski.}
\label{fig:VM:tJeff}
\end{figure}

Finally, it is interesting to investigate whether $t$ can also
provide a hard scale for pQCD. ZEUS has precise data out to $-t \sim 
12$ GeV$^2$ for $\rho$ production and also to beyond 6 GeV$^2$ for
$\phi$ and $J/\psi$. The ratio of the cross sections for $\phi$ and
$\rho$ starts somewhat below the $SU(4)$ expectation but reaches it
quite quickly, by $-t \sim 3 - 4$ GeV$^2$. In contrast, the $J/\psi$ ratio
remains below the $SU(4)$ prediction for much longer, hardly reaching it
even for $-t \sim 6$ GeV$^2$. Figure~\ref{fig:VM:tJeff} shows the ZEUS
data for all three vector mesons as a function of $-t$ compared to a
two-gluon exchange model and to a model in which a BFKL gluon ladder
is exchanged~\cite{JeffBFKL}. It can be seen that the two-gluon model completely fails to reproduce the data both in magnitude and in shape, whereas the
BFKL model, which has been fit to the ZEUS data, gives an excellent
fit for all three mesons. Although some of the model assumptions, such
as a fixed $\alpha_s$ and a $\delta$-function wave-function for the
light vector mesons, are somewhat questionable, the fact that this
BFKL model fits the data whereas conventional pQCD fails is very striking.
It can thus be concluded that $t$ does form an appropriate
hard scale for pQCD calculations.
However, the ratios of ZEUS vector-meson cross sections behave
differently as a function of $t$ compared to $Q^2$, so that
either these ratios are different at asymptotic values of the
two variables, or $t$ and $Q^2$ are not equivalent hard scales.   

\subsection{Deeply virtual Compton scattering}
\label{sec:DVCS}

Deeply virtual Compton scattering (DVCS) is an interesting process in
that it is the simplest possible non-elastic diffractive process.
It consists of diffractive scattering of the virtual
photon from the proton, putting the virtual photon onto mass shell so 
that the final state consists of the initial proton and positron plus
a photon. As such, no fragmentation or hadronic wave-functions
complicate the process. 
The Bethe-Heitler QED process also leads to the
same final state, which is both a blessing and a curse. The curse is 
that is necessary to separate the two processes; the blessing is that
the fact of identical final states leads to interference.
Since the amplitude and phase of the QED process is calculable, in principle
this opens the door to the determination of the unknown QCD amplitude
that governs DVCS. However, we are still some considerable distance
from this goal.

Given that a virtual photon participates in the collision with
the parton in the proton and that it emerges on mass shell, it is clear
that the scattered parton that must be re-integrated into the final-state
proton must undergo a change in its four-momentum corresponding
to a change in its $x$ value. Thus the DVCS
process is  sensitive to the so-called ``skewed parton'' distributions
inside the proton, which can be thought of as the cross-correlation function
between partons of fractional momentum $x_1$ and $x_2$. Thus much unique
information can be obtained by studying these processes.

\begin{figure}[ht]
\begin{center} 
\centerline{{\vspace{1cm}\epsfig{file=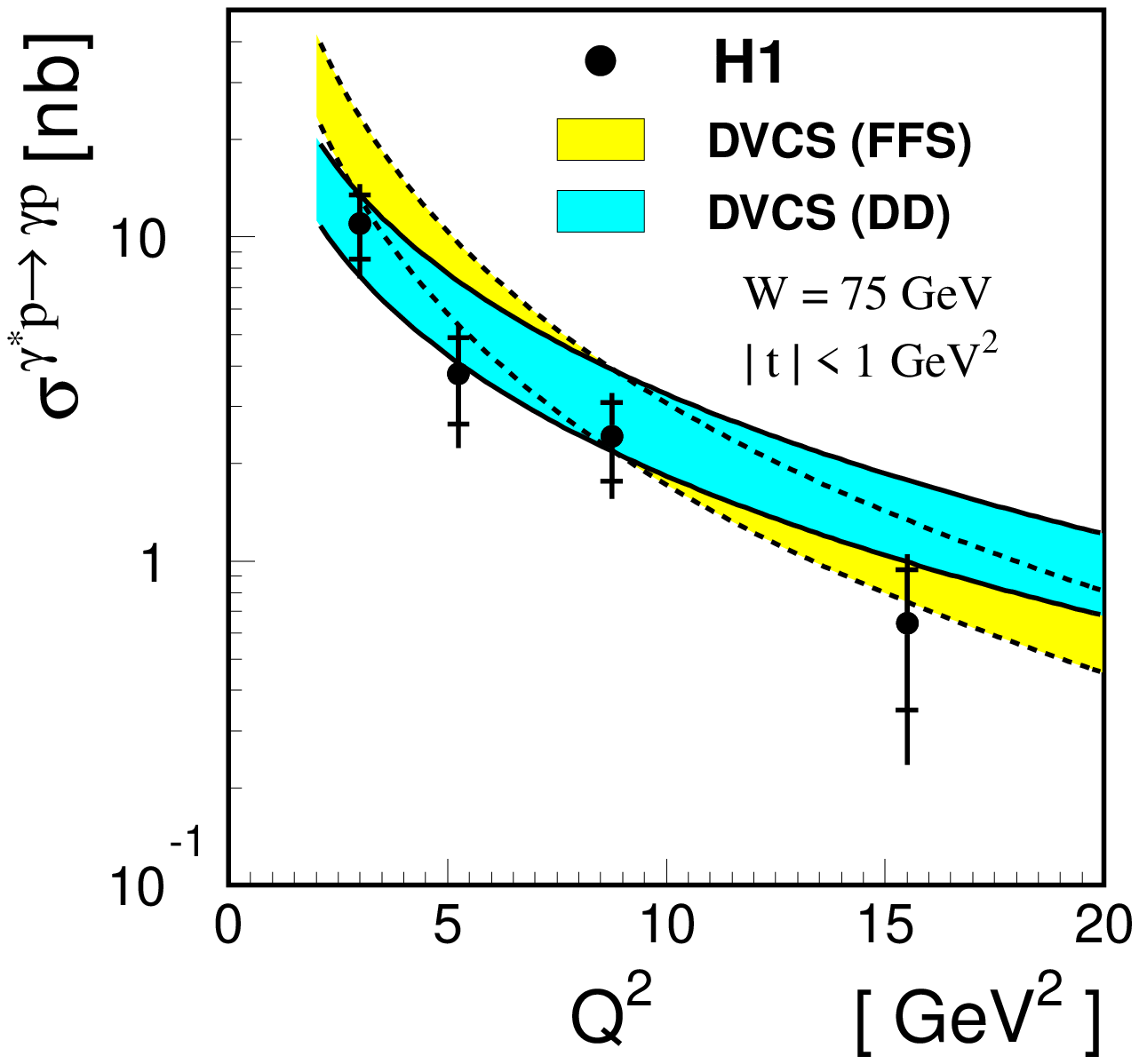,%
      width=6cm,%
      height=6cm%
         }}
\epsfig{file=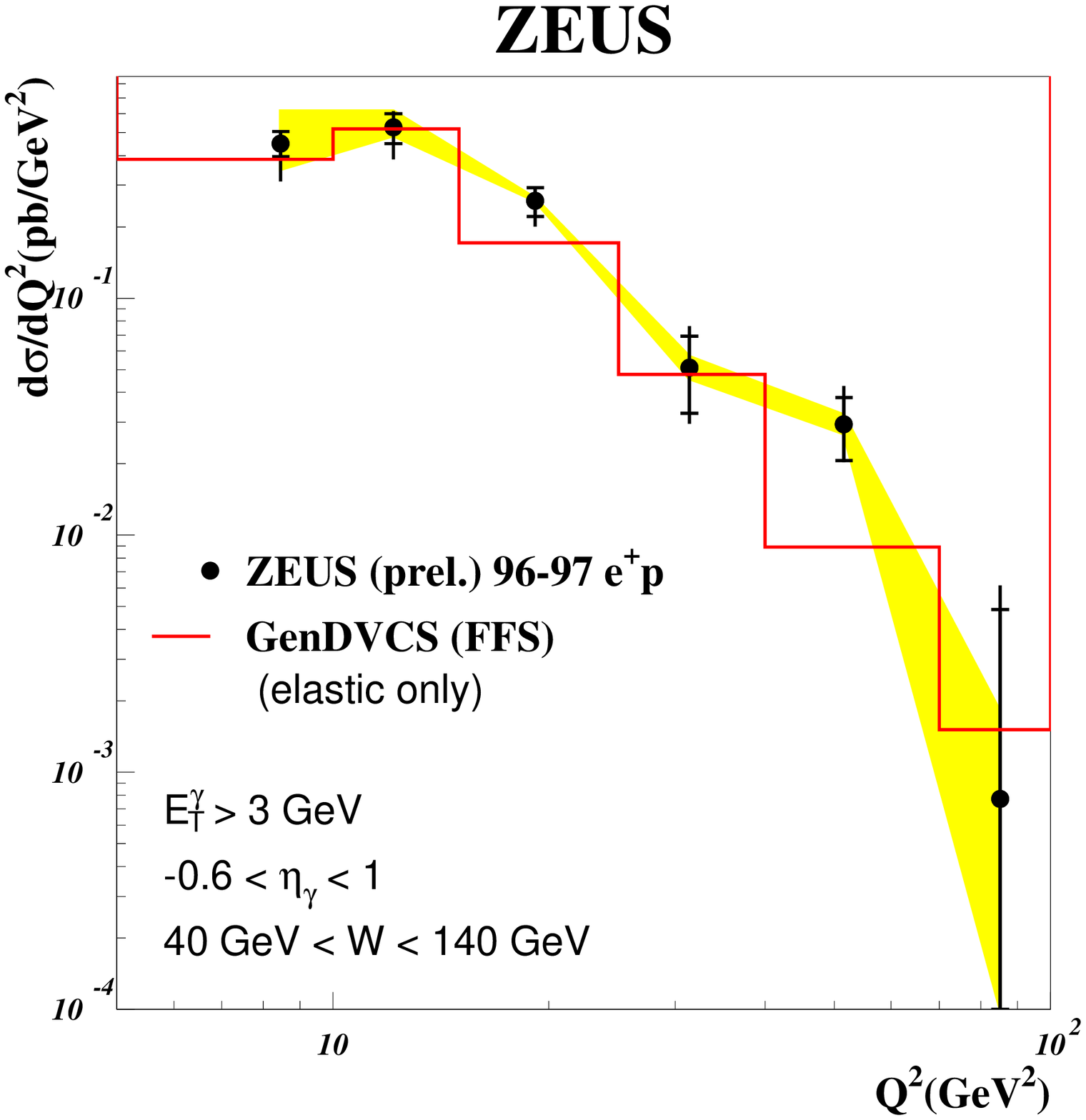,%
      width=5.5cm,%
      height=5.5cm,%
      clip=%
       }}
\end{center}
\caption{a) The cross section for deeply virtual Compton scattering
as a function of $Q^2$ as measured by the H1 collaboration. The
dotted area corresponds to the prediction of Frankfurt, Freund and Strikman,
while the solid area is the prediction of Donnachie and Dosch.
b) The DVCS cross section as measured by ZEUS. The curve shows
the prediction for the elastic process by Frankfurt, Freund and Strikman.}
\label{fig:H1:DVCS}
\end{figure}

The DVCS process was first seen at HERA by ZEUS~\cite{ZEUS-DVCS-conf}. The
H1 collaboration has published results on the observation of this process
and the first measurement of the cross section~\cite{H1-DVCS-pub}. The background from the Bethe-Heitler process can be subtracted
by utilising the kinematic characteristics of the two processes. In
DVCS, the photon is normally produced at a large angle to the incident
beam directions and the positron at a small angle, whereas, for the
Bethe-Heitler process, the reverse is the case. The sum
of Monte Carlo generators describing the two processes gives
a good description of the kinematic quantities of the data, giving confidence
that the DVCS cross section can be extracted. The
H1 and ZEUS cross sections are shown in Fig.~\ref{fig:H1:DVCS}.
It can be seen that the cross sections are completely dominated by statistical
errors and that both collaborations find good agreement with theoretical
models, notably that of Frankfurt, Freund and Strikman~\cite{FFS} and
Donnachie and Dosch~\cite{DD}. As more data is collected, the possibility of using this process
to determine skewed parton distributions will make it a fruitful area of study at HERA II.  

There are many more areas of diffraction, in particular
the diffractive structure functions, which I have not been able to
cover. The study of diffraction has turned out 
to be one of the most exciting and rich
areas of HERA physics. It will continue at HERA II and will hopefully
allow us to make progress in our understanding of soft interactions
and the related, and fundamental, question of understanding confinement in QCD. 

\section{Heavy quark production}
\label{sec:HQ}

\begin{figure}[ht]
\begin{center} 
\epsfig{file=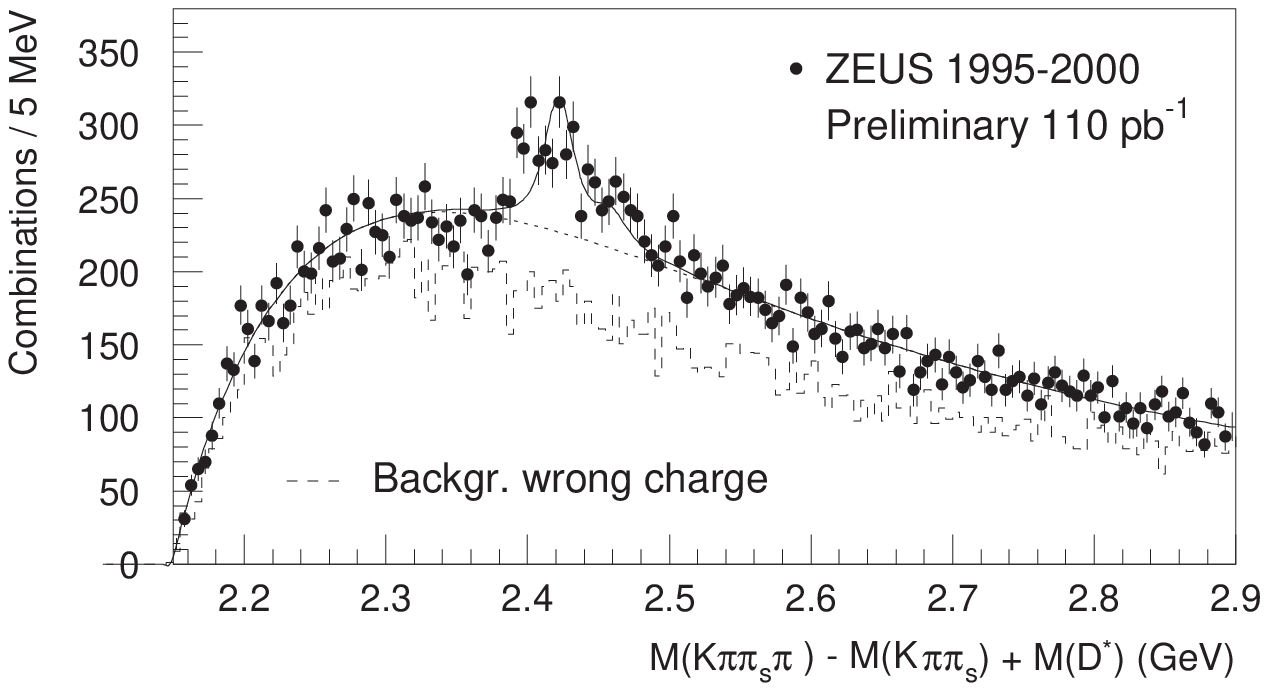,%
      width=11cm,%
      height=8cm%
        }
\end{center}
\caption{The $D^*\pi$ spectrum from ZEUS, where the $D^*$ decays via $D\pi$ and
the $D$ via $K\pi$. To improve the accuracy of the reconstruction, the
effective mass of the $K\pi\pi\pi$ system minus the effective mass of the $K\pi\pi$ system that forms the $D^*$ plus the $D^*$ mass is plotted. The curves show a polynomial background plus Gaussians for the known $D^0_1$ resonance at 2.420 GeV and the $D_2^*$ at
2.460 GeV. Another structure at just above 2.4 GeV also seems to be visible.}
\label{fig:ZEUS:charmbumps}
\end{figure}

We have already touched upon the production of charm quarks at HERA in
the discussion of the $F_2^c$ in Section~\ref{sec:f2charm}. In fact, copious
amounts of charm are produced in photoproduction at HERA, so that for many channels, HERA is a ``charm factory''. The ZEUS collaboration has made
several contributions to charm spectroscopy results: as an example,
Fig.~\ref{fig:ZEUS:charmbumps} shows the $D^*\pi$ spectrum from 110 pb$^{-1}$
of data. A rather complex set of structures can be observed between 2.4 and 2.5
GeV, some of which correspond to known resonances and some of which do not. Work
continues to understand this complex area; the sensitivity of ZEUS in
this sort of investigation is similar to that of the LEP experiments and CLEO.

\begin{figure}[ht]
\begin{center} 
\epsfig{file=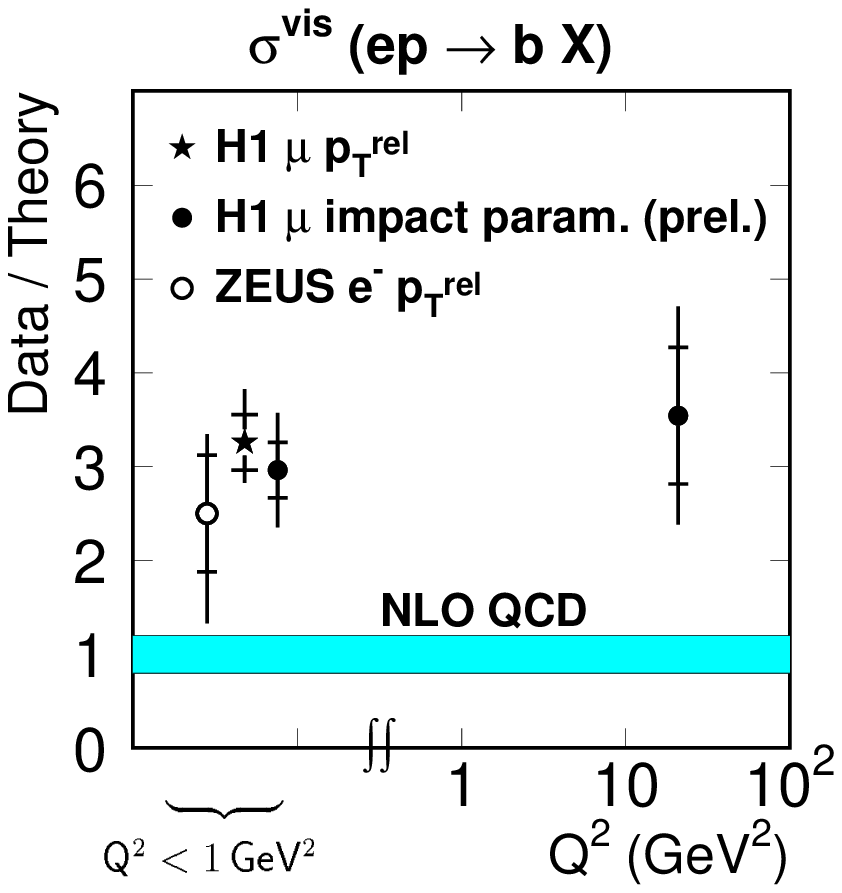,%
      width=7cm,%
      height=7cm%
        }
\end{center}
\caption{The ratio of the measured cross section and the theoretical prediction,
plotted against $Q^2$, from H1 on beauty production in photoproduction and DIS and from ZEUS in photoproduction,. The band shows the
uncertainty on the predictions of perturbative QCD.}
\label{fig:H1ZEUS:b}
\end{figure}

Both H1 and ZEUS have observed beauty production in photoproduction; H1 has
also published a cross section in DIS. The identification of the $b$
signal is based on the use of high-transverse-momentum leptons for ZEUS; H1
also use tracks with large impact parameter as measured in their silicon vertex detector. The results as a function of $Q^2$ are shown in Fig.~\ref{fig:H1ZEUS:b}. It can be
seen that the QCD predictions are substantially below the data for all $Q^2$,
thus joining a pattern also seen in proton-antiproton and photon-photon
collisions. The tendency for pQCD to fail to predict $B$ cross sections, which
naively would be thought to be an area in which it should work well, is becoming
increasingly interesting. The advent of HERA II will make an enormous difference
to the precision of this type of measurement and should allow a stringent test
of the theory of heavy-quark production.  

\section{High-$Q^2$ phenomena}
\label{sec:highQ2}

\begin{figure}[ht]
\begin{center}
\epsfig{file=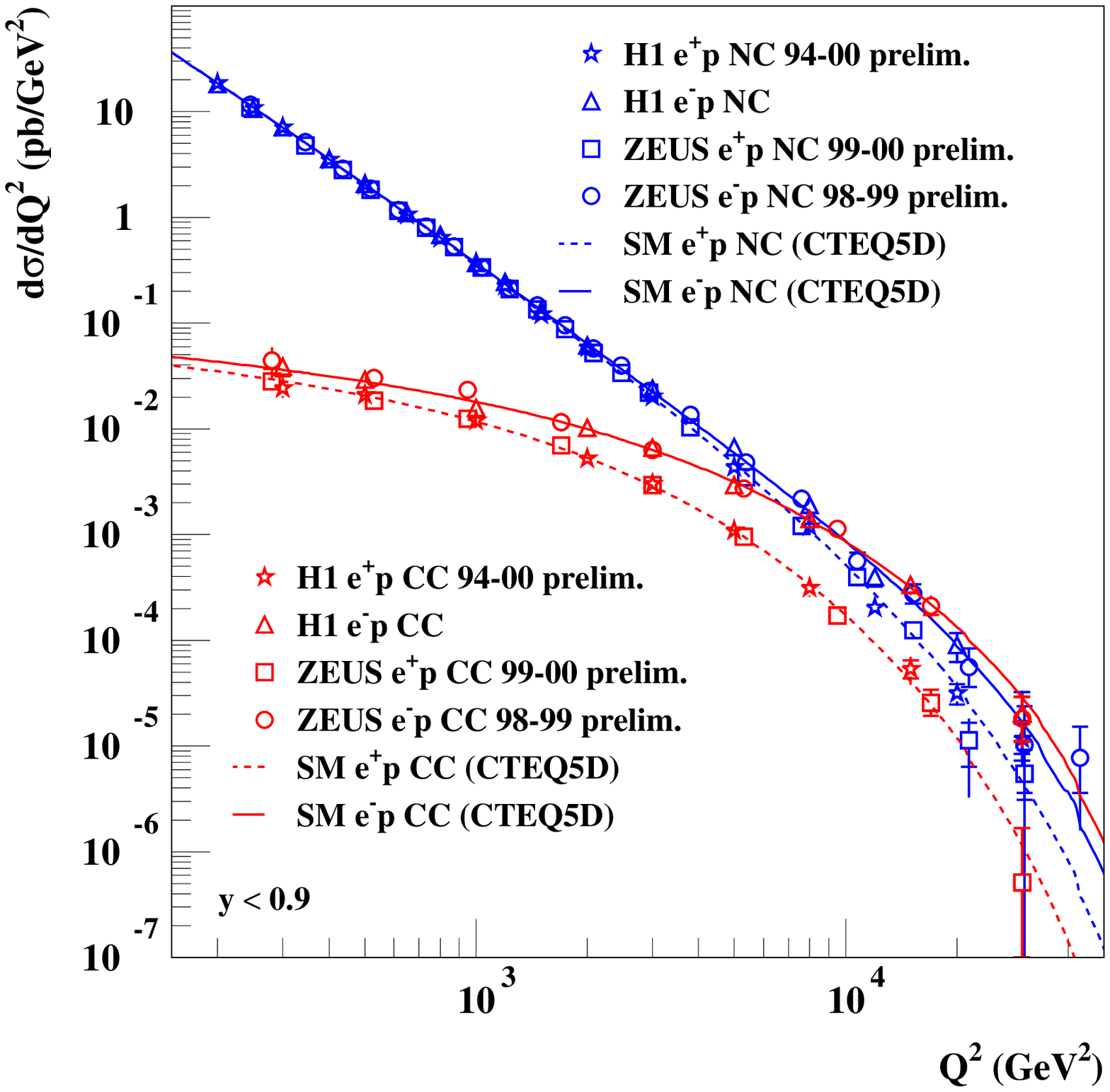,%
           width=11cm,%
           height=11cm%
        }
\end{center}
\caption{Charged and neutral current differential cross sections for
$e^\pm$ scattering as a function of $Q^2$ from H1 and ZEUS.}
\label{fig-H1ZEUS-CCNC}
\end{figure}

HERA provides an unique opportunity to study the electroweak 
interaction at $Q^2$ sufficiently high that the charged and neutral 
currents are of similar strength. 
Figure~\ref{fig-H1ZEUS-CCNC} shows the differential cross-sections for the 
charged and neutral currents as a function of $Q^2$ from H1 and ZEUS. 
It can be seen that, for $e^-p$ interactions, 
these two processes become of equal strength at $Q^2 \sim M_Z^2 \sim 10^4$ GeV$^2$. For $e^+p$ interactions, 
the charged current cross-section approaches the neutral current 
cross-section, but remains below it. The features of
this plot can be explained  
by inspection of Eq.~\ref{eq:Fl:sigma}, together with
Eqs.~\ref{eq-CCem} and \ref{eq-CCep} below:
\begin{eqnarray} 
\left. \frac{d^2 \sigma}{dx dQ^2} \right|_{\emin}^{CC} & = &   
\frac{G_F^2}{2\pi} \left( \frac{\MWtwo}{\MWtwo+Q^2}\right)^2 \cdot
 \nonumber \\ 
 & & 2x \{u(x) + c(x) + (1-y)^2(\overline{d}(x) + \overline{s}(x))\} 
\label{eq-CCem}  
\end{eqnarray} 

\begin{eqnarray} 
\left. \frac{d^2 \sigma}{dx dQ^2} \right|_{\eplus}^{CC} & = &   
\frac{G_F^2}{2\pi} \left(\frac{\MWtwo}{\MWtwo+Q^2}\right)^2 \cdot
\nonumber \\ 
 & & 2x\{\overline{u}(x) +\overline{c}(x) + (1-y)^2(d(x) + s(x))\} 
\label{eq-CCep} 
\end{eqnarray}

For the charged current case, the smaller 
size of the $e^+p$ cross-section compared to $e^-p$ is related to the
fact that, at high $Q^2$, Eq.~\ref{eq:yq2sx}
implies that both $ x, y \rightarrow  
1$.  There are two main contributory factors to the cross-section
difference that flow from this. First, there are twice as many $u$ valence quarks inside the proton that can couple to $W^-$ as $d$ quarks that
can couple to $W^+$. Secondly, 
the $(1-y)^2$ terms in Eqs.~\ref{eq-CCem} and 
\ref{eq-CCep}, which 
arise from the $V - A$ helicity structure of the charged weak current, 
imply that the valence-quark contribution, which is dominant at high $Q^2$,
is suppressed for the positron case but not for electrons.

\begin{figure}[ht]
\begin{center}
\epsfig{file=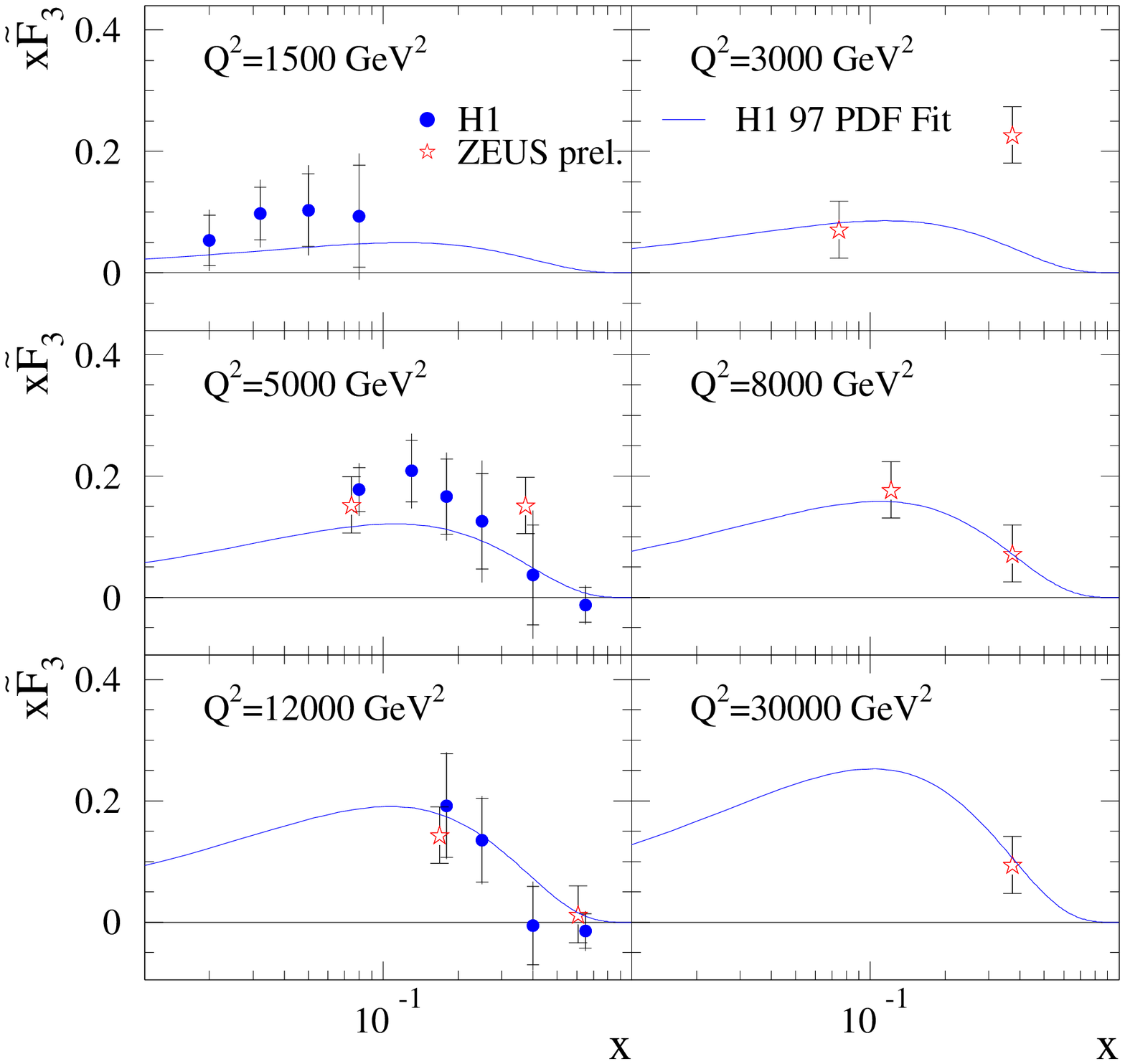,%
      width=11cm,%
      height=11cm%
        }
\end{center}
\caption{The $xF_3$ structure function as determined by
H1 and ZEUS as a function of $x$ in six bins of $Q^2$.}
\label{fig-H1ZEUS-xF3}
\end{figure}

The difference between the electron and positron neutral current
cross sections shown in Eq.~\ref{eq:Fl:sigma} allows the determination
of the parity-violating structure function $xF_3$ by taking the difference
of the cross sections. The results~\cite{H1-xF3} are shown in 
Fig.~\ref{fig-H1ZEUS-xF3}. Since its determination requires the subtraction of
two quantities that are almost equal, it is dominated by
statistical uncertainties, which are in turn dominated by the fact that
the electron data sample that has so far been obtained at HERA is much smaller than that for positrons.

\begin{figure}[ht]
\begin{center}
\epsfig{file=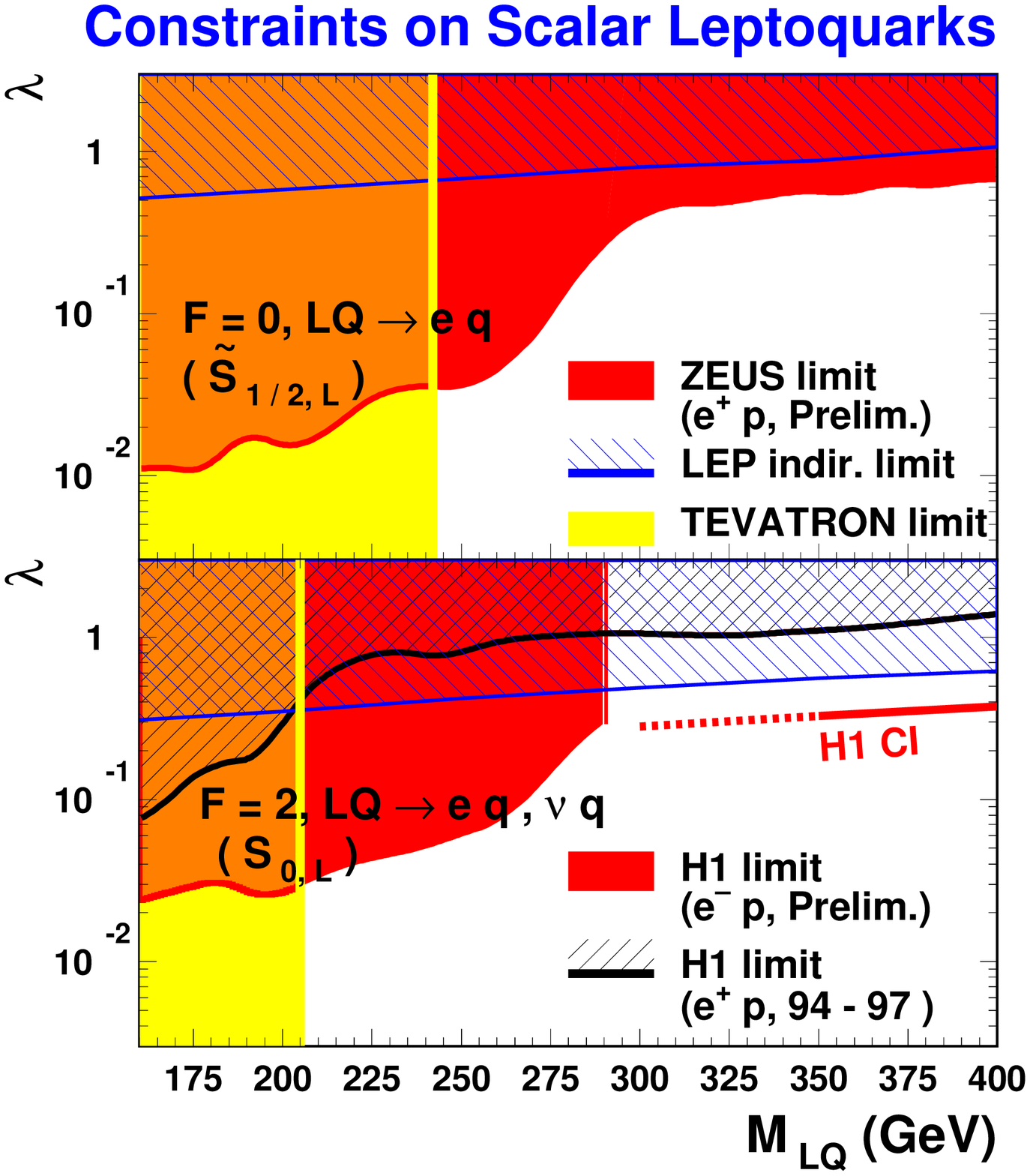,%
      width=8cm,%
      height=8cm%
        }
\end{center}
\caption{Limits on coupling strength $\lambda$ versus mass $M_{LQ}$  
for leptoquarks. The top plot shows limits
for fermion number = 0 leptoquarks decaying into the $eq$ final state from ZEUS. The lower plot shows limits from H1 for fermion 
number = 2 leptoquarks decaying into both $eq$ and $\nu q$ final states.
Also shown
are limits obtained from the Tevatron (yellow shaded area) and LEP
(blue striped area). These leptoquark species have identical quantum numbers
to squarks that violate R-parity.}
\label{fig-ZEUSH1-LQ}
\end{figure}

The high-$Q^2$ regime is also interesting since possible
new states from electron-quark fusion (e.g.\
leptoquarks) have masses given by $M^2 \sim sx$ and since the sensitivity
to the effects of new currents is maximised. An example of the
sensitivity that can be obtained at HERA is shown in Fig.~\ref{fig-ZEUSH1-LQ},
which shows the mass against coupling limits for two varieties of scalar leptoquark. Both H1 and ZEUS have comparable 
limits for a whole range of
such states with differing quantum numbers. It can be seen from
Fig.~\ref{fig-ZEUSH1-LQ}, and it is generally the case, that
for some states, in particular in $R$-parity-violating supersymmetry
models or leptoquarks, HERA has higher sensitivity than either LEP or the
Tevatron.  

Limits on excited leptons and quarks have also been obtained by ZEUS and
H1, which extend the limits from the LEP experiments considerably
beyond the LEP II centre-of-mass energy.

\begin{figure}[ht]
\begin{center}
\epsfig{file=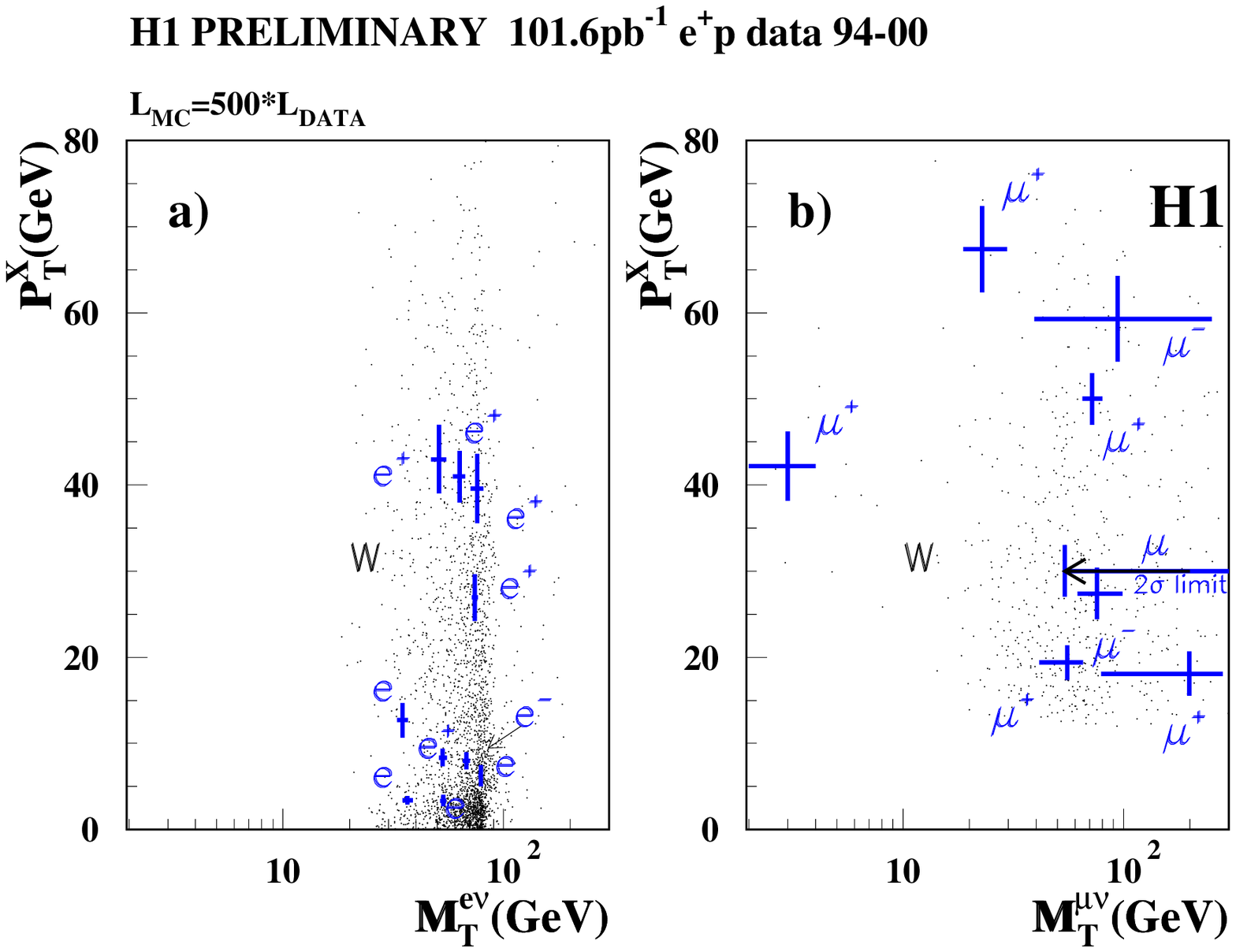,%
      width=11cm,%
      height=9cm%
        }
\end{center}
\caption{Distribution of transverse mass versus the $p_T$ of the
hadronic system for H1 events containing a) isolated electrons and b) isolated
muons. The dots show the distribution of Standard Model
$W$ Monte Carlo events corresponding to a 
luminosity 500 times that of the data.}
\label{fig-H1leptons}
\end{figure}

As well as stringent limits on new phenomena, the HERA data also show
intriguing features which may be signatures for new physics. 
The H1 collaboration has observed a class of
events that have isolated charged leptons with
large missing transverse momentum. Figure~\ref{fig-H1leptons} shows the distribution of the transverse
momentum of the hadronic system, $p_T^X$, against its transverse mass, 
separately for electrons (or positrons) and muons in such events. Also shown are the expectations from the Standard Model background, which is dominated by
single $W$ production. 

It can be seen that the distribution of the events
is rather different to the Standard Model expectation. Furthermore,
for the transverse mass of the
hadronic system greater than 25 GeV, H1 sees four electron and 
six muon events, compared to Standard Model expectations of
1.3 and 1.5 events, respectively. Unfortunately, this exciting
observation is not confirmed by ZEUS, which, for the same cut in
$p_T^X$, sees one event in each category compared to the Standard
Model expectation
of 1.1 and 1.3, respectively. Intensive discussions between the two
experiments have not revealed any reason why H1 might artificially
produce such an excess nor why ZEUS should not observe it. It would therefore seem that there must be an unlikely fluctuation: either the H1
observation is an upward fluctuation from the Standard Model, or
ZEUS has suffered a downward fluctuation from a signal for new physics. 
More data from HERA II will be required to resolve this puzzle.

\begin{figure}[ht]
\begin{center}
\epsfig{file=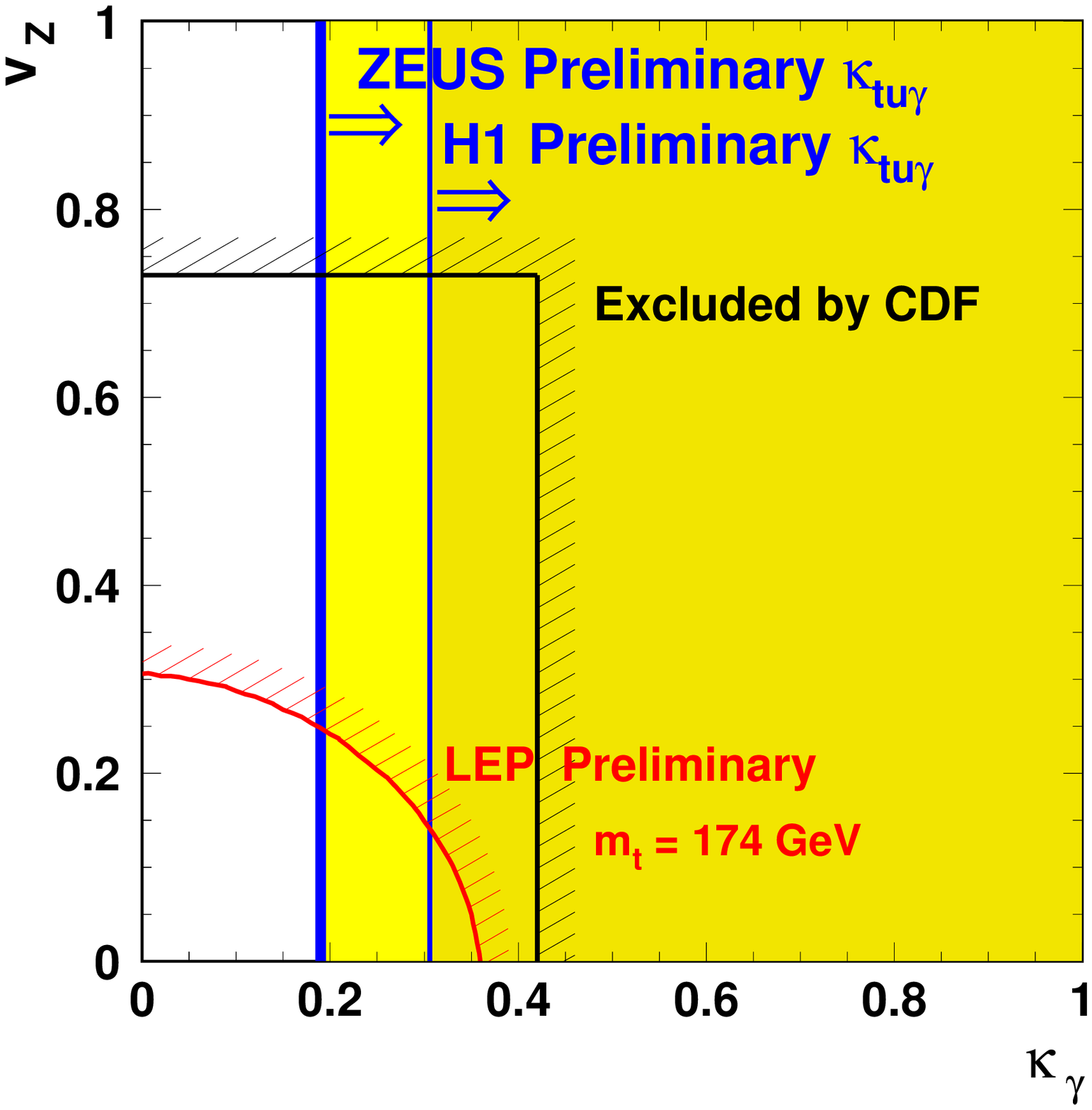,%
      width=8cm,%
      height=8cm%
      }
\end{center}
\caption{Limits on flavour-changing neutral current coupling strength for
single top production. Limits from H1 and ZEUS are plotted 
on the two-dimensional space of the photon and $Z$ coupling strengths.
Also shown are similar limits from the combined LEP experiments 
(to the right of the shaded curve) and 
from CDF (to the right of the black shaded lines).}
\label{fig-H1ZEUS-FCNC}
\end{figure}

One possible source of an excess of events with isolated leptons with
missing transverse momentum would be from a flavour-changing neutral
current process producing single top quarks. Both H1 and ZEUS have
used the samples described above to put limits on the FCNC couplings
of the $\gamma$ to light quark-top quark vertices. The
results are shown in Fig.~\ref{fig-H1ZEUS-FCNC}. Also shown are
the limits from LEP and CDF, which are complementary to those
from HERA, in the sense that, since the $Z$-exchange cross section at
HERA is so much smaller than that for $\gamma$ exchange, the HERA data limit only the photon coupling.
\section{HERA II physics}
\label{sec:HERAII}
Since many of the physics results discussed above, particularly those at
high $Q^2$, are statistics limited, there is a clear physics case for a significant increase in integrated luminosity for H1 and ZEUS. There are
also other interesting physics investigations possible at HERA that have
not yet been carried out. For example,
there is a natural build-up of
transverse polarisation of the lepton beam in HERA that occurs through the Sokholov-Ternov effect~\cite{S-Teffect}. As very successfully
demonstrated at HERMES using gas targets, this transverse polarisation can be
rotated into the longitudinal direction and utilised to do physics. The
installation of spin rotators in H1 and ZEUS would allow polarisation
studies to be carried out at very much higher $Q^2$. This is particularly
interesting to study the chiral properties of the electroweak interaction.
For these and several other reasons, it was decided to embark on a major
upgrade of both the HERA accelerator and the H1 and ZEUS detectors. The
aim of the HERA II programme is to produce a factor of approximately five increase in luminosity and accumulate 1 fb$^{-1}$ of data with 
both electron and positron collisions in both longitudinal polarisation states.

The changes to the HERA accelerator include the replacement of 480 meters of the vacuum system and the design and installation of almost 80 magnets in the region around the H1 and ZEUS interaction points. In particular, superconducting
quadrupole focussing elements were inserted inside both detectors to reduce
the beam emittance and spin rotators were installed on either side
of the H1 and ZEUS interaction regions. 

Both the ZEUS and H1 detectors have undergone a massive programme of 
consolidation and repair work, as well as major detector upgrades. 
As an example, I discuss briefly the changes made to ZEUS; the general
thrust of the upgrade is similar in the two detectors, although the details are different.

\subsection{Upgrades to ZEUS for HERA II}
\label{sec:ZEUSupgrade}
The ZEUS upgrades have concentrated in three main areas: the vertex region;
the forward direction; and the luminosity monitoring.

\subsubsection{The vertex region}
\label{sec:ZEUSvertex}

\begin{figure}[ht]
\begin{center}
\epsfig{file=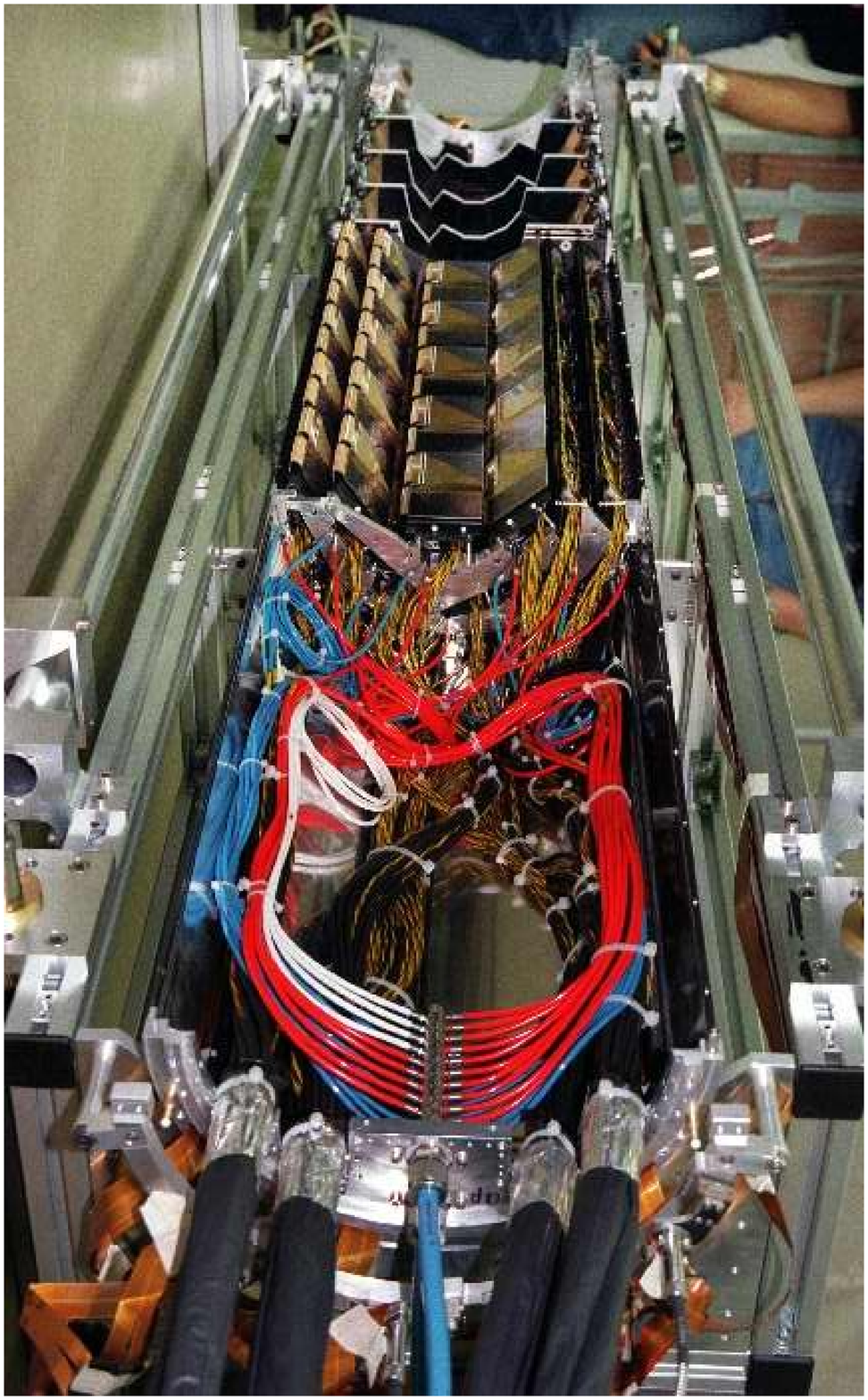,%
      width=6cm,%
      height=10cm%
        }
\end{center}
\caption{A photograph of one half of the MVD, showing the barrel ladders,
one half of each of the four forward wheels and the cables and services.}
\label{fig:MVD-barrel-photo}

\end{figure}

The tagging of the large flux of heavy quarks (charm and beauty) produced at HERA II can be greatly enhanced by the installation of a high-precision
charged-particle detector as close as possible to a thin beampipe. 
The ZEUS MVD~\cite{nim:a435:34,nim:a473:26} 
consists of 20 $\mu$m pitch $n$-type silicon-strip detectors 
with $p^+$-type implants. The readout pitch is 120 $\mu$m, leading
to more than 200,000 readout channels, which are digitised by a
custom-built clock, control and ADC system. 
The detectors are organised in two main groups: a ``barrel'', which surrounds the elliptical 2 mm-thick ($\sim 1.1$\% of a radiation length) 
aluminium-beryllium beam-pipe; and four ``wheels'', consisting
of wedge-shaped detectors mounted 
perpendicular to the beam-line in the forward direction from the
interaction point. 
Figure~\ref{fig:MVD-barrel-photo} shows one half
of the MVD before installation at DESY. In the barrel region, the ladders, each of which consists of five silicon detectors, and halves of the
four forward ``wheels'', can be seen, as can the dense array of readout and
services cables and the cooling system. The complete MVD was installed in ZEUS
in April 2001 and has been fully integrated with the ZEUS DAQ system;
both cosmic-ray and beam-related data have been taken. 

\begin{figure}[ht]
\begin{center}
\epsfig{file=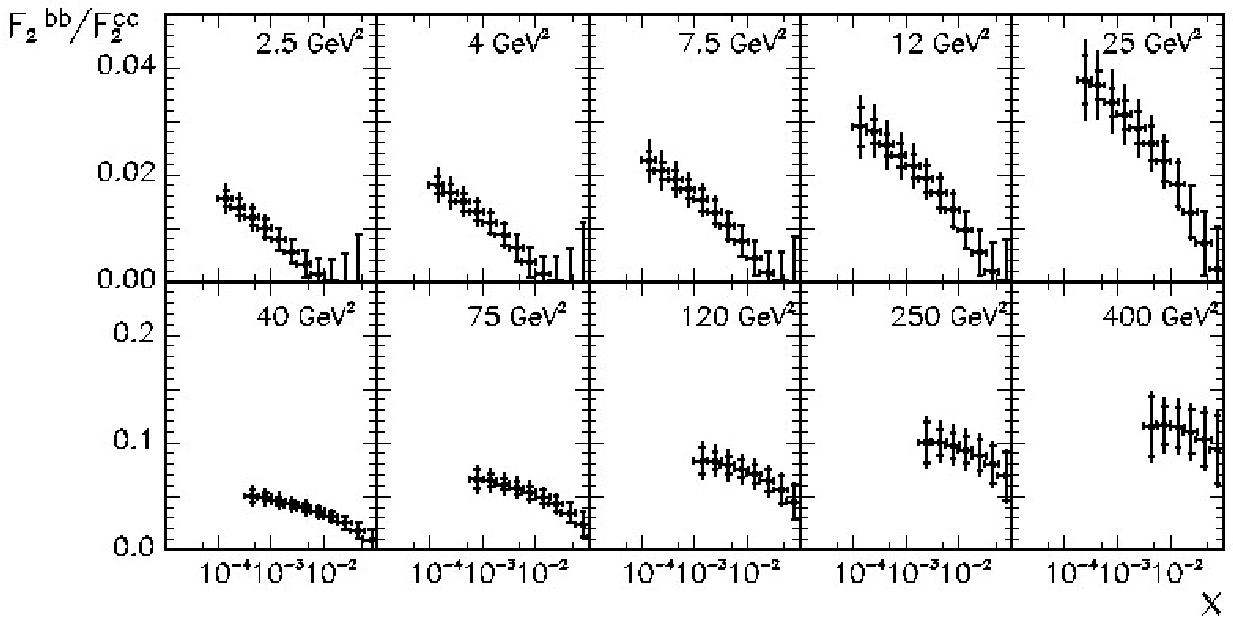,%
      width=12cm,%
      height=6cm%
        }
\end{center}
\caption{The MC prediction for the ratio of the contribution to $F_2$
of $b$-quark to $c$-quark production
in $Q^2$ bins as a function of $x$ after 500 pb$^{-1}$ of data at
HERA II.}
\label{fig:F2b-c-ratio}
\end{figure}

The physics programme addressed by the MVD is that of the flavour decomposition of the proton and photon and the search for physics beyond the Standard Model.
The large increase in luminosity of HERA II, together with the ability to
tag heavy-quark decays in the MVD, should greatly improve 
the measurement of $F_2^{c}$ discussed in Section~\ref{sec:f2charm}. After about 500~pb$^{-1}$, 
an uncertainty of around the 2\% currently measured on $F_2$ should
be obtained. In addition, $b$-quark production can be measured precisely; a
Monte Carlo simulation~\cite{proc:hera:1995:89} of a measurement of
$F_2^{b}/F_2^c$ after 500 
pb$^{-1}$ is shown in Fig.~\ref{fig:F2b-c-ratio}. 

It should also be possible, from a combination of neutral and charged
current measurements, to separate out the $u,d,s,c,b$ and $g$ contribution to $F_2$. 

\subsubsection{Charged-particle tracking in the forward direction}
\label{sec-stt}

The higher luminosity expected at HERA II will increase the number of very
high-$Q^2$ events in which the electron or positron is scattered into the forward direction. It will also give access to rare processes, including
possible physics beyond the Standard Model, which tend to have 
forward jets and/or leptons. The pattern-recognition capabilities of the ZEUS Forward Tracker have therefore been improved by the replacement of two layers of transition-radiation detector by layers of straw tubes. The straws are approximately 7.5 mm in diameter and
range in length from around 20 cm to just over 1 m. They are constructed from
two layers of 50 $\mu$m kapton foil coated with a 0.2 $\mu$m layer of aluminium, surrounding a 50 $\mu$m wire
at the centre. The straws are arranged in
wedges consisting of three layers rotated with respect to each other to give three-dimensional reconstruction. Each of the two ``supermodules'' consists of four layers of such wedges. 

\subsubsection{Luminosity monitor}
\label{sec-lumi}

The measurement of luminosity at HERA II must cope with the greatly
increased synchrotron-radiation background and the higher probability for multiple bremsstrahlung photons in one beam crossing. To compensate for this, two devices, with very different systematic uncertainties, have been constructed. Both devices use the information from a small calorimeter placed around 6 m from the interaction point which detects the radiating electron.
It is hoped that the
reduction of systematic error that can be obtained from independent
luminosity measurements using very different techniques will allow
a precision of around 1\% to be attained. 

\subsection{Polarisation}
\label{sec-pol}

Polarisations of around 65\% have been
achieved at HERA I. It is hoped to increase the
accuracy with which the polarisation can be measured to $\delta P/P \sim$ 2\%
per bunch per minute. 
This will be achieved by a collaboration between H1, HERMES,
ZEUS and the HERA machine in the POL2000 project. The collaboration
has constructed two instruments, one to measure
the longitudinal polarisation and the other to measure the transverse polarisation. Both detect
asymmetries in back-scattered light from high-intensity polarised lasers.

\begin{figure}[ht]
\begin{center}
\epsfig{file=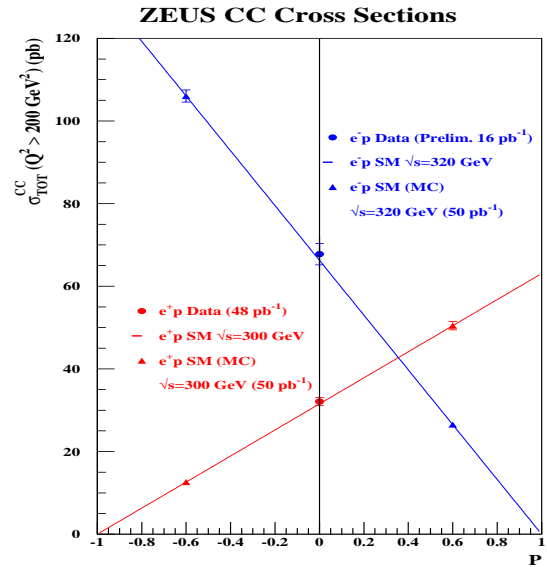,%
      width=8cm,%
      height=8cm,%
       clip=%
        }
\end{center}
\caption{ The cross section for charged current interactions. The points at
$P=0$ are obtained from ZEUS preliminary results at the indicated centre-of-mass
energies, while those at non-zero polarisation are Monte Carlo simulations of 
the expected accuracy in ZEUS assuming the Standard Model cross section for an integrated luminosity of 50 pb$^{-1}$ per point.}
\label{fig:e+-pol}
\end{figure}

The combination of high-precision measurements of both luminosity and
polarisation will be important in a wide range of HERA II physics,
particularly in the electroweak sector.
The charged current cross section should vanish for the appropriate
combinations of lepton charge and polarisation. A measurement at three
polarisations, such as shown in Fig.~\ref{fig:e+-pol}, even with
an integrated luminosity of only 50 pb$^{-1}$ per point, will provide
an accurate test of this prediction and thereby give sensitivity
to possible new currents outside the Standard Model.

Strong polarisation effects are also predicted at high $Q^2$ in
neutral current interactions, where, e.g. at $Q^2 = 10^4$ GeV$^2$ and 
$x = 0.2$, there is a factor of two difference between the 
predicted cross sections for left- and right-handed electrons.

In addition to the use of precise luminosity and polarisation information
in the study of electroweak processes, polarisation also offers an invaluable
tool in the study of possible signals beyond the Standard Model. Varying the polarisation to reduce the
cross sections of Standard Model processes can improve the
signal to background for new physics signals, such as leptoquarks or supersymmetric particles that violate $R$ parity, for which
HERA will be competitive with the
Tevatron for the next few years. 

\section{Summary}
\label{sec:summary}

We have seen that the study of deep inelastic scattering has been seminal in
the development of our current understanding of particle physics.
In the last decade, HERA I has changed our perception of QCD out
of all recognition. In many cases the precision of the data 
mandate NNL, or even high order, QCD predictions. The study
of diffraction and the transition region between soft and
hard physics may be the beginning of the era of quantitative study
of the central problem of the strong interaction, confinement.
The significant increase in expected luminosity and the use of new tools
such as polarisation promise that HERA II physics should continue
to provide both new results and surprises in the years to come.
          
\section*{Acknowledgements}
I am grateful to George Zoupanos and his colleagues
for organising a most stimulating summer school
in beautiful surroundings and with outstanding hospitality.

\end{document}